\newif\ifislncs
\islncsfalse

\ifislncs 
\documentclass{../llncs/llncs}
\else
\documentclass[11pt]{article}
\usepackage[margin=1.0in]{geometry}
\fi

\usepackage{ifthen}

\newboolean{lncs}   
\ifislncs 
\setboolean{lncs}{true}
\else
\setboolean{lncs}{false}
\fi

\newboolean{linenumbers}   
\setboolean{linenumbers}{false}

\ifthenelse{\boolean{lncs}}{
\usepackage{amsfonts}
\let\doendproof\endproof
\renewcommand\endproof{~\hfill$\qed$\doendproof}
}{
  \usepackage{amsmath,amssymb} 
  \usepackage{amsthm}
}

\usepackage[utf8]{inputenc}
\usepackage{graphicx,dblfloatfix}
\usepackage{subcaption}
\usepackage{url}

\usepackage{cite}
\usepackage{mathtools}
\usepackage[ruled,vlined,linesnumbered]{algorithm2e}
\usepackage{paralist}
\usepackage{todonotes}
\usepackage{thmtools, thm-restate}
\usepackage{xcolor}
\usepackage{multirow}

\usepackage{hyperref}

\usepackage[capitalize]{cleveref}
\crefname{theorem}{Theorem}{theorems}
\crefname{definition}{Definition}{definitions}
\crefname{proposition}{Proposition}{propositions}
\crefname{lemma}{Lemma}{lemmas}
\crefname{corollary}{Corollary}{corollaries}
\crefname{claim}{Claim}{claims}
\crefname{observation}{Observation}{observations}
\crefname{fact}{Fact}{facts}
\crefname{dfn}{Definition}{definitions}
\crefname{obs}{Observation}{observations}
\crefname{pb}{Problem}{problems}
\crefname{algo}{Algorithm}{algorithms}

\usepackage{paralist} 

\usepackage[shortlabels]{enumitem}

\newlist{proofcases}{enumerate}{3} 
\setlist[proofcases,1]{%
  label=\arabic*),ref=\arabic*,
  leftmargin=0pt,labelsep=0pt,align=left,%
  labelwidth=\widthof{(8)~},
  itemindent=\widthof{(8)~},
  listparindent=\parindent}
\setlist[proofcases,2]{%
  label=\theproofcasesi.\alph*),ref=\theproofcasesi.\alph*,
  leftmargin=2em,labelsep=0pt,align=left,%
  labelwidth=\widthof{(8.m)~},
  itemindent=\widthof{(8.m)~},
  listparindent=\parindent}
\if@cref@capitalise
  \crefname{proofcases}{Case}{Cases}
\else
  \crefname{proofcases}{case}{cases}
\fi
\crefalias{proofcasesi}{proofcases}
\crefalias{proofcasesii}{proofcases}

\usepackage{xifthen}
\newcommand{\ifthen}[2]{\ifthenelse{#1}{#2}{}}
\newcommand{\ifnot}[2]{\ifthenelse{#1}{}{#2}}

\ifthenelse{\boolean{lncs}}{
}{

}

\ifthenelse{\boolean{linenumbers}}{ 
\usepackage[left, pagewise]{lineno}
\linenumbers
\newcommand*\patchAmsMathEnvironmentForLineno[1]{%
  \expandafter\let\csname old#1\expandafter\endcsname\csname #1\endcsname
  \expandafter\let\csname oldend#1\expandafter\endcsname\csname end#1\endcsname
  \renewenvironment{#1}%
     {\linenomath\csname old#1\endcsname}%
     {\csname oldend#1\endcsname\endlinenomath}}%
\newcommand*\patchBothAmsMathEnvironmentsForLineno[1]{%
  \patchAmsMathEnvironmentForLineno{#1}%
  \patchAmsMathEnvironmentForLineno{#1*}}%
\AtBeginDocument{%
\patchBothAmsMathEnvironmentsForLineno{equation}%
\patchBothAmsMathEnvironmentsForLineno{align}%
\patchBothAmsMathEnvironmentsForLineno{flalign}%
\patchBothAmsMathEnvironmentsForLineno{alignat}%
\patchBothAmsMathEnvironmentsForLineno{gather}%
\patchBothAmsMathEnvironmentsForLineno{multline}%
}
}{
}

\makeatletter
\newdimen\commentwd
\let\oldtcp\tcp
\def\alignedtcp*[#1]#2{
\setbox0\hbox{#2}%
\ifdim\wd\z@>\commentwd\global\commentwd\wd\z@\fi
\oldtcp*[r]{\leavevmode\hbox to \commentwd{\box0\hfill}}}

\let\oldalgorithm\algorithm
\def\algorithm{\oldalgorithm
\global\commentwd\z@
\expandafter\ifx\csname commentwd@\romannumeral\csname c@\algocf@float\endcsname\endcsname\relax\else
\global\commentwd\csname commentwd@\romannumeral\csname c@\algocf@float\endcsname\endcsname
\fi
}
\let\oldendalgorithm\endalgorithm
\def\endalgorithm{\oldendalgorithm
\immediate\write\@auxout{\gdef\expandafter\string\csname commentwd@\romannumeral\csname c@\algocf@float\endcsname\endcsname{%
\the\commentwd}}}

\DeclareMathOperator{\dist}{dist} 

\renewcommand{\O}{\ensuremath{\mathcal{O}}}
\newcommand{\noise}{\ensuremath{\operatorname{N}}}

\usepackage{placeins}
\usepackage[font=footnotesize]{caption}
\newcommand{\tablefontsize}{\footnotesize}

\usepackage{etoolbox}
\newtoggle{draft}
\togglefalse{draft} 

\usepackage{tikz}
\usetikzlibrary{spy}

\usepackage{siunitx}
\usepackage{multirow}
\usepackage{booktabs}
\usepackage{array}
\newcolumntype{L}[1]{>{\raggedright\let\newline\\\arraybackslash\hspace{0pt}}m{#1}}
\newcolumntype{C}[1]{>{\centering\let\newline\\\arraybackslash\hspace{0pt}}m{#1}}
\newcolumntype{R}[1]{>{\raggedleft\let\newline\\\arraybackslash\hspace{0pt}}m{#1}}

\newcommand{\randvd}{{\sc Rand4DColor}}
\newcommand{\randrespect}{{\sc Rand4DRespColor}}
\newcommand{\randfinal}{{\sc Rand4DFinalColor}}
\newcommand{\randed}{{\sc Rand1DColor}}

\newcommand{\ColorReduction}{{\sc ColorRed}}
\newcommand{\colorred}{\ColorReduction}
\newcommand{\colorrand}{{\sc CRRandColor}}
\newcommand{\colorcor}{{\sc CRRCor}}

\newcommand{\mwcolor}{{\sc MWColor}}
\newcommand{\mwcor}{{\sc MWCor}}

\newcommand{\yucolor}{{\sc YuColor}}
\newcommand{\yucor}{{\sc YuCor}}

\newcommand{\txConst}{{\sc txConst}}
\newcommand{\duration}{{\sc duration}}
\newcommand{\factor}{{\sc factor}}
\newcommand{\Rt}{\ensuremath{r^T}}
\newcommand{\Rb}{\ensuremath{r^B}}

\SetCommentSty{mycommfont}

\begin{document}
\newgeometry{margin=1.5in}

\title{Experimental Evaluation of Distributed Node Coloring Algorithms for Wireless Networks}

\ifthenelse{\boolean{lncs}}{ 
\author{
  Fabian Fuchs\\
}
\institute{
  {\normalsize Karlsruhe Institute for Technology } \\
  {\normalsize Karlsruhe, Germany } \\
  {\normalsize fabian.fuchs@kit.edu } 
}
}{
\author{
{Fabian Fuchs}\\[0.75em]
Institute of Theoretical Informatics \\
Karlsruhe Institute of Technology\\
Karlsruhe, Germany\\[0.75em]
fabian.fuchs@kit.edu
}
}

\date{\vspace{-5mm}}
\maketitle

\graphicspath{{./plots/}{./}}

\begin{abstract}
  In this paper we evaluate distributed node coloring algorithms for
  wireless networks using the network simulator Sinalgo
  \cite{sinalgo}. All considered algorithms operate in the realistic
  signal-to-interference-and-noise-ratio (SINR) model of
  interference. We evaluate two recent coloring algorithms, \randvd\
  and {\sc ColorReduction} (in the following \colorred), proposed by
  Fuchs and Prutkin in \cite{fp-sddcs-15}, the MW-Coloring algorithm
  introduced by Moscibroda and Wattenhofer \cite{mw-curn-08} and
  transferred to the SINR model by Derbel and Talbi
  \cite{dt-dncsi-10}, and a variant of the coloring algorithm of Yu et
  al.  \cite{ywhl-ddcpm-11}.  We additionally consider several
  practical improvements to the algorithms and evaluate their
  performance in both static and dynamic scenarios.

  Our experiments show that \randvd\ is very fast, computing a
  valid $(4\Delta)$-coloring in less than one third of the time slots
  required for local broadcasting, where $\Delta$ is the maximum node
  degree in the network. Regarding other $\O(\Delta)$-coloring
  algorithms \randvd\ is at least $4$ to $5$ times faster.
  Additionally, the algorithm is robust even in networks with mobile
  nodes and an additional listening phase at the start of the
  algorithm makes \randvd\ robust against the late wake-up of large
  parts of the network.

  Regarding $(\Delta+1)$-coloring algorithms, we observe that
  \colorred\ it is significantly faster than the considered variant of
  the Yu et al. coloring algorithm, which is the only
  other~$(\Delta+1)$-coloring algorithm for the SINR model.  Further
  improvement can be made with an error-correcting variant that
  increases the runtime by allowing some uncertainty in the
  communication and afterwards correcting the introduced conflicts.
\end{abstract}

\clearpage
\restoregeometry

\newcommand{\figpath}{./}

\section{Introduction}
\label{ch:exp:sec:introduction}

Distributed node coloring is the underlying problem for many
fundamental issues related to establishing efficient communication in
wireless ad hoc and sensor networks. We can, for example, reduce the
problems of establishing a time-, code-, or
frequency-division-multiple-access (TDMA, CDMA, FDMA) schedule to a
node coloring problem \cite{r-aufaca-99}. In this work we study and
experimentally evaluate distributed node coloring algorithms that were
designed for the realistic signal-to-interference-an-noise-ratio
(SINR) model of interference. This model is widely used for decades in
the electrical engineering community and was adopted by the
algorithmic community after a seminal work by Gupta and Kumar
\cite{gk-cwn-00}. In contrast to graph-based models, the SINR model
reflects both the local and the global nature of wireless
transmissions. However, to analytically prove guarantees on the
runtime and show an algorithms correctness becomes relatively
complex. Thus, over the past years techniques were developed to tackle
the complexity of the model. This, however, led to the introduction of
several constant factors in different parts of the algorithms. In this
paper we study four distributed node coloring algorithms in a more
practical setting.  We use the network simulator Sinalgo
\cite{sinalgo} to execute the algorithms in a variety of deployment
scenarios in the static and the dynamic setting.

Let us briefly consider the algorithms we evaluate in this paper. All
algorithms are designed for the realistic SINR model of interference
and can be used to establish a TDMA schedule using methods from
\cite{dt-dncsi-10,fw-olbsc-13}. We denote the number of nodes by~$n$
and the maximum degree in the network by~$\Delta$. The first
algorithm, which we denote by \mwcolor\ in this paper, was proposed by
Moscibroda and Wattenhofer in \cite{mw-curn-08} for the protocol model
and transfered to the SINR model by Derbel and Talbi
\cite{dt-dncsi-10}. \mwcolor\ computes an~$\O(\Delta)$-coloring
in~$\O(\Delta \log n)$ time slots by first selecting leader nodes,
which coordinate the color selection of other nodes so that only few
nodes in each neighborhood compete for the same color.  As second
algorithm we evaluate is a variant of the
distributed~$(\Delta+1)$-coloring algorithm proposed by Yu et
al.~\cite{ywhl-ddcpm-11}. Their algorithm computes leaders, which then
increase their transmission power to block nodes that could hinder the
leaders (original) neighbors from selecting a valid color.  We use a
variant of their original algorithm, as their algorithm operates in a
slightly different setting, however, the ideas are applicable in our
setting as well.  We denote the variant by \yucolor\ and introduce it
in \cref{ch:exp:sec:yu-intro}. Finally, we consider two algorithms
Fuchs and Prutkin proposed in \cite{f-fsncs-15,fp-sddcs-15}. \randvd\
is a very simple randomized coloring algorithm that computes a
$(4\Delta)$-coloring by simply selecting a new random color whenever a
conflict is detected. Finally, \colorred, uses an existing coloring to
coordinate the color selection process to compute a
$(\Delta+1)$-coloring. The nodes executing the algorithm first select
a set of leaders, which compute a medium access schedule based on the
existing colors. Due to this schedule only few nodes are active at a
time, which enables the active nodes to quickly win the competition
for their final colors.  All evaluated algorithms distributively
compute the valid node coloring in $\O(\Delta \log n)$ time slots.

\subsection{Related Work}
\label{ch:exp:sec:related-work}

In wireless networks, node colorings are particularly interesting as
they allow the nodes to establish more efficient communication for
example by computing a TDMA communication schedule (still using the
realistic SINR model) \cite{dt-dncsi-10,fw-olbsc-13}.  Although
distributed node coloring algorithms are widely applicable and their
study was initiated more than~$25$ years ago, only few experimental
evaluations are concerned with distributed node coloring algorithms
and none consider distributed algorithms for wireless networks.  To
the best of our knowledge the first experimental evaluation on
distributed node coloring algorithms is due to Finocchi, Panconesi,
and Silvestri \cite{fps-exadv-05}. They study very simple node
coloring algorithms, which are similar to our \randvd\ coloring
algorithm. Their algorithms and the evaluation are based on simpler
message-passing models, which do not consider interference. They
observed that such simple algorithms are very fast, and proposed some
practical improvements to the coloring algorithms. In an earlier
experimental study Marathe, Panconesi and Risinger \cite{mpr-aes-04}
considered simple edge colorings of the same randomized
trial-and-error flavour as the later considered vertex coloring
algorithms and found that such algorithms performed ``extremely
good''.  Pindiproli and Kothapalli \cite{pk-eadca-09} extend the study
on distributed node coloring algorithms by Finocchi, Panconesi and
Silvestri by considering the same randomized algorithms and compare it
to a similar algorithm that requires only~$\O(\sqrt{\log n})$ rounds
of transmitting a single bit. Hernandez and Blum \cite{hb-dgcwahn-11}
compare a distributed coloring algorithm inspired by Japanese tree
frogs to the algorithms studied by Finocchi, Panconesi, and Silvestri,
however, they focus on minimizing the number of colors used.


\subsection{Contribution}
\label{ch:exp:sec:contributions}

First, we show that the very simple \randvd\ coloring algorithm is
very fast, achieving a runtime an order of one magnitude faster than
its direct competitor, the MW-coloring algorithm. Interestingly, the
algorithm computes a valid~$(4\Delta)$-coloring in less time slots
than required for one round of local broadcasting.  We additionally
show that our \colorred\ algorithm is significantly faster than
\yucolor, our variant of the~$(\Delta+1)$-coloring algorithm by Yu et
al.

Second, we propose heuristic improvements for \colorred, \mwcolor, and
\yucolor. The improvements are inspired by \randvd\ and
allow the nodes to decrease the number of time slots accounted for the
transmission of a message. 
We show that they considerably improve the
runtime while keeping the number of conflicts in the network close to
zero.

Third, we study the correction variants in a network with mobile nodes
and in a network in which a large fraction of the nodes start the
algorithm after the remaining network has computed a valid
coloring. We observe that \randvd\ is most robust against mobility of
nodes. Regarding the late wake-up of some nodes, we 
show that simple measures are sufficient to make \randvd\ robust in
this scenario.


\bigskip
\noindent\textbf{Outline:}
The remainder of this paper is structured as follows. In the next
section we describe and introduce the algorithms we consider in our
experiments. In \cref{sec:sinalgo-settings} we describe the simulator
and the setting of our experiments before evaluating the algorithms in
\cref{ch:coloringexp:sec:setup}. We conclude this paper in
\cref{ch:exp:sec:conclussion}.

\section{Considered Algorithms}
\label{ch:exp:sec:algorithms}
\renewcommand{\figpath}{images/}

Node coloring is the problem of assigning a color to each node in the
network such that no two neighbors have the same color. In distributed
computing, a $(\Delta+1)$-coloring (which is always possible) is the
ultimate goal, as it is NP-hard to color a graph with the minimum
number of colors even in a centralized way \cite{gj-np-book-79}.  We
call the color of a node \emph{valid} if no neighbor selected the same
color and say that a node that selected the same color as one of its
neighbors to have a \emph{conflict} with this neighbor.  We use local
broadcasting \cite{gmw-lbpim-08} as the basic form of communication,
which allows successful communication within \duration\ time slots (we
determine this parameter in \cref{ch:coloringexp:sec:setup}).  We
illustrate the flow of each algorithm on an example network in 
\cref{sec:illustr-algos}.

\subsection{\randvd}
\label{ch:exp:sec:rand4d-intro}

\randvd\ is based on a very simple randomized algorithm well-known for
message-passing models since decades \cite[Chapter 10]{be-dcg-13}. It
has recently been proven to be efficient in the SINR model by Fuchs
and Prutkin \cite{fp-sddcs-15}. The phase-based algorithm computes a
valid $(\Delta+1)$-coloring in~$\O(\Delta \log n)$ time slots. During
each phase of the algorithm, the node may receive the colors of some
of its neighbors.  If, at the end of the phase, a conflict between its
current color and the color received by a neighbor is detected, the
node selects a new color at random.  The overall execution is exactly
as known from the message-passing models, however, the runtime is
competitive by reducing the length of the phases. This leads to
uncertainty in whether a conflict can be detected within each phases,
which is bounded by Fuchs and Prutkin.

Apart from \randvd\ we consider some variants of the algorithm. For
the first variant, \randrespect, we add a listening phase to the start
of the algorithm and require the node to store the latest received
color of each neighbor. When resetting the color the nodes do not
select a color that is currently stored for a neighbor and thereby
reduce the number of conflicts (especially in case of highly
asynchronous wake-up of nodes).  We do also consider a variant
\randfinal\ that finalizes the selected color after it did not receive
a color conflict for at least \duration\ time slots. This enables each
node of the algorithm to decide when the coloring algorithm is
completed.  In another variant we reduce the number of available
colors to~$\Delta+1$ and call the variant \randed. This variant
computes a~$(\Delta+1)$-coloring, although our theoretical guarantees
holds only for~$c>4\Delta$ colors.

\subsection{\colorred}
\label{ch:exp:sec:cr-intro}

The algorithm \colorred\ by Fuchs and Prutkin
\cite{fp-sddcs-15} computes a valid~$(\Delta+1)$-coloring in
$\O(\Delta \log n)$ time slots. \colorred\ computes two levels of
MISs, the first level MIS determines independent leaders, which then
coordinate the activity of the remaining nodes. After the first level
MIS all non-leaders request an active interval from the leaders,
during which the non-leader nodes then repeatedly execute the faster
second level MIS algorithm.  Once a non-leader node~$v$ is in the
independent set, it selects a color from the set of free colors~$F_v$,
transmits the color to all its neighbors and resigns from the
independent set. The active interval is based on the initial (valid)
color of the non-leader node and the schedule determined by the leader
(independent from other leaders). These schedules achieve that few
nodes compete in the second level MIS, which allows making these MISs
very fast.  We additionally consider a variants of \colorred\ that
does not use a valid node coloring but each node simply selects a
random number from the set of available colors. We denote this variant
by \colorrand.

\subsection{\mwcolor}
\label{ch:exp:sec:mw-intro}

We implement the MW-Coloring algorithm as described by Derbel and
Talbi \cite{dt-dncsi-10} and denote it by {\sc MWColoring}. The
algorithm computes an~$\O(\Delta)$-coloring and proceeds as follows:
First the nodes compete to be in an MIS to become leaders and select
color~$0$. The remaining nodes request a continuous block of colors
from a selected leader. Once this color block is received, the node
competes in another MIS against at most constantly many neighbors for
a color. If the MIS is won, it selects the color, otherwise it moves
on to the next color in its color interval and competes again.  A
notable difference between \colorred\ and \mwcolor\ is that the number
of time slots required for the color-competing MIS in \mwcolor\ is
significantly higher than the second level MIS in \colorred, however,
a lot less of these slower MISs are executed.

\subsection{\yucolor}
\label{ch:exp:sec:yu-intro}

The coloring algorithm by Yu et. al~\cite{ywhl-ddcpm-11} computes a
$(\Delta+1)$-coloring in~$\O(\Delta \log n + \log^2 n)$ time
slots. The main idea behind achieving~$\Delta+1$ colors in their
algorithm is to increase the transmission power in order to coordinate
the color selection process within a larger distance. To achieve this
the algorithm uses two transmission powers~$r_1$ and~$r_2$,
where~$r_1$ is the regular broadcasting range and~$r_2=3\cdot r_1$.
The algorithm itself works as follows: First, the nodes compute an MIS
with respect to~$r_2$. All nodes in the MIS transmit a so-called
DoNotTransmit-message to all nodes within $r_2$. Thereby the nodes
within the range~$r_2$ enter a blocked state S, which they only leave
once they receive a StartTransmit-message or a StartColoring-message.
The nodes in the MIS transmit a StartColoring-message, however, only
to the nodes within the smaller range~$r_1$. These nodes start with
the color selection process by transmitting an AskColor-message to
their MIS node. The MIS node coordinates the requests and allows one
after the other to select the smallest color not taken by a
neighbor. The colors can be selected without a conflict, as all
close-by nodes are either coordinated by the MIS node or are in the
blocked state S.  Naturally, once a color is selected by a node, the
node informs all its neighbors about its color selection.

In the setting Yu et. al designed the algorithm for, the nodes are not
given an estimate of the maximum degree~$\Delta$. Thus, to ensure
successful communication a slow-start mechanism is used for the
transmissions. To circumvent this length mechanism we adapt the
algorithm to the case of known~$\Delta$ and prove the following
theorem. 

\begin{restatable}{theorem}{yutheorem}
  \label{ch:exp:thm:yu}
  {\sc YuColoring} computes a~$(\Delta+1)$-coloring in~$\O(\Delta \log n)$
  time slots in our setting.
\end{restatable}

\begin{algorithm}
\DontPrintSemicolon
\textbf{Continuously:}\;
\ShowLn \lIf{Received DoNotTransmit$_u$} {$F_v \gets
  F_v \cup \{u\}$ and transit to $state \gets$ blocked}
\ShowLn \lIf{Received Color$_u$(c) from node $u$} {$C_v \gets
  C_v \cup \{c\}$}

\ShowLn \Switch{$state$} {

\ShowLn \Case{start}{
\ShowLn wait for $\O(\Delta\log n)$ time slots \;
\ShowLn transit to $state \gets$ MIS
}

\ShowLn \Case{blocked}{
\ShowLn \lIf{Received StartColoring$_u$} { transit to $state \gets$ C1}
}

\ShowLn \Case{MIS}{
\ShowLn  We use MIS($\ell=1$) from \colorred \cite{fp-sddcs-15}. 
Successful nodes transit to $state\gets$ leader
}

\ShowLn \Case{leader} {
\ShowLn Transmit DoNotTransmit$_v$
with range $r_2$ and prob. $p_{high}$ for $\O(\log n)$ slots\;
\ShowLn select color $0$\;
\ShowLn Transmit StartColoring$_v$ with range $r_1$ and \newline prob. $p_{high}$ for $\O(\log n)$ time slots\;
\ShowLn \eIf{Q not empty}{
  \ShowLn $u \gets$ Q.pop()\;
  \ShowLn Transmit Grant$_u$ with range $r_1$ and  prob.
  $p_{high}$ for $\O(\log n)$ time slots\;
} {
  \ShowLn Transmit StartColoring$_v$ with range $r_1$ and \newline prob. $p_{high}$ for $\O(\log n)$ slots\;
}
}

\ShowLn \Case {C1}{ 
\ShowLn Transmit AskColor$_v$ with range $r_1$ and \newline prob.
$p_{low}$ for $\O(\Delta \log n)$ time slots\;
\ShowLn \lIf{ Received Grant$_u$} { transit to $state \gets$ C2}
}

\ShowLn \Case{C2} {
\ShowLn select smallest color $c$ not in $C_v$\;
\ShowLn Transmit Color$_v(c)$ with range $r_1$ and prob.
$p_{high}$ for $\O(\log n)$ time slots\;
}

}
\caption{{\sc  YuColoring} for node $v$}
\label{algo:yucoloring}
\end{algorithm} 

The correctness essentially follows from the correctness of the
original algorithm. For the argument to be more concise, we elaborate
on the main points in the following and give a pseudocode of \yucolor\
in \cref{algo:yucoloring}.

\subparagraph{The coloring is valid:} Let us consider a node $v$ and
assume its coloring is not valid due to a conflict with its neighbor
$u$. If $v$ has color $0$, it is a leader node and the conflicting
node $u$ must have been one of the nodes $v$ dominated. Thus, with
high probability, $u$ received the DoNotTransmit and the
StartColoring, afterwards transmitted AskColoring itself and received
a Grant message - leading to the selection of another color
$c \not= 0$.  If $v$'s color is not $0$, it selected its color during
such a color selection process itself. As both $v$ and $u$
successfully transmit their color after selection with high
probability and neighbors respect this selection, it must be the case
that $v$ and $u$ selected the color simultaneously. Since the leader
nodes wait long enough between transmitting the Grant message to the
two nodes, this can only happen if $v$ and $u$ listen to two different
leaders.  This, however, is not possible as all nodes within at least
two broadcasting ranges of $v$ received the DoNotTransmit message of
$v$'s leader with high probability.

\subparagraph{All nodes get colored:} Essentially this holds as each
node $v$ is either in the MIS (with respect to $r_2$) at some point or one
of its neighbors is in the MIS and allows $v$ to select a color. 

\subparagraph{The runtime of the algorithm is $\O(\Delta \log n)$ time
  slots:} Let us consider the maximum time until a node $v$ or one of
its neighbors is in the MIS. Remember that the MIS is computed with
respect to the range $r_2$, while the neighborhood relation we
consider for the coloring is relative to $r_1$. As each
$r_1$-neighborhood of a MIS node is colored after~$\O(\Delta \log n)$,
this is also the asymptotic time that passes between the MIS node
transmits DoNotTransmit and StartTransmit. As at most $36$ nodes can
be independent regarding the range $r_1$ in the $r_2$-range of a node
$v$, after at most $36$ rounds of the MIS (and potentially the
following blocked/coloring state) all nodes in the $r_2$-range of~$v$
must have either been in the MIS or are neighbors of an
MIS-node. Thus, either~$v$ or one of its neighbors wins the MIS
competition and starts the coloring routine afterwards. Overall, this
results in a runtime of $\O(\Delta \log n)$ time slots.

\subsection{Correcting Variants}
\label{sec:algos:correcting-variants}

Apart from \randvd, the coloring algorithms do not account for errors,
as they do not happen if all transmissions are successful. We denote
the number of time slots required to ensure successful transmission by
all nodes by \duration, but even this does not guarantee that there
are no failures (due to the probabilistic nature of the
transmissions). Thus, we consider so-called \emph{correcting variants}
of the algorithms, in which we combine the algorithms with ideas of
\randvd. Namely, we decrease the time accounted for successful
transmission of messages, and resolve the introduced conflicts
afterwards. We resolve the conflicts by resetting non-leader nodes to
the last uncolored state that is still proper. For \colorred\ nodes
compute a new valid active interval based on the previous active
interval and the schedule length and wait for this new interval. In
\mwcolor\ the nodes reset to the first color competition of the color
block they received. For \yucolor\ the nodes must reset the initial
MIS, as the nodes leader might have resigned by now.  For leader nodes
we use the simpler strategy of reseting to a random color to prevent
issues that arise once leaders may resign from their duties. For
\colorred\ we use the set of unused colors~$F_v$ and for \mwcolor\ and
\yucolor\ we use~$[\Delta]$. We denote the correcting variants by
\colorcor, \mwcor, and \yucolor.

\subsection{Determining Parameters}
\label{sec:algos:determ-param}

There are only few parameters of the algorithms that must be
determined apart from those related to communication.  Let us
therefore first consider these parameters. Except for \randvd\ all
algorithms use local broadcasting and fast local broadcasting. While
local broadcasting might be used by all nodes simultaneously, fast
local broadcasting is restricted (by theoretical considerations) to a
constant number of nodes in each broadcasting range. We determine the
optimal transmission probability and the duration (denoted by
\duration) of local broadcast in \cref{sec:basic-comm-param} 
and determine a factor (denoted by \factor) between local broadcasting
and its fast variant for the algorithms. Although \randvd\ does not
use regular broadcasting, we still use \factor\ to determine the
length of the phases (this is not the same but related to a fast local
broadcast). Apart from this, there are no parameters required for
\randvd\ and \yucolor. For \colorcor\ and \mwcolor, one could set
additional parameters that determine the time slots accounted for a
second level MIS and the length of a color block, respectively. We do
not set those parameters but set the length of a second level MIS as a
fast local broadcast, and the color block to be~$8$ (as we do not rate
the algorithms based on the number of colors they require, using a
large enough value is sufficient).

\section{Sinalgo Settings}
\label{sec:sinalgo-settings}

\renewcommand{\figpath}{sinalgo-basic/}

We conduct our experiments using the current version 0.75.3 of Sinalgo
\cite{sinalgo}, an open-source simulation framework for networks
algorithms in
Java. 
%
Sinalgo has built-in support for a variety of communication and
interference models, and is implemented in a modular fashion, making
it easy to add customized models or algorithms.  The main simulation
framework offers both round-based and asynchronous or event-based
simulation. Apart from that connectivity, interference, mobility,
reliability, distribution, and message transmission models implement a
wide variety of settings.  

We shall briefly introduce the relevant
parts of the simulation framework in the following.  In the
round-based setting, in every time slot each node is considered
once. Thus, the node handles all successfully received messages and
performs one step. Both the message handling and the step depend on
the implementation of the network algorithm.  In the event-based
simulation, each action of the algorithm must be invoked by a
timer. Despite the absence of rounds or slots, we denote the time
required for one transmission as a time slot.
Mobility is not supported in this setting due to the high
number of events and the corresponding updates of all positions
(required for example for connectivity and interference computations).
Let us now consider the models used in our experiments. 

\textbf{Connectivity:} To determine the neighborhood relations we
implemented a new model, which calculates the broadcasting range
directly based on the SINR parameters used in the simulation. Based on
the broadcasting range, the neighbors of a node $v$ are determined as
the nodes within this range of $v$.

\textbf{Interference:} Successful reception of the nodes transmission
is determined by the standard geometric SINR model in all our
simulations. In the SINR model a transmission from a sender to a
receiver is \emph{feasible} if it can be decoded by the
receiver.  It depends on the ratio between the desired signal and the
sum of interference from other nodes plus the background noise whether
a certain transmission is successful. Let each node~$v$ in the network
use the same transmission power~$P$. Then a transmission from~$u$
to~$v$ is feasible if and only if
\begin{align*}
  \frac{\frac{P}{\dist(u,v)^\alpha}}{\sum_{w \in I}
    \frac{P}{\dist(w,v)^\alpha} + \noise} \geq \beta,
\end{align*}
where~$\alpha \in [2,6]$ is the attenuation coefficient, the constant
$\beta > 1$ depends on the hardware, $\noise$ denotes the
environmental noise,~$\dist(u,v)$ the Euclidean distance between two
nodes~$u$ and~$v$, and~$I \subseteq V$ is the set of nodes
transmitting simultaneously to~$u$.  The parameters for the geometric
SINR model of interference are set as follows. We use a value of
$\alpha=4$ for the attenuation coefficient, which is assumed to be
between $2$ for a free field environment and $6$ for buildings in
practice \cite{r-wcom-09}. We use a threshold of $\beta=10$ and an
environmental noise value of $10^{-9}$, which are both within the
ranges often reported \cite{su2007impact,zhang2007distributed}.  We
use a uniform transmission power, which is set to $P=1$ for all nodes.
The \emph{broadcasting range}~$\Rb$ of a node~$v$ defines the range
around~$v$ up to which $v$'s messages should be received.  Based on
the SINR constraint, the \emph{transmission
  range}~$\Rt \leq ( \frac{P}{\beta \noise} )^{1/\alpha}$ is an upper
bound for the broadcasting range (with~$\Rb < \Rt$ to allow multiple
simultaneous transmissions).  Our parameters lead to a transmission
range of \SI{100}{\meter}, an additional broadcasting range parameter
of $2$ leads to the broadcasting range of about \SI{84}{\meter}.

A flaw in the SINR-module delivered with Sinalgo in version $0.75.3$
drops some transmissions although they are feasible in the SINR
model. We show how to correct this in \cref{app:exp:sec:all-distributions}.  For more
details on the SINR model itself we refer, for example to
\cite{fp-sddcs-15}.

\textbf{Distribution:} We mostly deploy our nodes on a square area of
$\SI{1000}{\meter} \times \SI{1000}{\meter}$ using several
distribution strategies. We use the built-in random and grid model,
which deploy the nodes uniformly at random and according to a regular
two-dimensional grid, respectively. Custom models we use are a
perturbed grid, in which each nodes grid position is uniformly at
random drawn from a $\SI{1}{\meter^2}$ area centered at the original
grid position, and a cluster distribution which distributes all nodes
in a predefined number of clusters (we use $10$
clusters). Additionally, we use a combination of the cluster model
with the other models, in which \SI{50}{\percent} of the nodes are
distributed according to the cluster model and the remaining nodes
according to either the random, grid or perturbed grid model. Our
distribution models are illustrated in some example networks in
\cref{ch:prelim:fig:distribution-models} in \cref{app:illustr-depl}. To increase
comparability 
of our experiments we use a precomputed set of position files.

\textbf{Message transmission:} Messages are transmitted with a certain
probability in each time slot. We compute the transmission probability
based on a transmission constant \txConst\ and the maximum degree
$\Delta$ in the network as \txConst$/\Delta$ (apart from constant
factors this is as described by Goussevskaia et
al. \cite{gmw-lbpim-08}).  Regarding the time required for a message
transmission we assume constant time message transmission. As all
messages are of size at most $\O(\log n)$ in our algorithms, this fits
the algorithms requirements. To ensure that a message can be
transmitted in one time slot, which is of length $1$, we use a
transmission time of $0.999$ time slots. We do neither allow
simultaneous transmission and reception nor the simultaneous reception
of several packets.

\textbf{Reliability:} Throughout our simulations we use so-called
reliable transmission, which implies that transmitted messages are
received unless they are discarded by the interference model.

\textbf{Mobility:} Sinalgo supports mobility based on either random
waypoints, or random directions. We use the latter model, as it
consistently provides a balanced distribution of the nodes on the
area, while the random waypoint model would lead to a high
concentration of nodes in the center of the deployment
area \cite{bh-mobility-06}.

Our algorithms are implemented in subclasses of the node class, which
plugs into the described models and provides standard transmission and
reception features.  To measure the number of nodes with a valid and
an invalid color, each node notifies the simulator whenever it selects
a new color, which is then checked for validity with colors selected
by the neighbors. This is done within the simulation framework, thus
we do neither use messages nor tell the nodes about the result of this
color inspection.

For each experiments we use~$100$ runs on the same pre-computed
deployments.  Apart from the overall results we report only results
for the random deployment due to space constraints, other results are
deferred to \cref{app:exp:sec:all-distributions}.  
We measure the time required to compute a valid coloring and the
number of nodes that were not able to select a valid color.  We use
one time slot as the time required for one transmission. To measure
the runtime of our algorithms, we deploy the nodes simultaneously in
the area and start the algorithms asynchronously after a waiting
period that is chosen uniformly at random between~$0$ and~$10$ time
slots (using real numbers) for each node. The runtime measurement
starts with the deployment of the nodes and ends once all algorithms
are in a finished state or all nodes have selected a valid color.
Note that the time slots of the nodes are not synchronized and may
overlap partially.

\subsection{Basic Communication Parameters}
\label{sec:basic-comm-param}

To implement the message transmission in the algorithms, we use local
broadcasting with known~$\Delta$, and therefore use the parameters
transmission constant \txConst\ and broadcast duration \duration.  To
compute the transmission probability used by the nodes we divide a
parameter \txConst\ by the maximum degree~$\Delta$.  Other parameters
are the \duration, which is set to the number of time slots required
for reliable communication in the network, and \factor, which is part
of the ratio between between regular and fast local broadcasting:
$\Delta\cdot$\factor.  Although \randvd\ does not use local
broadcasting we use the same \txConst, to increase comparability of
the algorithms.
%
%
We show the relation of the parameters in calculating the values used
for the simulation in \cref{ch:exp:tab:lb-flb}
\begin{table}[hbt]
\centering
\tablefontsize
\caption{Transmission probabilities and durations based on the simulation parameters}
\label{ch:exp:tab:lb-flb}
\begin{tabular}{L{5.2cm}L{3.5cm}}
\toprule
Simulation parameter & Value \\\midrule
Local broadcast \\ 
\;\;\;\; transmission probability &  $\frac{\text{\txConst}}{\Delta}$  \\
\;\;\;\; transmission duration & \duration \\
Fast local broadcast \\
\;\;\;\; transmission probability & $\frac{\text{\txConst}}{\Delta} \times \Delta
                      \times$\factor\newline$=$\;\txConst$\times$\factor \\
\;\;\;\; transmission duration & $\frac{\text{\duration}}{\Delta
                                 \times \text{\factor}}$ \\
\bottomrule
\end{tabular}
\end{table}

To determine the parameter \txConst\ and \duration\ we execute local
broadcasting for several values of \txConst\ and report the optimal
parameter and the runtime in \cref{ch:prelim:tab:lb-parameters}.
Additionally we report some characteristics of the networks such as
the average maximum degree $\Delta$ and the average degree, the
transmission probability (which is computed as
\txConst$/\Delta$). Note that we set \duration\ so that it is slightly
larger than the average maximum runtime of local broadcasting. This
should achieve successful transmission with a high probability.
\begin{table}[hbt]
\centering
\tablefontsize
\caption{Parameters that achieve successful local broadcasting in the
  different distributions. R=Random, G=Grid, PG=PerturbedGrid, C=Cluster}
\label{ch:prelim:tab:lb-parameters}
\begin{tabular}{L{3.0cm}R{1.05cm}R{1.05cm}R{1.05cm}R{1.15cm}R{1.15cm}R{1.15cm}R{1.15cm}R{1.15cm}}
\toprule
   Distribution      & R & G & PG & C & C\&R & C\&G & C\&PG \\ \midrule
\txConst  &    \num{0.15}    &  \num{0.15}    &  \num{0.10}    &    \num{0.30}     &        \num{0.25}       &        \num{0.20}     &  \num{0.20}            \\ 
Maximum degree $\Delta$  &    \num{36.6}    &  \num{20.0}    &  \num{27.9}    &    \num{182.0}     &        \num{106.0}       &        \num{101.8}     &  \num{100.8}            \\ 
Average degree   &    \num{20.6}    &  \num{18.6}    &  \num{20.9}    &    \num{93.8}     &        \num{39.4}       &        \num{38.7}     &  \num{39.9}            \\ 
Transmission probability &   $4.11 \times 10^{-3}$   &  $7.5 \times 10^{-3}$ &  $3.59 \times 10^{-3}$ &  $1.71 \times 10^{-3}$ &  $1.46 \times 10^{-3}$ &  $2.02 \times 10^{-3}$   &  $2.55 \times 10^{-3}$\\ 
Avg. runtime of a local broadcast &    \num{4592}    &  \num{3345}    &  \num{4845}    &    \num{12903}     &        \num{8062}       &        \num{8183}     &  \num{8089}  \\
\duration &    \num{4600}    &  \num{3400}    &  \num{4900}    &    \num{12900}     &        \num{8100}       &        \num{8200}     &  \num{8100}  \\
\bottomrule
\end{tabular}
\end{table}

\section{Experiments}
\label{ch:coloringexp:sec:setup}

In this section we evaluate the distributed node coloring algorithms
described in \cref{ch:exp:sec:algorithms} using the simulation
framework described in \cref{sec:sinalgo-settings}. Therefore we
determine a final parameter \factor, which determines the difference
between regular local broadcasting and a fast local broadcasting
regarding both the duration and transmission probability. Afterwards
we consider the algorithms and their variants separately in
\cref{sec:randvd-its-variants,ch:exp:sec:cr-exp,ch:exp:sec:correcting-variants}. In
\cref{sec:progress-algos} we compare the progress of the algorithms
before comparing the algorithms themselves in
\cref{ch:exp:sec:coloring-comparison}.  In \cref{ch:exp:sec:mobility}
we study the performance of the algorithms under network dynamics such
as moving nodes and the highly asynchronous wake-up of nodes.

\subsection{Determining \factor}
\label{sec:determining-factor}

We use the parameters \txConst\ and \duration\ to achieve reliable
local broadcast. In this section we determine the parameter \factor,
which we use to increase the transmission probability and decrease the
time accounted for the \emph{fast} local broadcast (by multiplying and
dividing by~$\Delta \cdot$ \factor, respectively).

In contrast to the other coloring algorithms that we evaluate,
\randvd\ does not use regular local broadcasting but only a variant
that achieves success with a certain probability.  Therefore, we use
\duration~$\times$ \factor\ as the length of each phase, while using
the regular \txConst\ from local broadcasting, as all nodes transmit
during each phase. We study the length of the phases in
\cref{sec:randvd-length-phases} 
and observe that the shorter the
phases, the faster the algorithm (without any disadvantages). Thus, we
use \factor\ $=0.001$ in the following. 
For \colorred, \mwcolor, and \yucolor,~$\Delta\cdot$\factor\ is the
ratio between local broadcasting with all nodes and fast local
broadcasting with a small subset of the nodes. As the optimal value of
\factor\ may depend on the size of the sets, how often this mode of
transmission is used, and possibly other factors, we determine this
parameter for each of the algorithms separately.  The results using
\num{1000} nodes and the random deployment scheme are given in
\cref{ch:exp:tab:factor}.
\begin{table}[hbt]
\centering
\tablefontsize
\caption{\tablefontsize Average number of conflicts and average runtime using different
  parameters \factor.}
\label{ch:exp:tab:factor}
\begin{tabular}{llrrrrrrr}
\toprule
\factor &      & {\bf \num{0.05}} & {\bf \num{0.1}} & {\bf \num{0.2}}  & {\bf \num{0.3}} & {\bf \num{0.4}} & {\bf \num{0.6}} & \num{0.8} \\ \midrule
\multirow{2}{*}{\colorred}  & conflicts & \num{0.00}& \num{0.04}& \num{0.10}& \num{0.00}& \num{0.12}& $\mathbf{0.51}$& \num{2.47}\\ 
& runtime & \num{339013}& \num{171099}& \num{87924}&
                                                            \num{59995}& \num{46266}& $\mathbf{32224}$& \num{25384}\\ 
\multirow{2}{*}{\mwcolor} & conflicts & \num{0.1}   & \num{0.1}   & $\mathbf{0.4}$   & \num{1.0}  & \num{1.5} & \num{2.7} &- \\
                            & runtime   & \num{81195} & \num{44700} & $\mathbf{27982}$ & \num{23995} & \num{22807} & \num{21870} &- \\
\multirow{2}{*}{\yucolor} & conflicts & \num{0.6}   & \num{0.7}   & $\mathbf{1.4}$   & \num{2.9}  & \num{6.4} & \num{22.4} &-\\
                            & runtime   & \num{286167} & \num{160088} & $\mathbf{99946}$ & \num{82707} & \num{72660} & \num{67131} &-\\ 
\bottomrule
\end{tabular}
\end{table}

We observe that the number of conflicts increases with the parameter
\factor, while the runtime decreases. We select the parameter \factor\
$= 0.6$ for \colorred\ and \factor\ $= 0.2$ for \mwcolor\ and
\yucolor. This results in an average number of conflict of~$0.51$,
$0.4$ and~$1.4$ after an average runtime of~\num{32224},~\num{27982}
and~\num{99946} time slots for \colorred, \mwcolor\ and \yucolor.

\subsection{\randvd\ and its Variants}
\label{sec:randvd-its-variants}

Let us now compare the different variants of \randvd\ we described in
\cref{ch:exp:sec:rand4d-intro}. We use \factor\ $= 0.001$ as
determined previously. The result of this comparison is given as
\cref{ch:exp:tab:rand-runtimes} for the random
distribution. 
Clearly, the basic \randvd\ algorithm is the fastest with a runtime of
only~$\num{1256}$ time slots. This was expected as the variants either
improve the resulting coloring or make the algorithm more robust
regarding a specific setting. For \randrespect\ the nodes wait
\duration\ time slots before selecting a color in order to learn
colors already selected by neighbors, while \randfinal waits for
\duration time slots after it detected the last conflict before
finalizing its color. Thus both variants require approximately the
runtime of \randvd plus \duration:~\num{5668} and~\num{5865} time
slots, respectively. \randed\ reduces to set of available colors,
which leads to more conflicts. Thus the runtime of~\num{4174} time
slots is surprisingly good.  For these algorithms the number of
finished nodes corresponds to the number of nodes with a valid color,
as these variants do never finalize their color. Only for \randfinal\
the value corresponds to the number of nodes that finalized their
color.

\subsection{Does \colorred\ Require Valid Colors?}
\label{ch:exp:sec:cr-exp}
\renewcommand{\figpath}{plots/cr/}

The initial color of \colorred\ must correspond to a valid node
coloring (which is pre-computed at no cost). The variant \colorrand,
however, simply draws a random number from the a set of colors. We
study in this section whether this potentially invalid initial
coloring can still be used to compute a valid~$(\Delta+1)$
coloring. We measure the average runtime and number of conflicts for
the sets~$\{0,1\dots,c\cdot\Delta\}$, with~$c=1,2,3$\footnote{For
  $c>1$ we write~$c\Delta$ instead of~$c(\Delta+1)$ for brevity.}. We
report the results in \cref{ch:exp:tab:cr-colorrand} for the random
deployment.
\begin{table}[hbt]
\begin{minipage}{0.5\textwidth} 
\centering
\tablefontsize
\caption{\tablefontsize Comparison of average runtime and average number of conflicts of our \randvd\ variants }
\label{ch:exp:tab:rand-runtimes}
\begin{tabular}{lrr}
\toprule
  & runtime &  conflicts\\ 
  \midrule
  \randvd        &  $\num{1256}$  & $\num{0.00}$              \\ 
  \randrespect & $\num{5668}$        & $\num{0.00}$              \\ 
  \randfinal   &  $\num{5865}$        & $\num{0.00}$              \\ 
  \randed   & $\num{4174}$      & $\num{0.00}$
  \\ 
  \bottomrule
\end{tabular}
\end{minipage}
\begin{minipage}{0.5\textwidth} 
\centering
\tablefontsize
\caption{\tablefontsize Comparing \colorred\ and \colorrand\ for a varying 
  number of colors in the initial coloring.}
\label{ch:exp:tab:cr-colorrand}
\begin{tabular}{lrrrrrrr}
\toprule
Number of initial colors      & $\Delta+1$ & $2\Delta$ & $3\Delta$ \\ \midrule
\colorred \\
\;\;\;\; conflicts & \num{3.29}& \num{0.86}& \num{0.55}\\ 
\;\;\;\; runtime & \num{18638}& \num{24824}& \num{32197}\\ 
\colorrand \\
\;\;\;\; conflicts & \num{6.52}& $\mathbf{0.93}$& \num{0.65}\\ 
\;\;\;\; runtime & \num{19766}& $\mathbf{24758}$& \num{32287}\\ 
\bottomrule
\end{tabular}
\end{minipage}
\end{table}

The results indicate that \colorred\ does not require a valid coloring
to perform well in practice. We see a significant increase in the
number of conflicts for \colorrand\ only for colorings of
cardinality~$\Delta+1$, however, assuming a valid~$\Delta+1$ coloring
to be given renders executing the algorithm unnecessary.  For
colorings of size larger than~$2\Delta$ the difference in the number
of conflicts and the runtime of the algorithms is negligible. Also, we
observe that the smaller the color set the faster the algorithm. As
computing a valid coloring additionally requires some effort, we focus
on the variant \colorrand\ with a random~$(2\Delta)$-coloring in the
following.

\subsection{Correcting Variants}
\label{ch:exp:sec:correcting-variants}

In this section we consider heuristic improvements to \colorrand,
\mwcolor, and \yucolor, namely \colorcor, \mwcor, and \yucor. With
these heuristics, we aim at making the algorithms both faster and more
robust towards failures in the communication. Instead of ignoring
detected color conflicts as it is mostly done in the basic algorithms,
we actively deal with them and try to resolve the conflicts,
cf. \cref{sec:algos:correcting-variants}.  It is obvious that,
although the number of conflicts may be reduced, the runtime of the
algorithm increases for these variants. However, as the algorithms are
able to detect and resolve conflicts, we reduce the time accounted for
a successful transmission to decrease the runtime while introducing
some (hopefully temporary) conflicts.  We do this by reducing the
parameter \duration\ to a fraction of the value and denote the reduced
parameter by \duration'. We report the results for the random
deployment in \cref{ch:exp:tab:correcting-runtime-conflicts}.
\begin{table}[hbt]
\centering
\tablefontsize
\caption{\tablefontsize  Average runtime and conflicts by the correcting variants. We used varying fractions of the
  \duration\ parameter (denoted by \duration') and the random deployment. }
\label{ch:exp:tab:correcting-runtime-conflicts}
\begin{tabular}{llrrrrrr}
  \toprule
  \multicolumn{2}{l}{Fraction of \duration} & $1/32$  & $1/16$ & $1/8$   & $1/4$   & $1/2$   & $1$     \\ 
  \multicolumn{2}{l}{Resulting \duration'} & $143$  & $287$ & $575$
                                                                         & $1150$   & $2300$   & $4600$     \\ \midrule
  \multirow{2}{*}{\colorcor} & conflicts & \num{0.00}& \num{0.00}& $\mathbf{0.00}$& \num{0.00}& \num{0.00}& \num{0.00}\\ 
& runtime & \num{8965}& \num{6984}& $\mathbf{6489}$& \num{8883}& \num{14348}& \num{25218}\\ 
  \multirow{2}{*}{\mwcor}& conflicts & \num{0.13}& \num{0.23}& $\mathbf{0.10}$& \num{0.02}& \num{0.02}& \num{0.00}\\ 
& runtime & \num{11065}& \num{7688}& $\mathbf{6834}$& \num{9105}& \num{15762}& \num{31027}\\ 
  \multirow{2}{*}{\yucor} & conflicts & \num{4.62}& $\mathbf{1.23}$& \num{0.28}& \num{0.09}& \num{0.00}& \num{0.00}\\ 
& runtime & \num{4370}& $\mathbf{9807}$& \num{16652}& \num{29635}& \num{58189}& \num{116646}\\
  \bottomrule
\end{tabular}
\end{table}

For the standard case of unchanged \duration\ we observe that the
average number of conflicts decreases from~$0.93$,~$0.4$ and~$1.4$
(cf. \cref{ch:exp:tab:cr-colorrand,ch:exp:tab:factor}) to~$0.00$ for
\colorcor, \mwcor\ and \yucor, while the runtime increases as
expected. Using a smaller parameter \duration', however, the
correcting variants are able to improve upon the basic algorithms. For
\colorrand\ the correcting variant achieves to compute a coloring
without conflicts even for very small values of \duration'. The best
runtime is obtained using \duration' $= 575$, for which the algorithm
computes a~$(\Delta +1)$-coloring in~\num{6489} time slots. Similarly,
\mwcor\ achieves to~$\O(\Delta)$-color the network even for the
smallest considered \duration' values essentially without a
conflict. The (very) small number of average conflicts we observed is
probably due to not-yet detected conflicts. As \mwcolor\ does not
store the neighbors colors new conflicts occur more frequently than in
\colorcor.The best runtime of \mwcor\ is also achieved for \duration'
$= 575$, resulting in~\num{6834} time slots.  \yucor, on the other
hand, does not achieve a coloring without conflict for the smaller
\duration' values. This is due to the blocking of nodes due to the
DoNotTransmit messages. For small \duration' values, some nodes are
not able to receive the StartTransmit message and remain blocked for
remaining algorithm. The best runtime is achieved for \duration'
$=287$ with only~\num{4370} time slots, however, leading to an average
number of conflicts of~$4.62$. Thus we select \duration'~$=287$,
resulting in an~$(\Delta+1)$-coloring algorithm with an average number
of~$1.23$ conflicts after~\num{9807} time slots (which is still an
order of magnitude faster than the basic algorithm).

\subsection{Comparing the Progress of the Algorithms}
\label{sec:progress-algos}

Before comparing the runtime of all algorithms, we briefly study the
progress of  \randvd, the algorithms \colorrand, \mwcolor, and \yucolor, and
their correcting variants. The average progress (i.e., the average
number of finished nodes over time) of all algorithms is illustrated
in \cref{ch:exp:fig:correcting-progress}.
\renewcommand{\figpath}{plots/correcting/}
\begin{figure}[hbt]
  \centering 
\begin{subfigure}[b]{0.31\textwidth}
    \centering
    \renewcommand{\figpath}{plots/rand4d/}
    \includegraphics[width=1\textwidth, trim=10mm 10mm 25mm 20mm,
    clip]{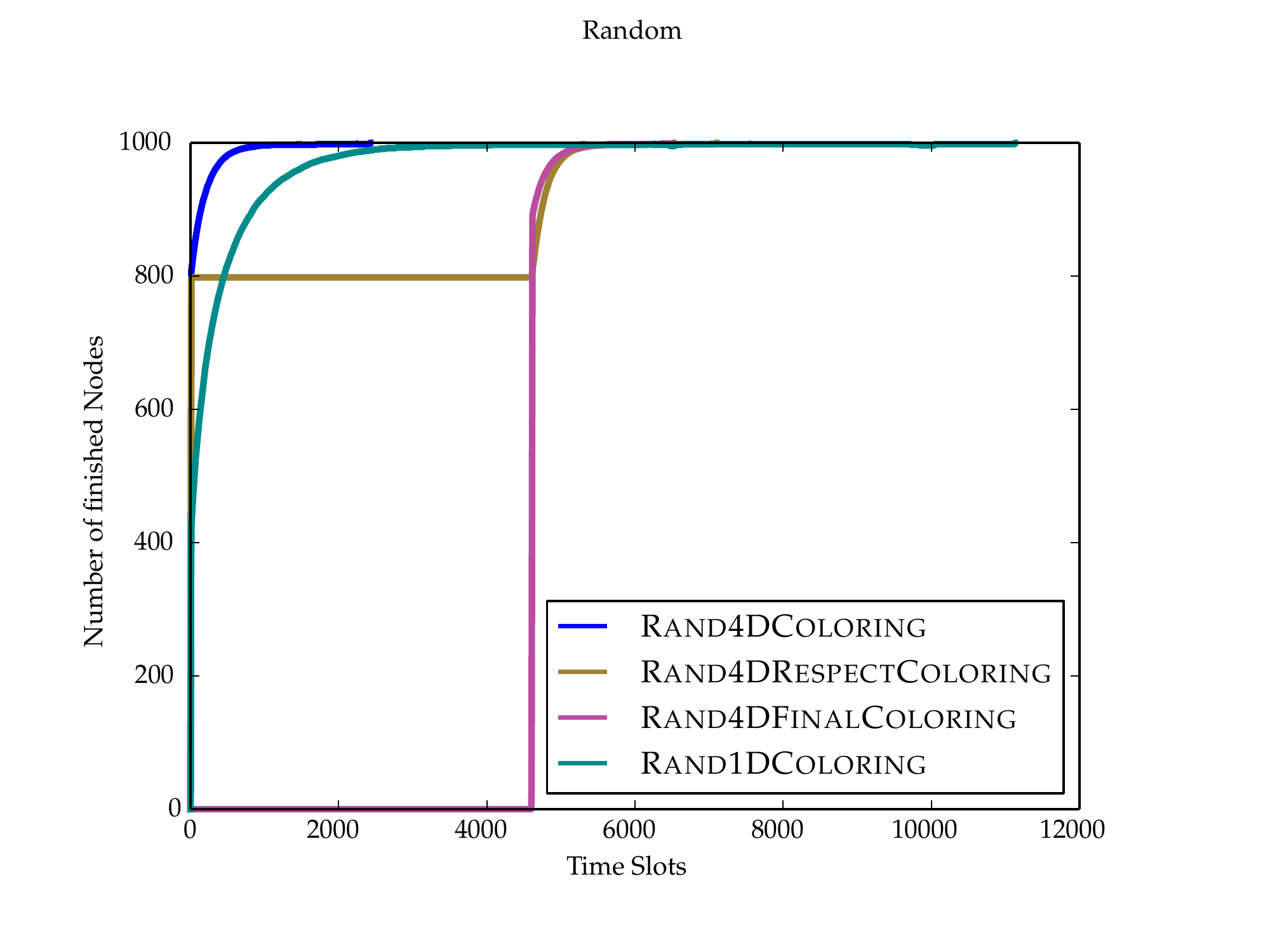}
  \caption{Progress of the {\sc Rand}-variants}
\label{ch:exp:fig:rand-progress}
 \end{subfigure}
  \hfill
  \begin{subfigure}[b]{0.31\textwidth}
    \centering
    \includegraphics[width=1\textwidth, trim=10mm 10mm 25mm 20mm,
    clip]{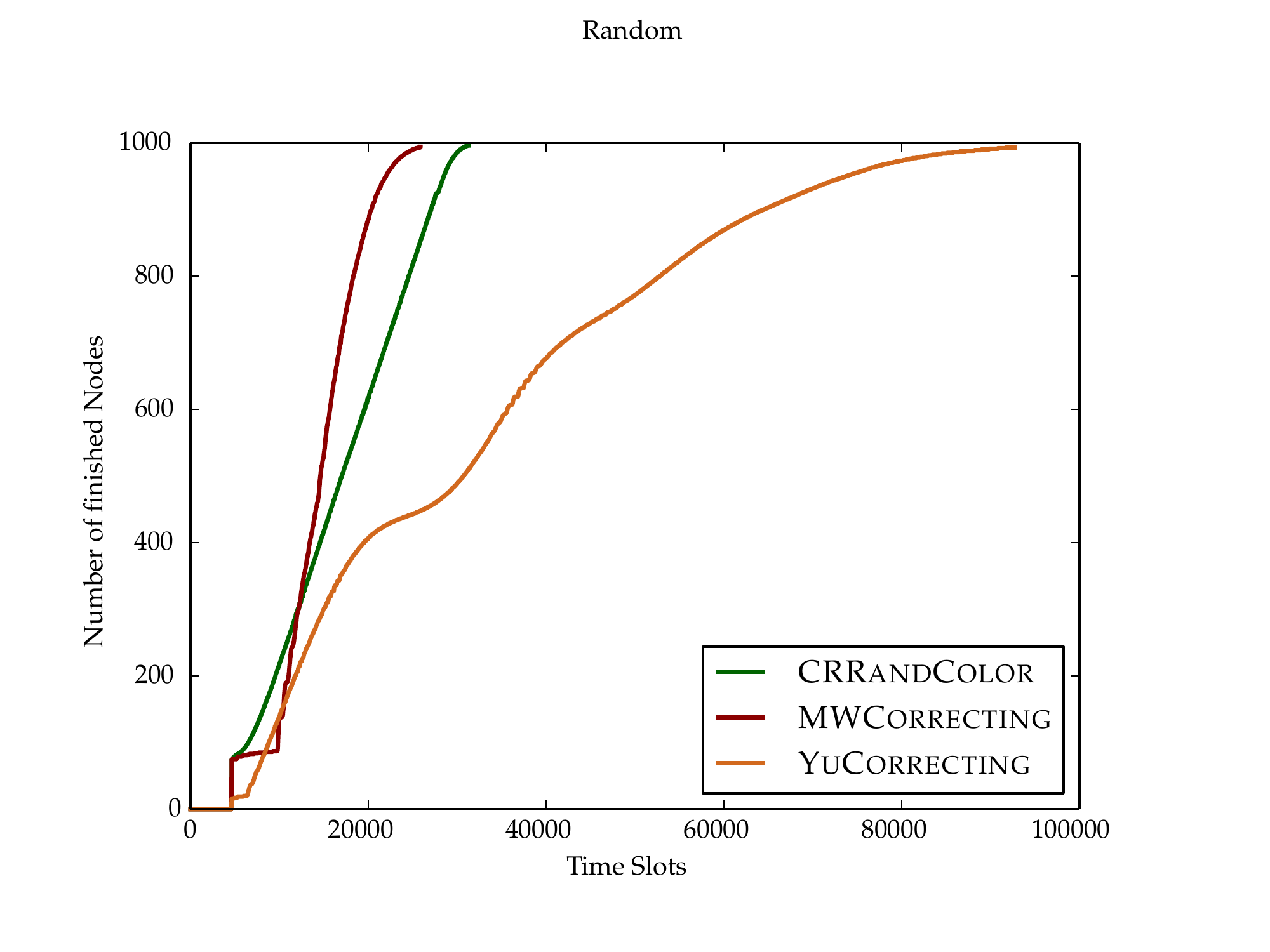}
  \caption{Progress of basic algorithms}
\label{ch:exp:fig:correcting-base-progress}
 \end{subfigure}
  \hfill
  \begin{subfigure}[b]{0.31\textwidth}
    \centering
    \includegraphics[width=1\textwidth, trim=10mm 10mm 25mm 20mm,
    clip]{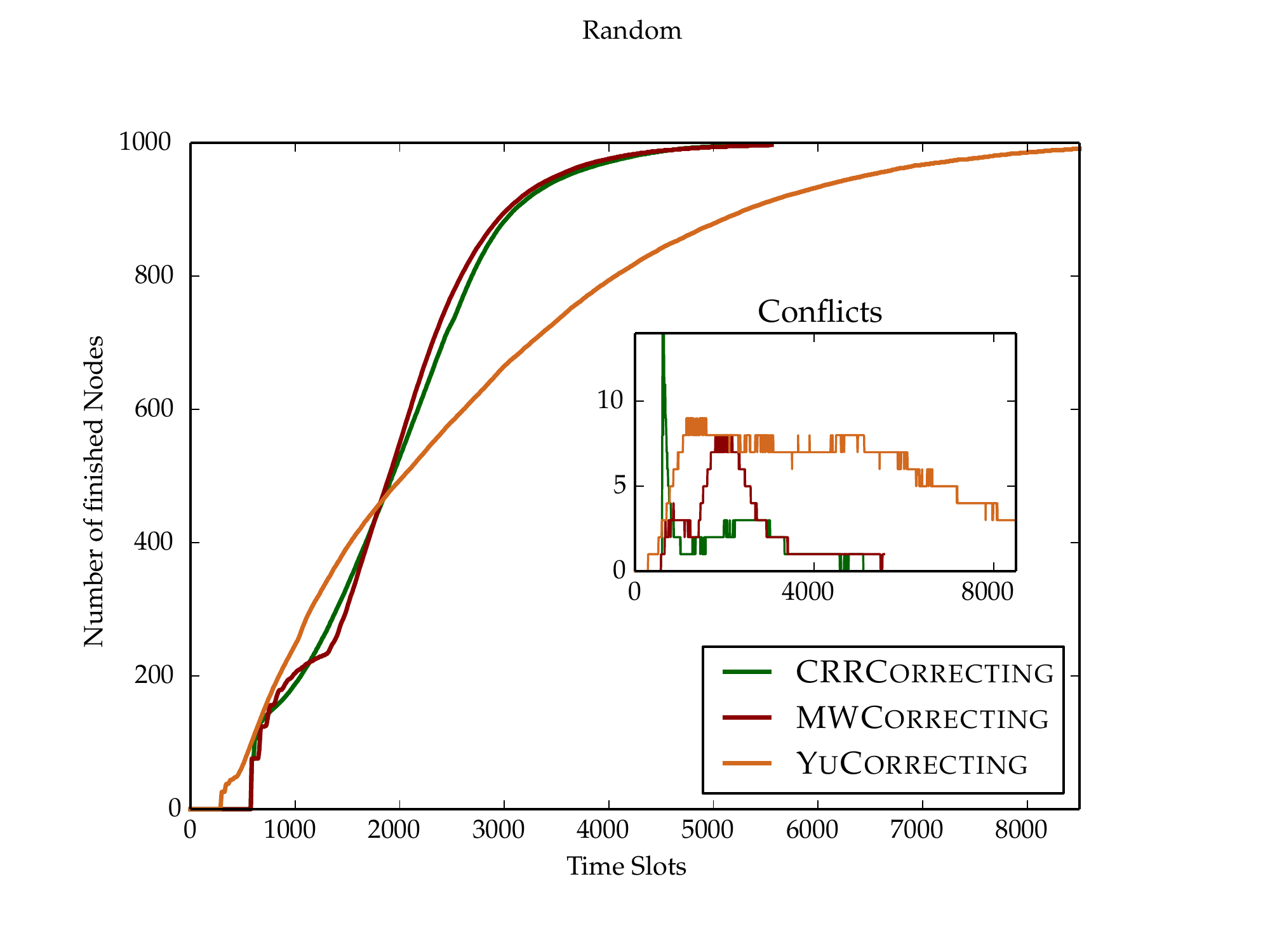}
   \caption{Progress of correcting variants}
\label{ch:exp:fig:correcting-variants-progress}
 \end{subfigure}
 \caption{Progress of the algorithms \colorred, \mwcolor, \yucolor\
   (left) and their correcting variants (right). For the correcting
   variants we additionally depict the average number of conflicts.}
\label{ch:exp:fig:correcting-progress}
\end{figure}

We observe in \cref{ch:exp:fig:rand-progress} that while \randvd\ and
\randed\ are able to make fast progress after the algorithms started,
the two remaining variants do not resolve a conflict or finalize a
color for a little more than $\num{4000}$ time slots before finishing
within the next $\num{1000}$-$\num{2000}$ time slots.  This behaviour
is due to two reasons. For \randrespect, which is expected to be more
robust in the case of heterogenous wake-up patterns, this is caused by
a listening period, during which the colors of neighbors are received
and stored. Afterwards, this variant selects its color so it does not
coincide with the latest received color from any of the neighbors. For
\randfinal\ it is due to a waiting period before finalizing the
current color. This waiting period of a node is reset with each
conflict it detects and allows the node to decide when a selected
color can be finalized. In both algorithms the waiting or listening
period is set to exactly \duration\ time slots.  Finally, \randed\
decreases the number of used colors to $\Delta+1$, resulting in a
runtime of $\num{4174}$ time slots. Note that reducing the number of
colors to $\Delta+1$ results in a higher variance in the runtime, as
the probability to select a color of one of the neighbors increases.

Regarding the remaining algorithms, the first, sudden increase is
exactly after \duration\ (or \duration') time slots.  This is when the
first nodes enter the MIS, become leader and select their color. As
the MIS is computed regarding a three times larger broadcasting range
in \yucolor, a lot less nodes enter the MIS in this algorithm. Some
time after the MIS computation, the nodes around the leaders select
their colors. While this allows both \colorrand\ and \mwcolor\ to
gradually color all nodes in the network, several plateaus can be
observed for \yucolor. This is due to the fact that in \yucolor\ not
all nodes are able to select a color after the first MIS execution, as
the MIS is computed regarding a larger broadcasting range than the
neighborhood around the leaders which is allowed to select a
color. Thus, we see at least two more MIS executions leaving their
traces in this progress around time slots \num{20000} and \num{40000}
for \yucolor\ in \cref{ch:exp:fig:correcting-base-progress}. Although
this happens also in \yucor, we cannot clearly observe the moment it
happens in the average progress depicted in
\cref{ch:exp:fig:correcting-progress}.
 
For \colorrand\ and \mwcolor, the leaders dominate the whole network,
resulting in each node requesting an active interval or a color
interval from its respective leader.  Let us compare the progress of
\colorrand\ and \mwcolor. Recall that leaders in \colorrand\
coordinate the time interval in which dominated nodes compete to
select a final color, while in \mwcolor\ the leaders coordinate which
colors the dominated nodes compete for.  This fundamental difference
allows \colorrand\ to use fast local broadcasting for the second level
MIS (which can be seen as a competition for being allowed to select a
valid color). The coordination of active nodes based on the nodes'
initial color leads to an almost perfectly linear increase of the
number of finished nodes over time.  In \mwcolor, the nodes compete in
fewer but slower MIS-executions for the colors, with leads to the
rapid increase once the threshold is reached (in
\cref{ch:exp:fig:correcting-base-progress} at around~\num{9000} time
slots). Note that the rapid increase once the threshold is met is due
to the requirement of achieving a valid node coloring with very few
conflicts, however, it also hints that faster progress is possible, as
shown by our correcting variants.

Regarding our correcting variants, we observe that the progress of
\colorcor\ and \mwcor\ is very similar. One reason for this is that
both use the same \duration' value of $575$ time slots. The second
reason is that by reducing the time accounted for local broadcasting,
all the slack is removed from the algorithms. As both algorithms elect
leaders, which then allow a dominated node to either be active or
compete for certain colors (depending on the algorithm), the remaining
progress essentially shows this similarity.  

Inside \cref{ch:exp:fig:correcting-variants-progress} we additionally
show the number of conflicts occurring in the respective algorithms
over time. We observe that in general, the number of simultaneous
conflicts is relatively low with well below $15$ conflicts at each
time.
For \colorcor\ there is a peak at or around the leader election phase,
indicating that too many nodes entered the MIS. A smaller peak is also
visible for \mwcolor, however, here more nodes fail to select a valid
color in the following color competition. For \yucor, the number of
conflicting nodes is relatively stable at around $8$ simultaneous
conflicts. Although the number of conflicts decreases, not all can be
corrected, as some conflicts are due to blocked nodes\footnote{Recall
  that we set the leader to a quit-state and select color $0$ once the
  leader of a blocked node resigns from its leader functionality
  (cf. \cref{ch:exp:sec:yu-intro}).}.

\subsection{Performance Comparison of Coloring Algorithms}
\label{ch:exp:sec:coloring-comparison}

To compare the performance of the algorithms on the different
deployment strategies, we show the runtime and the number of conflicts
of the algorithms on the different deployments in
\cref{ch:exp:tab:comparison-runtime-conflicts}. For the random
deployment, the values are the best values from
\cref{ch:exp:tab:correcting-runtime-conflicts,ch:exp:tab:factor,ch:exp:tab:cr-colorrand,ch:exp:tab:rand-runtimes}. For
the remaining distributions we selected the values analogously,
cf. \cref{app:exp:sec:all-distributions} for detailed results. (To get
even more insight in the behaviour of the algorithms we study their
progress, also in the full version.)
\begin{table}[!hbt]
\centering
\tablefontsize
\caption{\tablefontsize Average runtime and number of conflicts for
  all considered algorithms in all considered deployment strategies.}
\label{ch:exp:tab:comparison-runtime-conflicts}
\begin{tabular}{llrrrrrrr}
  \toprule
  \multicolumn{2}{l}{Distribution} & R  & G & PG   & C   &  C\&R  & C\&G &
  C\&PG \\ \midrule
\multirow{2}{*}{\randvd}  & 
conflicts     & \num{0.00}&\num{0.00}&\num{0.00}&\num{0.00}&\num{0.00}&\num{0.00}  & \num{ 0.00} \\ 
 &     runtime       &\num{1256}&\num{974}&\num{1372}&\num{3321}&\num{2316}&\num{2186} &\num{2016} \\ 
\multirow{2}{*}{\randed} &
     conflicts     & \num{0.00}&\num{ 0.00  }&\num{ 0.00   }&\num{ 0.00   }&\num{ 0.00   }&\num{ 0.00}& \num{ 0.00}  \\
 &  runtime &\num{4174}&\num{3358}&\num{4548}&\num{11822}&\num{8153}&\num{8211} &\num{6627}\\ 
\multirow{2}{*}{\colorrand} &
     conflicts     & \num{0.93   }&\num{ 0.82  }&\num{ 0.10   }&\num{ 1.05   }&\num{ 1.83   }&\num{ 1.03}  & \num{3.23}\\
 &  runtime       &\num{ 24758 }&\num{20367}&\num{27817}&\num{67881}&\num{42034}&\num{42692} &\num{42385}\\ 
\multirow{2}{*}{\colorcor} &
     conflicts     & \num{ 0.00 }&\num{0.00}&\num{0.00}&\num{0.00}&\num{0.00}&\num{0.00}&  \num{0.02}\\ 
 &  runtime        &\num{ 6489}&\num{7017}&\num{7017}&\num{17802}&\num{11884}&\num{11333} &\num{10896}\\ 
                                                                                                        
\multirow{2}{*}{\mwcolor} &
     conflicts     & \num{0.42}&\num{0.26}&\num{0.28}&\num{0.56}&\num{0.44}&\num{0.34}  &\num{1.06} \\ 
  & runtime       &\num{ 27982}&\num{25456}&\num{32812}&\num{73670}&\num{46363}&\num{47041} &\num{46142}\\ 
                                                                                                        
\multirow{2}{*}{\mwcor} &
     conflicts     & \num{ 0.10  }&\num{0.21}&\num{0.12}&\num{0.45}&\num{0.44}&\num{0.36}  & \num{0.42}\\ 
  & runtime        &\num{ 6834}&\num{5741}&\num{7567}&\num{23531}&\num{12780}&\num{13535} &\num{13353}\\ 
                                                                                                        
\multirow{2}{*}{\yucolor} &
     conflicts     & \num{ 1.39  }&\num{2.01}&\num{1.12}&\num{0.9}&\num{0.88}&\num{0.94} & \num{1.30} \\
  & runtime        &\num{ 99946 }&\num{113105}&\num{129054}&\num{164839}&\num{122267}&\num{141003} &\num{126488}\\ 
                                                                                                        
\multirow{2}{*}{\yucor} &
     conflicts     & \num{  1.23 }&\num{1.77}&\num{2.15}&\num{0.74}&\num{0.81}&\num{1.23}  & \num{0.64}\\
 &  runtime        &\num{ 9807}&\num{8654}&\num{11479}&\num{20849}&\num{15060}&\num{15835} &\num{14620}\\ 
\bottomrule
\end{tabular}
\end{table}

Regarding the conflicts we tried to select the parameters so that the
number of conflicts are low. Only for \yucolor\ and \yucor\ we allowed
a slightly higher number of average conflicts to reduce the runtime,
if possible. Thus, we focus on the runtime in the following. The
results of the algorithms are very consistent throughout the different
deployments. This indicates that the communication parameters are
sufficiently well chosen to allow the algorithms to deliver their
performance without being constrained by, for example, congestion
problems.

We do not depict the basic variant \colorred\ in
\cref{ch:exp:tab:comparison-runtime-conflicts} as the performance is
essentially the same as \colorrand\
(cf. \cref{ch:exp:tab:cr-colorrand}). 
As \colorrand\ and \colorcor\ compute~$(\Delta+1)$-colorings,
\yucolor\ and \yucor\ are its main competitors. A valid coloring of
the same size is also computed by our variant \randed\ that
heuristically reduces the number of available colors in \randvd\ to
$(\Delta+1)$, however, we discuss this variant later.  \colorrand\
computes a~$(\Delta+1)$-coloring using between~\num{20367} and
\num{67881} time slots. The correcting variant \colorcor\ reduces the
runtime to values between~\num{6489} and~\num{17802} time slots. This
is at par with \mwcolor\ and \mwcor, and significantly less than
\yucolor\ and \yucor. The basic algorithm \yucolor\ requires
between~\num{99946} and~\num{164839} time slots, while \yucor\ reduces
the runtime to values between~\num{8654} and~\num{20849} time slots.

Our other algorithm, \randvd, computes a~$(4\Delta)$-coloring, hence
\mwcolor\ and its variant \mwcor\ are its main competitors. Depending
on the deployment strategy, \randvd\ achieves a runtime between
\num{974} and~\num{3321} time slots. This is by far superior to the
runtime achieved by both \mwcolor\ and \mwcor. \mwcor\ requires
between~\num{5741} and~\num{23531} time slots, which is between~$4$
and~$5$ times the runtime of \randvd.  \mwcolor\ achieves a runtime
between~\num{25456} and~\num{73670} time slots.  Note that \randvd\
does not finalize the colors, however, even\footnote{Tentative
  experiments indicate faster runtimes if a reduced
  parameter \duration' is used.} using a \duration\ for finalizing the color,
the variant \randfinal\, which finalizes colors, achieves a runtime
between~\num{4354} and~\num{16110} time slots
(cf. \cref{ch:exp:tab:rand-runtimes}).

Reducing the number of available colors in \randvd\ to
only~$(\Delta+1)$ colors yields a~$(\Delta+1)$-coloring heuristic,
which we denote by \randed.  The heuristic selects a random color from
$[\Delta]$ by resolving conflicts once detected and requires only
between~\num{3358} and~\num{11822} time slots.  Thus it achieves the
lowest runtimes of all considered~$(\Delta+1)$-coloring algorithms. As
mentioned, however, \randvd\ and its variants do not finalize their
colors. Hence, if finalization of the colors is required \colorcor\
achieves the best results for~$(\Delta+1)$-coloring algorithms.

We conclude from our comparison that the best performance for
$\O(\Delta)$-colorings is achieved by \randvd\, and for
$(\Delta+1)$-colorings by \randed\ and \colorcor, depending on the
exact setting.

\subsection{Coloring in Dynamic Networks}
\label{ch:exp:sec:mobility}

In this section we consider two scenarios, in the first one we allow
the nodes to move, which forces the algorithms to maintain the
validity of the coloring in a dynamically changing network. The
algorithms cannot maintain the coloring valid at each node, as the
neighborhood changes in an unforeseen manner. Thus, we study how large
the fraction of nodes is that maintains a valid color despite the
dynamic changes.  In the second scenario a fraction of the nodes wake
up after the remaining part of the network has already selected a
color. Thereby we evaluate how well the algorithms can cope with
highly asynchronous wake-up schemes. We restrict ourselves to the
random deployment for these experiments and do not use pre-computed
position files for the wake-up scenario.  

In this section we consider only $250$ nodes on a deployment
area of~$\SI{500}{\meter} \times \SI{500}{\meter}$, which results in a
similar density as in the previously considered settings. We reduce
the number of nodes, as mobility increases the time required for the
simulation significantly. Due to the high complexity of updating the
relevant positions for each single event the used simulation framework
Sinalgo additionally requires the synchronous mode, in which time
slots of the different nodes are perfectly synchronized. Thus, it is
sufficient to update the positions once per time slot. The nodes are
deployed uniformly at random and move according to the RandomDirection
model, cf. \cref{sec:sinalgo-settings}. The nodes
alternate between moving and waiting time, both times are drawn
randomly and follow a Gaussian distribution with mean value of $100$
time slots and a variance of $50$. The speed of a node is also drawn
at random according to a Gaussian distribution with a mean between~$0$
and $1$ meter per time slot and a variance of $2$. For simplicity we
refer to the mean values as the \emph{node speed}. 

\begin{figure}[hbt]
  \centering
  \begin{subfigure}[b]{0.24\textwidth}
    \centering
    \includegraphics[width=1\textwidth, trim=10mm 11mm 26mm 20mm,
    clip]{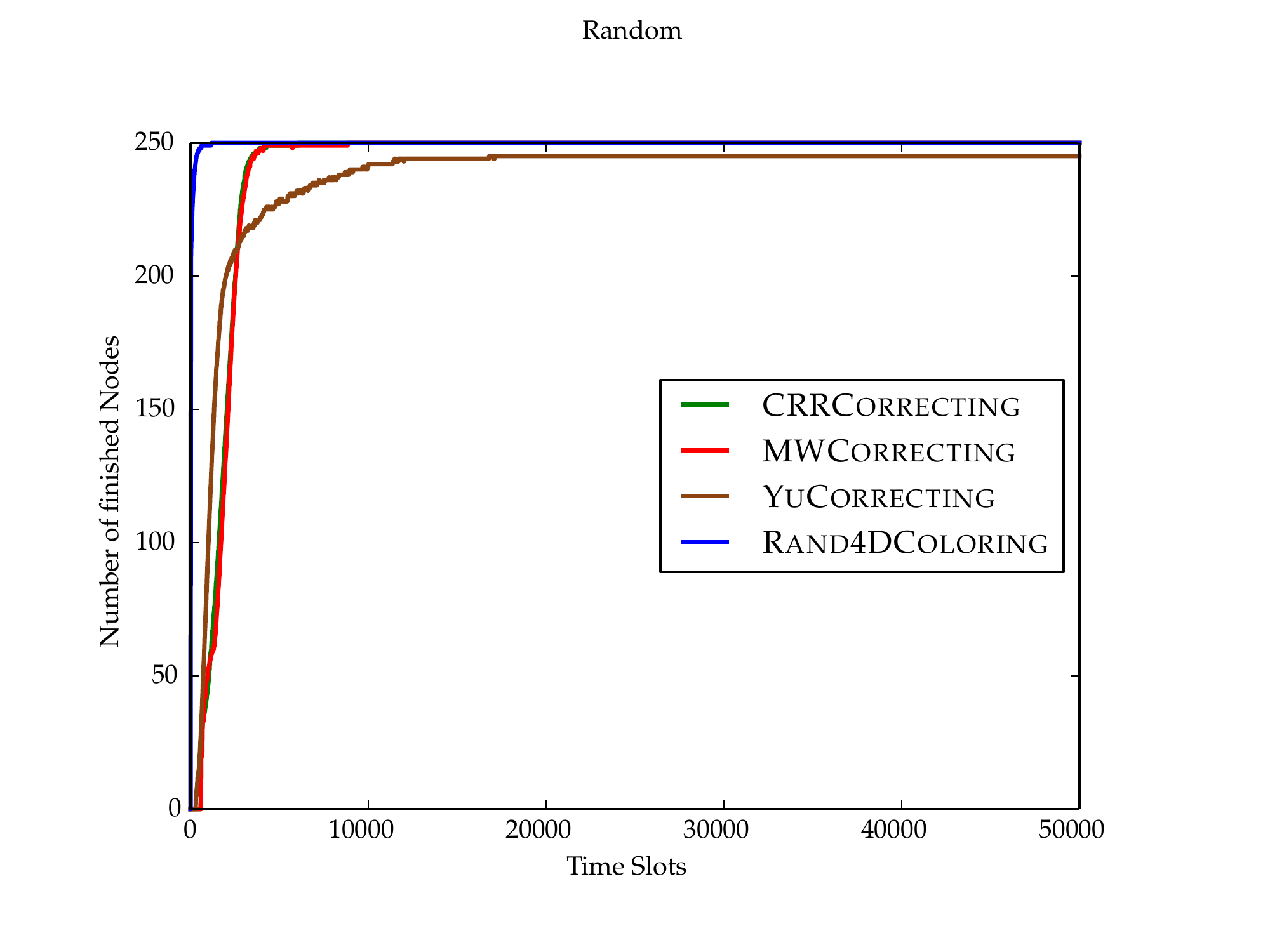}
  \caption{No mobility}
  \label{ch:exp:fig:mobility-progress-nm}
 \end{subfigure}
  \hfill
  \begin{subfigure}[b]{0.24\textwidth}
    \centering
    \includegraphics[width=1\textwidth, trim=10mm 11mm 26mm 20mm,
    clip]{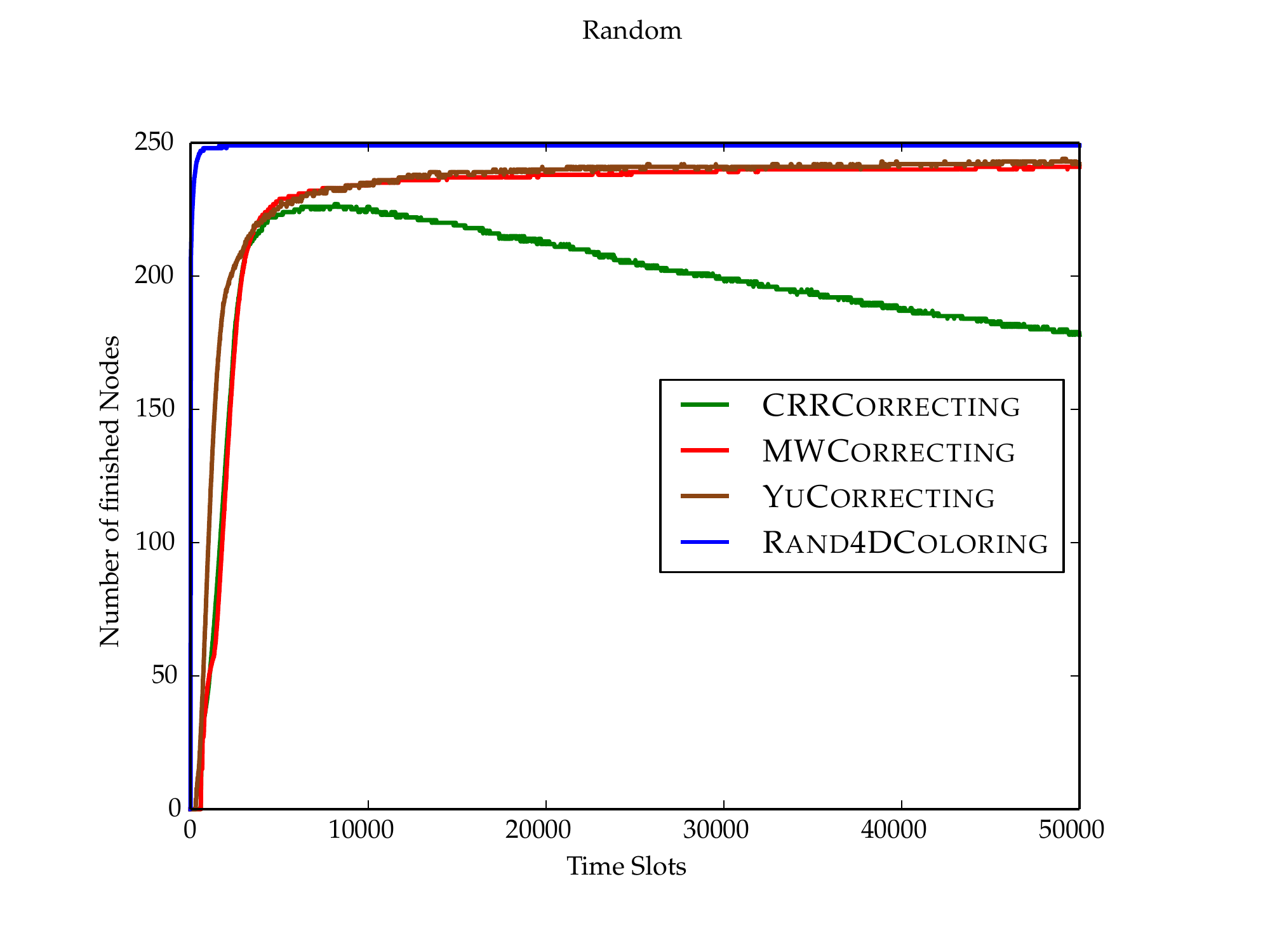}
   \caption{Node speed~$0.1$}
 \end{subfigure}
  \begin{subfigure}[b]{0.24\textwidth}
    \centering
    \includegraphics[width=1\textwidth, trim=10mm 11mm 26mm 20mm,
    clip]{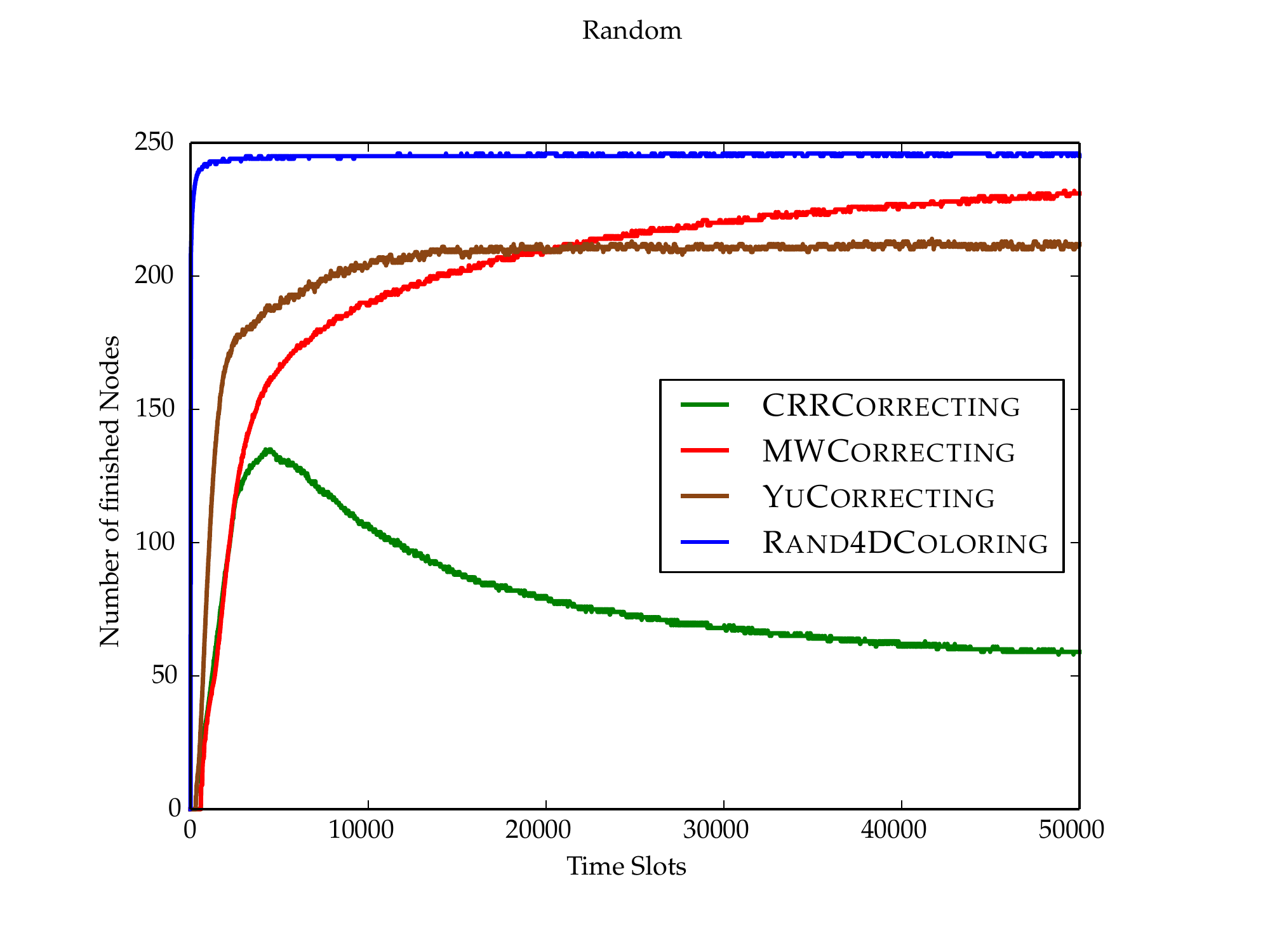}
   \caption{Node speed~$0.5$}
 \end{subfigure}
\hfill
  \begin{subfigure}[b]{0.24\textwidth}
    \centering
    \includegraphics[width=1\textwidth, trim=10mm 12mm 26mm 20mm,
    clip]{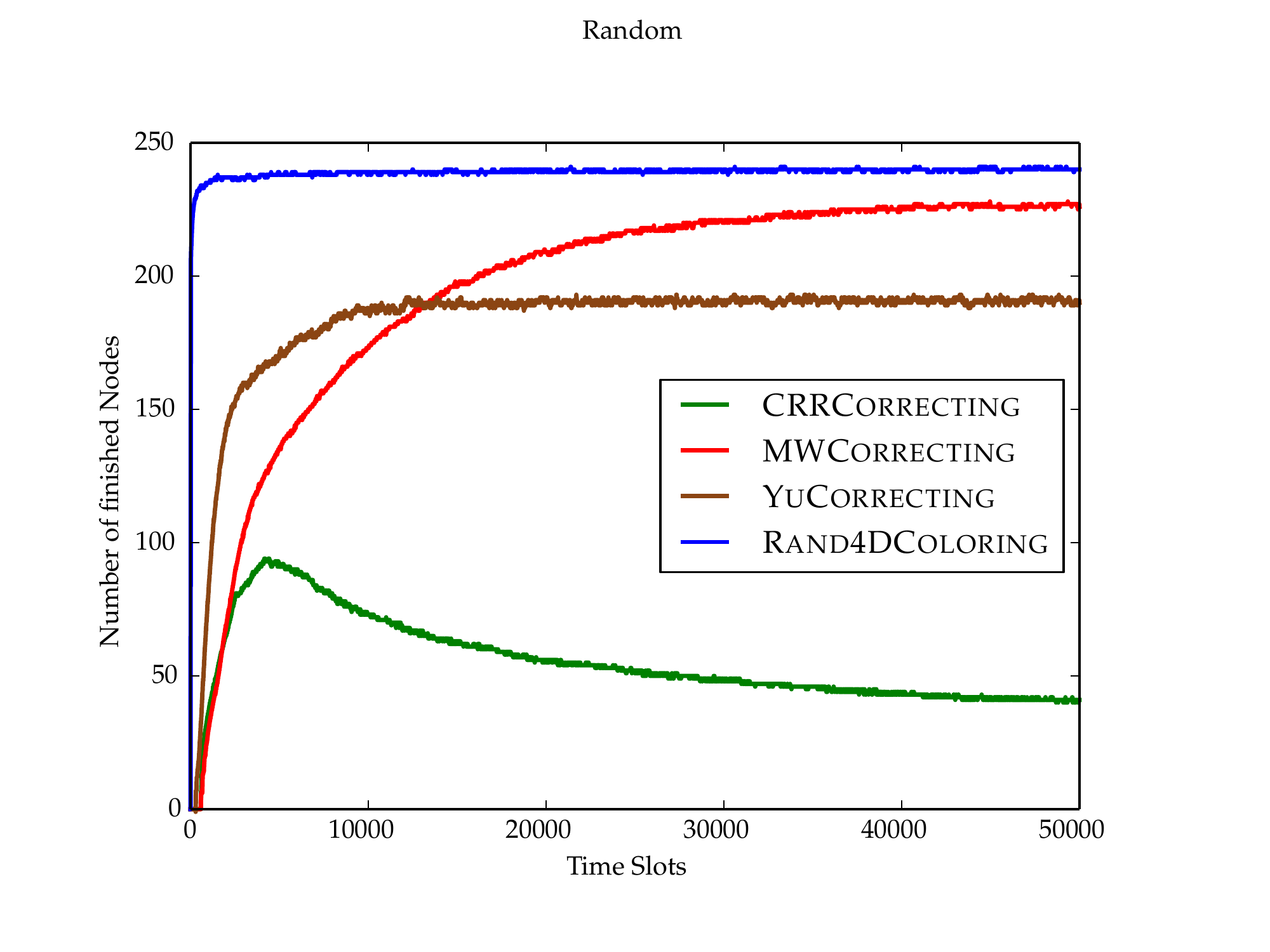}
   \caption{Node speed~$1.0$}
 \end{subfigure}
 \caption{The progress of the coloring for the different
   algorithms. We depict the average number of finished nodes over
   time for different nodes speed values drawn according to a Gaussian
   distribution with mean values between~$0$ and~$1$.}
\label{ch:exp:fig:mobility-progress}
\end{figure}

We show the number of finished nodes over time for \randvd, \colorcor,
\mwcor, and \yucor\ for node speed values of~$0$,~$0.1$,~$0.5$, and
$1.0$ in \cref{ch:exp:fig:mobility-progress}.
\renewcommand{\figpath}{plots/mobility/}

\randvd, \mwcor, and \yucor\ maintain a relatively high fraction of
valid nodes, while \colorcor\ performs poorly. The main reason for the
poor performance of \colorcor is that a relatively long time passes
between detecting the conflict and selecting a new color. Conflicting
non-leader nodes compute the next valid active interval based on the
previous active interval and the schedule length. Due to mobility,
however, the active interval does not guarantee that few nodes are
active in each interval. Instead it mainly leads to many nodes waiting
to compete for a color. This is different for both \yucor\ and \mwcor,
which reset to a state in which they can compete for a valid color
right immediately. Thus, while the schedule helps coordinating the
color selection scheme in the static setting, this is not the case for
the dynamic setting.

\yucor\footnote{In this setting we do not select color $0$ for blocked
  nodes, as nodes may leave the blocked state. As this does not happen
  without mobility, there is a small fraction of nodes not finishing
  the algorithm in \cref{ch:exp:fig:mobility-progress-nm}}, maintains
a significantly higher fraction of valid colors than \colorcor.
Non-leader nodes may get blocked and separated from the blocking node,
which could also lead to a low performance. However, the algorithm
prevents such a case. Even if a node is blocked by one leader it may
receive the StartColoring message by another leader and continue to
request a color by this other leader.  Another source of error in the
dynamic setting is, as in \colorcor, the request state, in which the
nodes depend on the one leader they selected.  However, less nodes are
affected, as the nodes are allowed to leave the request state if they
get blocked by another node.  Finally, the algorithm may benefit from
the smaller deployment area more than the other algorithms due to its
increased transmission range, however, we expect this to be not
significant.  Overall, the algorithm achieves a solid performance in
the mobile setting.

\mwcolor\ performs even a little better than \yucolor, which is
probably mainly due to the increased number of colors and thereby the
lower probability for a conflict. As before, some nodes may get stuck
in the request state, as dominated nodes select one leader and keep
trying to contact this leader, however, resetting the nodes to their
first color competition state leads to significantly less time
required to re-color the nodes than in \colorcor.

The best performance in the mobile setting is achieved by \randvd,
which maintains a very high fraction of validly colored nodes
throughout the execution. This is due to the fast runtime of the
algorithm and the fact that whenever a conflict is detected the nodes
simply try to resolve it immediately. Thus, a very high percentage of validly
colored nodes is maintained. Even for a node speed of~$1.0$ less than
\SI{4}{\percent} of the nodes have an invalid color.  This performance
can be achieved as no structure needs to be build or maintained and
detected conflicts are treated immediately by selecting a new random
color.


\subsubsection{Asynchronous Wake-up}
\label{sec:asynchronous-wake-up}

Let us now examine the robustness of the algorithm with respect to
some nodes in the network waking-up later than large parts of the
network.  We consider how many of the~$500$ already colored nodes are
disturbed by another~$100$ to~$500$ nodes waking up in the network and
executing the algorithm.  We use the random deployment on an area
of~$\SI{1000}{\meter} \times \SI{1000}{\meter}$ as in most parts of
this paper. We consider the algorithms \randvd, \randrespect\ and the
correcting variants \colorcor\ and \mwcor\ and measure the number of
already colored nodes that detect a conflict. We do not consider
\yucor, as \yucolor\ does explicitly not support the setting of nodes
waking up in late stages of the algorithm (which also resulted in
worse results in preliminary experiments).  The results of the
experiment are shown in
\cref{ch:exp:tab:wakeup}. Note that the
basic algorithms \colorred\ and \mwcolor\ are expected to produce no
or a lot less disturbance as they support late wake-up of nodes and
the reliable communication ensures that the nodes are aware of colors
selected by neighbors. However, we consider the performance of the
variants we optimized towards achieving a high performance regarding
the runtime and the number of conflicts, thus, already colored nodes
may be disturbed as we do not have reliable local broadcasting.

\renewcommand{\figpath}{plots/wakeup/}
\begin{figure}[hbt]
    \centering
    \includegraphics[width=0.7\textwidth, trim=20mm 0mm 35mm 10mm,
    clip]{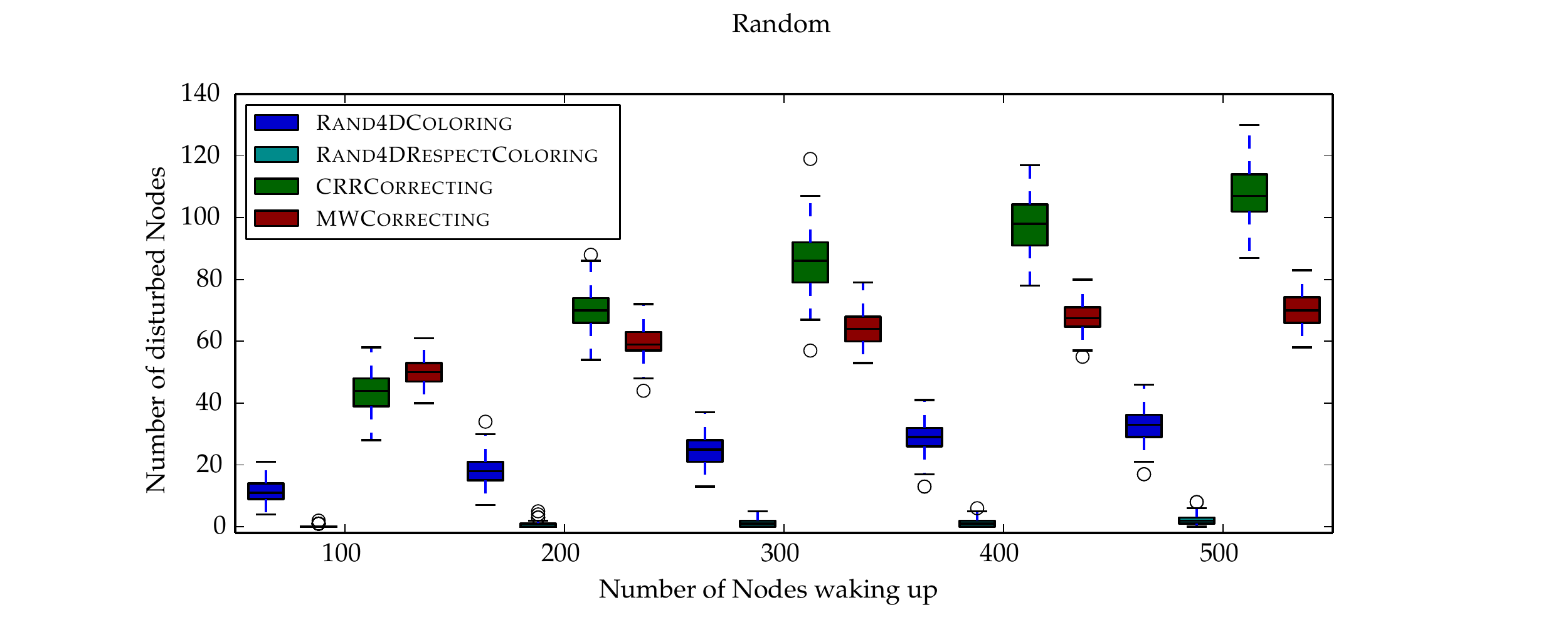}
    \caption{After~\num{500} nodes finished and agreed on a valid
      coloring a varying number of additional nodes wake up
      (x-axis). We count the number of previously finished nodes that
      are disturbed, i.e. they detected a conflict after the
      additional nodes woke up. }
\label{ch:exp:fig:wakeup}
\end{figure}
\begin{table}[htb]
\centering
\tablefontsize
\caption{The average number of disturbed nodes for a varying number
  of nodes waking up after the~$500$ pre-deployed nodes selected a
  valid color.}
\label{ch:exp:tab:wakeup}
\begin{tabular}{L{5cm}lllll}
\toprule
Number of nodes \newline waking up late & \num{100}  & \num{200}   & \num{300}   & \num{400}   & \num{500}
  \\ \midrule

\randvd & \num{11.24}& \num{18.05}& \num{24.97}& \num{28.51}& \num{32.39}\\ 
\randrespect & \num{0.14}& \num{0.72}& \num{1.12}& \num{1.54}& \num{2.34}\\ 
\colorcor & \num{43.79}& \num{69.84}& \num{85.89}& \num{98.05}& \num{107.54}\\ 
\mwcor & \num{50.12}& \num{59.89}& \num{64.55}& \num{67.93}& \num{70.15}
  \\ \bottomrule
\end{tabular}
\end{table}

For \randvd\ the results indicate that even if additional~$500$ nodes
wake up, only~$32$ already colored nodes detect a conflict.  If we
reduce the number of nodes starting that late, the number of disturbed
nodes decreases further to~$10.9$, which corresponds to about
\SI{2}{\percent} of the pre-colored nodes. Thus, although atheoretical
considerations in \cite{fp-sddcs-15} indicate that late wake-up of few
nodes may introduce many conflicts we do not observe this here.  The
\randrespect\ variant of the algorithm, which additionally requires
the nodes to listen for \duration\ time slots before transmitting
color messages, achieves essentially optimal results with only~$0.2$
to~$2.3$ disturbed nodes on average.  The remaining correcting
variants perform a lot worse, with roughly between~$40$ and~$180$
disturbed nodes. However, we expect that adding an additional
listening phase and preventing the nodes from selecting colors taken
by neighbors would significantly decrease the number of disturbed
nodes, similar to how \randrespect\ improved upon
\randvd.  

\section{Conclusion}
\label{ch:exp:sec:conclussion}

In this paper we experimentally evaluated several distributed node
coloring algorithms designed for wireless ad-hoc and sensor
networks. All algorithms operate under the realistic SINR model of
interference and compute the valid colorings in $\O(\Delta \log n)$
time slots. We used the network simulator Sinalgo \cite{sinalgo} to
study the runtime and the number of conflicts in the computed
colorings on several deployment scenarios.
We conclude that our simple~$(4\Delta)$-coloring algorithm \randvd\ is
very fast, requiring significantly less time than any other considered
coloring algorithm. Our experiments additionally show that the
algorithm is or can be made robust towards dynamic networks.
Regarding $(\Delta+1)$-colorings, both \colorred\ and \randed\ are
faster than the competing \yucolor\ algorithm. Our correcting variants
improve the runtime for all algorithms, while preserving the relative
ordering.


{\small
\bibliographystyle{IEEEtran}
\bibliography{../../bib/abbrv,../../bib/bib}
}

\appendix

A note on printing this Appendix: Printing
\cref{ch:prelim:fig:distribution-models} and 
\cref{ch:exp:fig:randvd-images,ch:exp:fig:colorred-images,ch:exp:fig:mwcolor-images,ch:exp:fig:yucolor-images}
is relatively time consuming due to the complexity of the
illustrations, however, it should still be printable in less than a
minute per affected page. 

\section{Illustration of the Deployments}
\label{app:illustr-depl}
\renewcommand{\figpath}{sinalgo-basic/jpg/}

Visualizations of the deployments are depicted in \cref{ch:prelim:fig:distribution-models}.

\newcommand{\tikzzpycolor}{yellow}
\begin{figure}[htb]
  \centering
  \begin{subfigure}[b]{0.3\textwidth}
\iftoggle{draft}{\includegraphics[width=1\textwidth,draft]{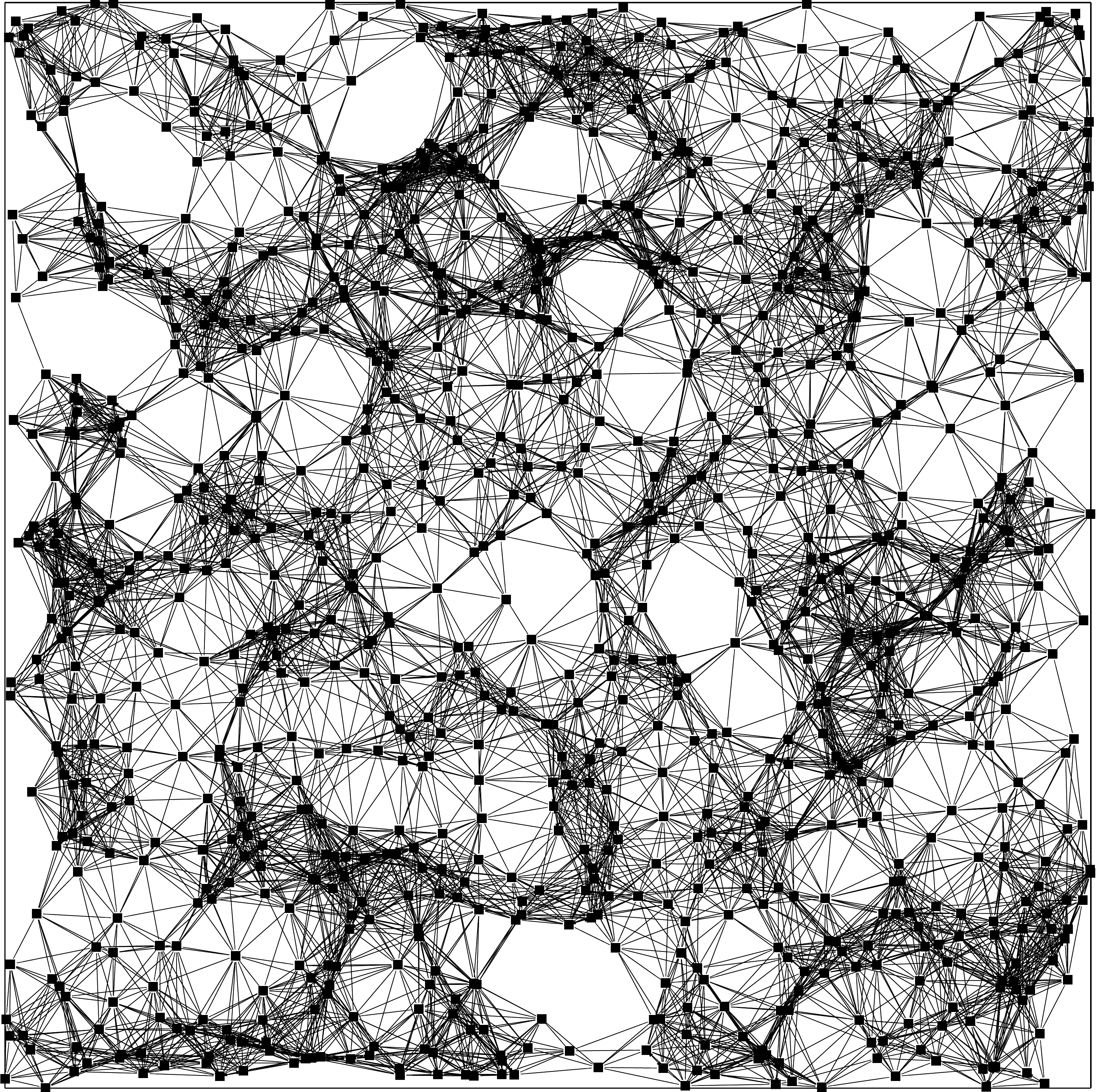}}{
\begin{tikzpicture}[spy using
  outlines={circle,\tikzzpycolor,magnification=3.0,size=2.5cm, connect
    spies, every spy on node/.append style={ultra thick}}]
\node {\includegraphics[width=1\textwidth]{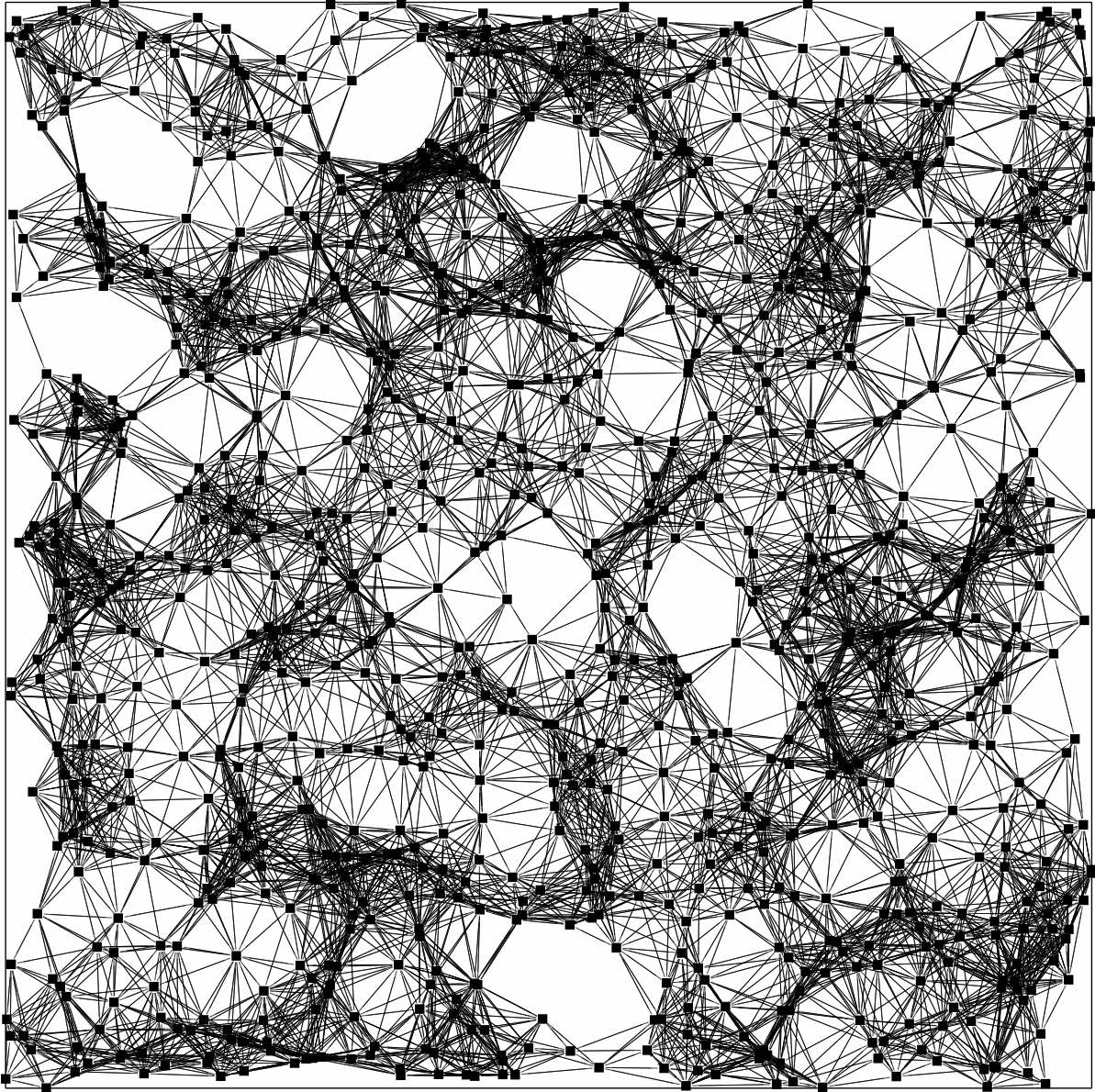}};
\spy on (0,-1) in node [fill=white,left] at (2.5,1.0);
\end{tikzpicture}
}
   \caption{Random deployment}
 \end{subfigure}
  \begin{subfigure}[b]{0.3\textwidth}
\iftoggle{draft}{\includegraphics[width=1\textwidth,draft]{\figpath pos-draft}}{
\begin{tikzpicture}[spy using
  outlines={circle,\tikzzpycolor,magnification=3.0,size=2.5cm, connect
    spies, every spy on node/.append style={ultra thick}}]
\node {\includegraphics[width=1\textwidth]{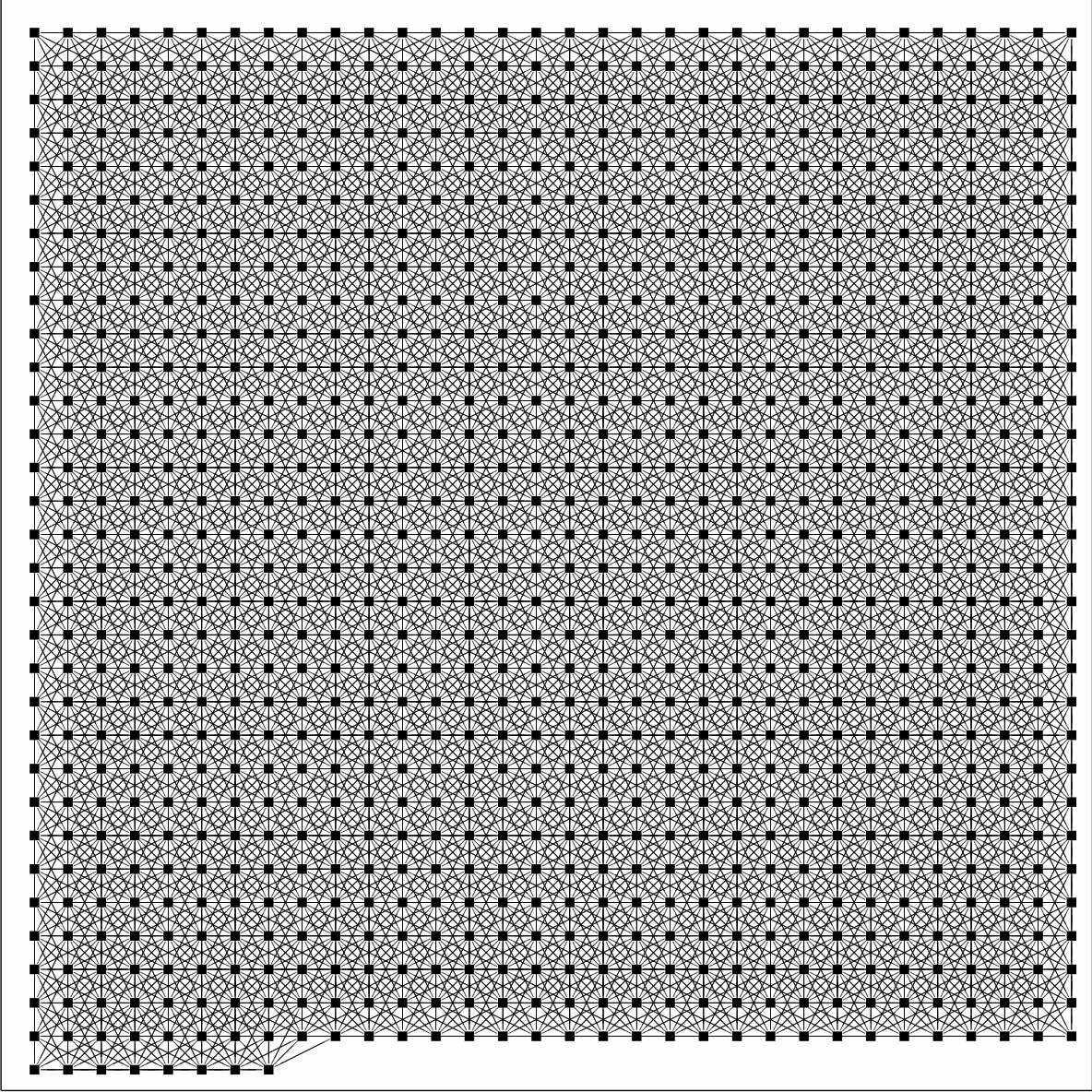}};
\spy on (0,-1) in node [fill=white,left] at (1.5,1.0);
\end{tikzpicture}
}
   \caption{Grid deployment}
 \end{subfigure}
  \begin{subfigure}[b]{0.3\textwidth}
\iftoggle{draft}{\includegraphics[width=1\textwidth,draft]{\figpath pos-draft}}{
\begin{tikzpicture}[spy using
  outlines={circle,\tikzzpycolor,magnification=3.0,size=2.5cm, connect
    spies, every spy on node/.append style={ultra thick}}]
\node {\includegraphics[width=1\textwidth]{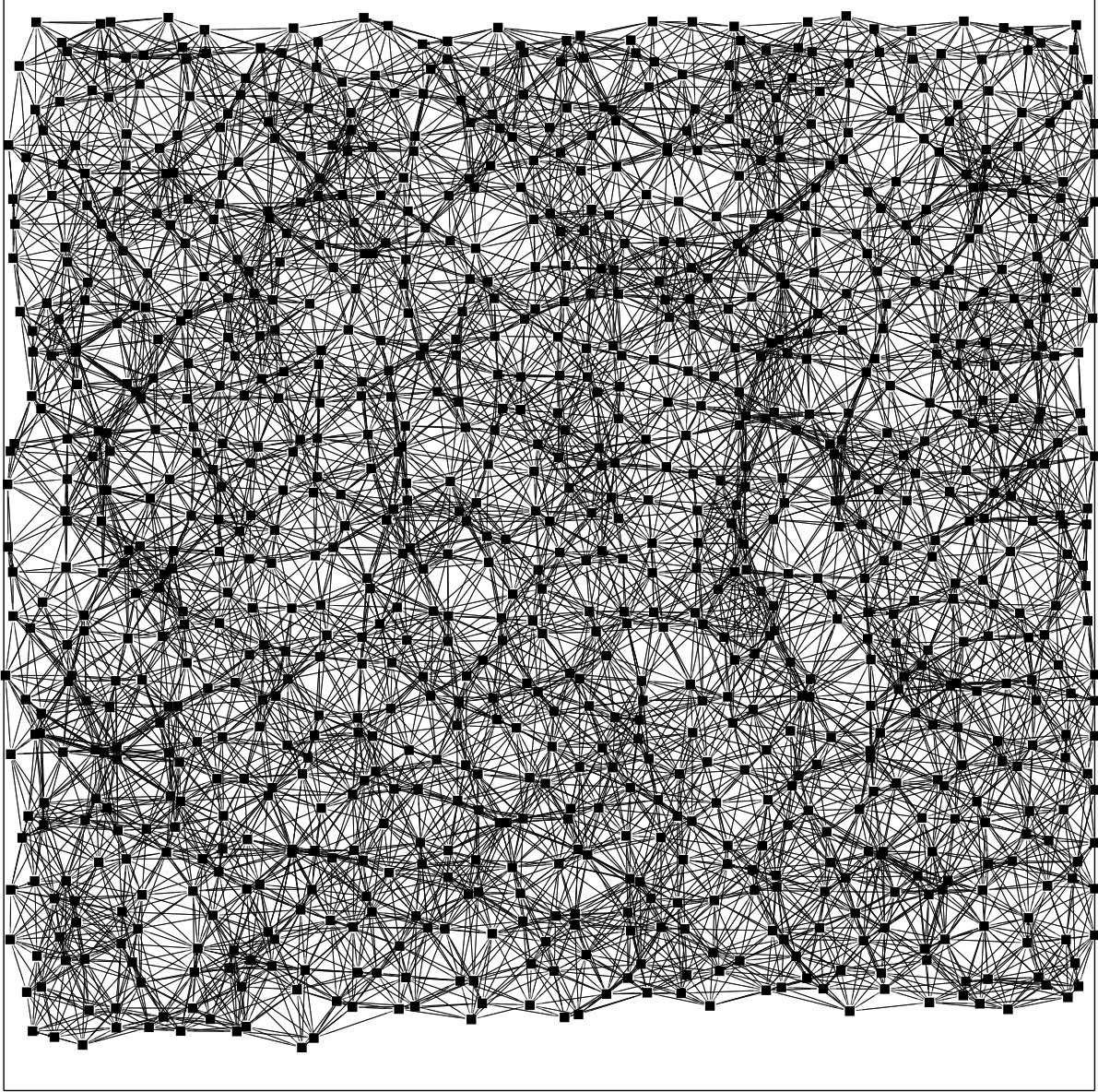}};
\spy on (0,-1) in node [fill=white,left] at (0.0,1.0);
\end{tikzpicture}
}
   \caption{Perturbed Grid deployment}
 \end{subfigure}
  \begin{subfigure}[b]{0.24\textwidth}
\iftoggle{draft}{\includegraphics[width=1\textwidth,draft]{\figpath pos-draft}}{
\begin{tikzpicture}[spy using
  outlines={circle,\tikzzpycolor,magnification=3.0,size=1.5cm, connect
    spies, every spy on node/.append style={ultra thick}}]
\node {\includegraphics[width=1\textwidth]{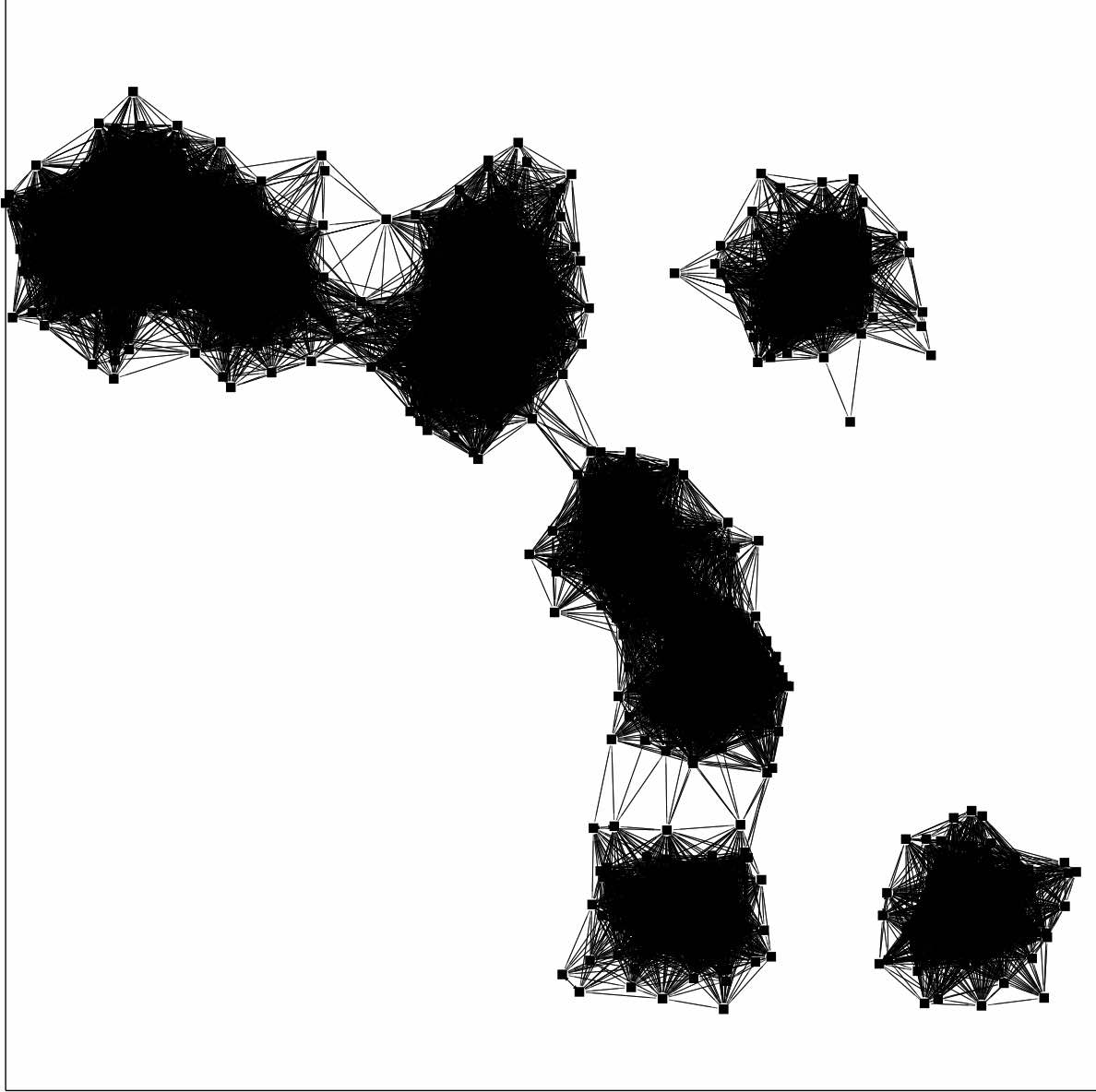}};
\spy on (0.05,0.5) in node [fill=white,left] at (-0.3,-1.2);
\end{tikzpicture}
}
   \caption{Cluster deployment \newline \newline}
 \end{subfigure}
  \begin{subfigure}[b]{0.24\textwidth}
\iftoggle{draft}{\includegraphics[width=1\textwidth,draft]{\figpath pos-draft}}{
\begin{tikzpicture}[spy using
  outlines={circle,\tikzzpycolor,magnification=3.0,size=1.5cm, connect
    spies, every spy on node/.append style={ultra thick}}]
\node {\includegraphics[width=1\textwidth]{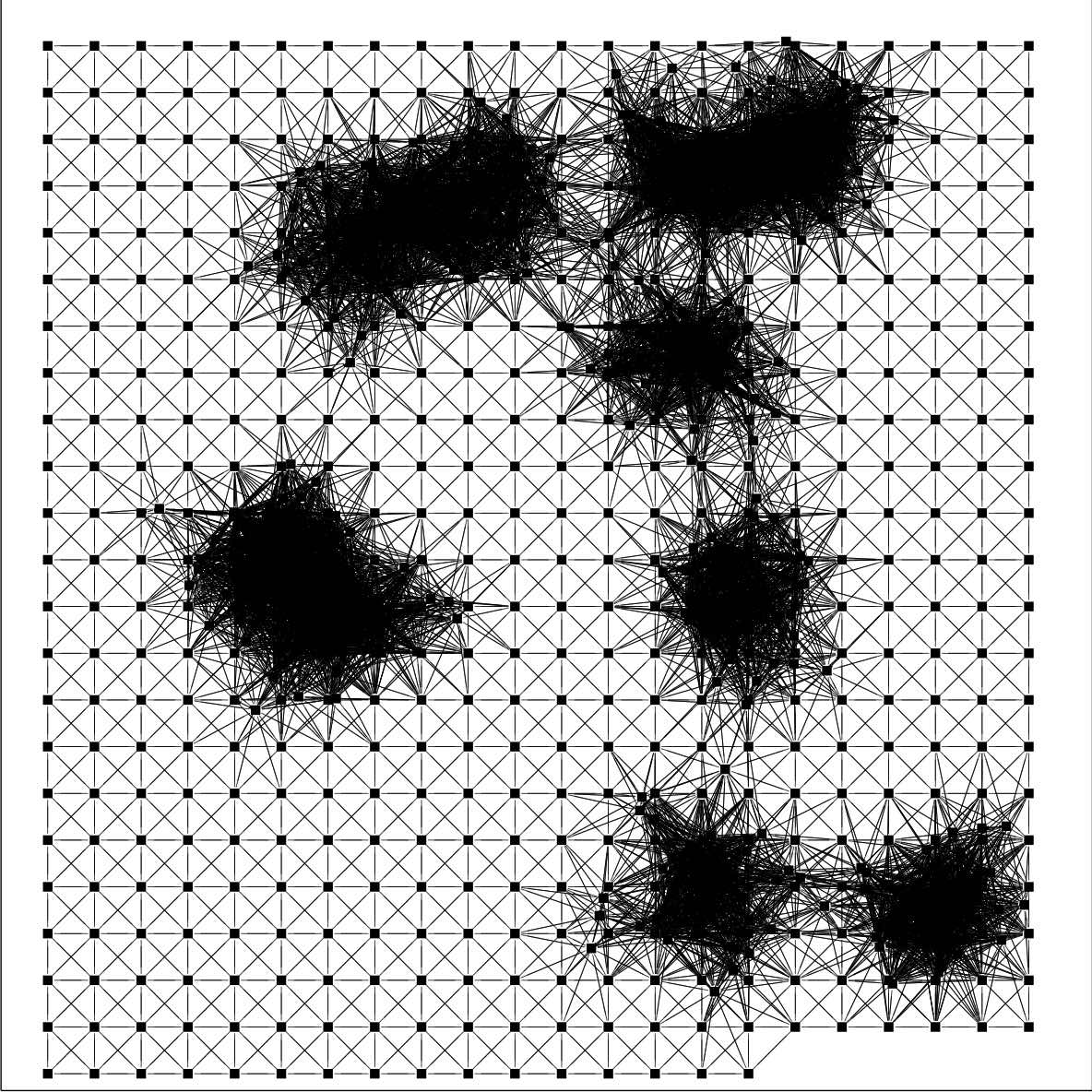}};
\spy on (0.27,0.45) in node [fill=white,left] at (-0.3,-1.2);
\end{tikzpicture}
}
\caption{\SI{50}{\percent} of nodes in cluster deployment,
  \SI{50}{\percent} as two-dimensional grid}
 \end{subfigure}
  \begin{subfigure}[b]{0.24\textwidth}
\iftoggle{draft}{\includegraphics[width=1\textwidth,draft]{\figpath pos-draft}}{
\begin{tikzpicture}[spy using
  outlines={circle,\tikzzpycolor,magnification=3.0,size=1.5cm, connect
    spies, every spy on node/.append style={ultra thick}}]
\node {\includegraphics[width=1\textwidth]{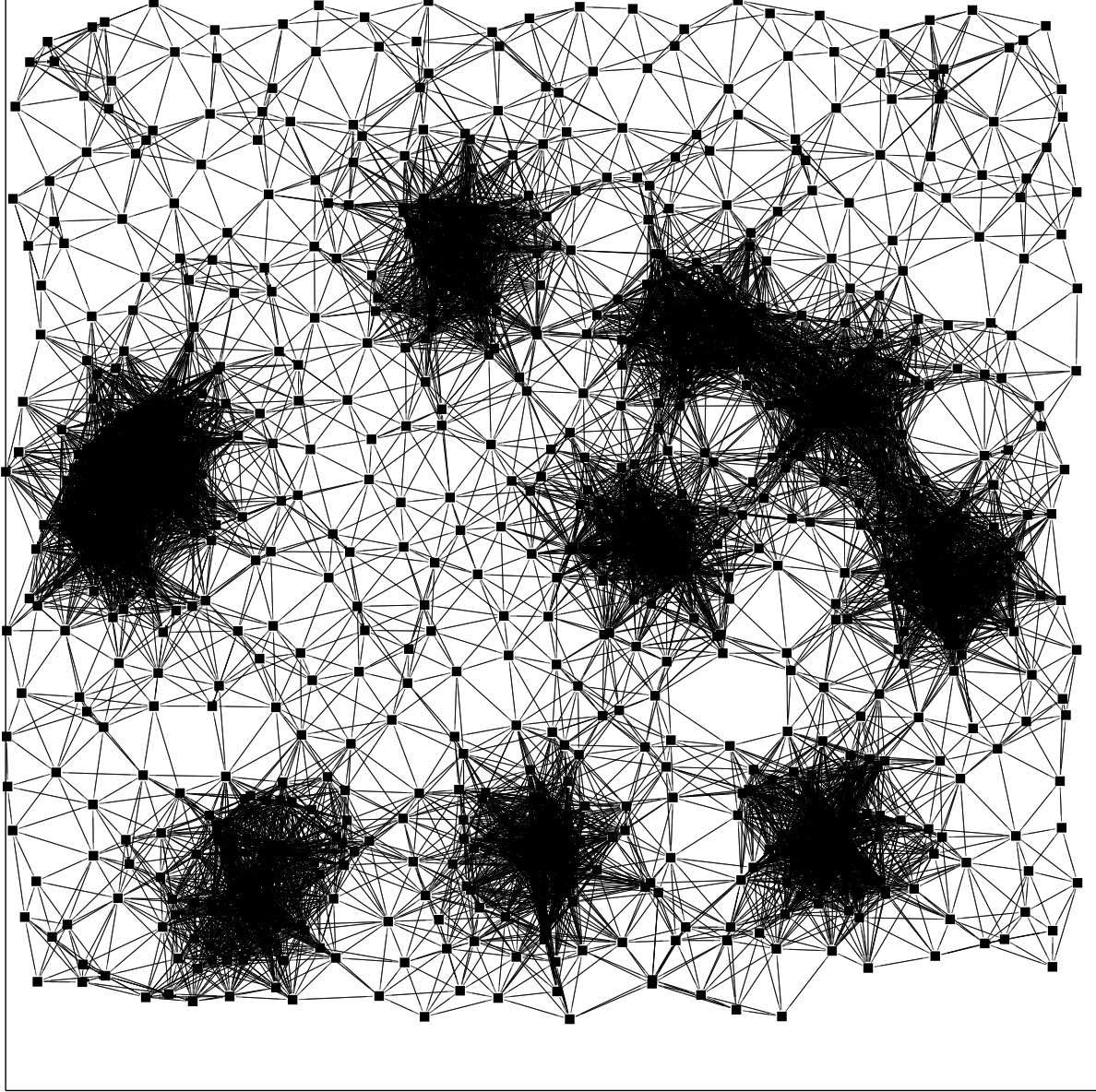}};
\spy on (0,0.4) in node [fill=white,left] at (2.0,-1.2);
\end{tikzpicture}
}
   \caption{\SI{50}{\percent} of nodes in cluster deployment, \SI{50}{\percent} as perturbed
               grid}
 \end{subfigure}
  \begin{subfigure}[b]{0.24\textwidth}
\iftoggle{draft}{\includegraphics[width=1\textwidth,draft]{\figpath pos-draft}}{
\begin{tikzpicture}[spy using
  outlines={circle,\tikzzpycolor,magnification=3.0,size=1.5cm, connect
    spies, every spy on node/.append style={ultra thick}}]
\node {\includegraphics[width=1\textwidth]{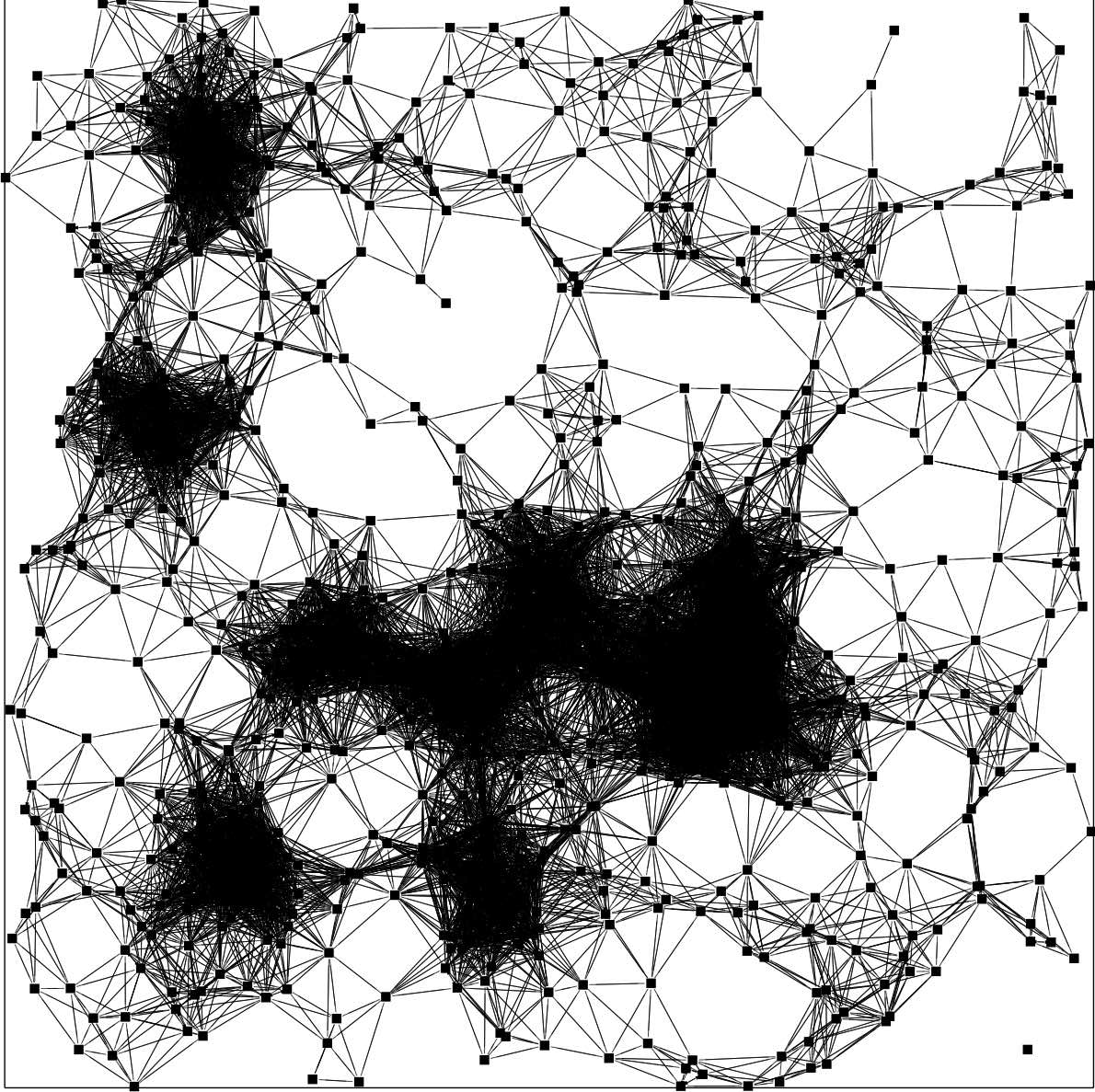}};
\spy on (0,0.4) in node [fill=white,left] at (2.0,-1.2);
\end{tikzpicture}
}
   \caption{\SI{50}{\percent} of nodes in cluster deployment, \SI{50}{\percent} as
     random deployment}
 \end{subfigure}
\caption{Sample networks illustrating our deployment strategies.}
\label{ch:prelim:fig:distribution-models}
\end{figure}

\section{\randvd\ and the Length of Phases}
\label{sec:randvd-length-phases}
\renewcommand{\figpath}{plots/rand4d/}

\begin{figure}[htb]
  \centering 
  \hfill
  \begin{subfigure}[b]{0.4\textwidth}
    \centering
    \includegraphics[width=1\textwidth, trim=5mm 0mm 35mm 20mm,
    clip]{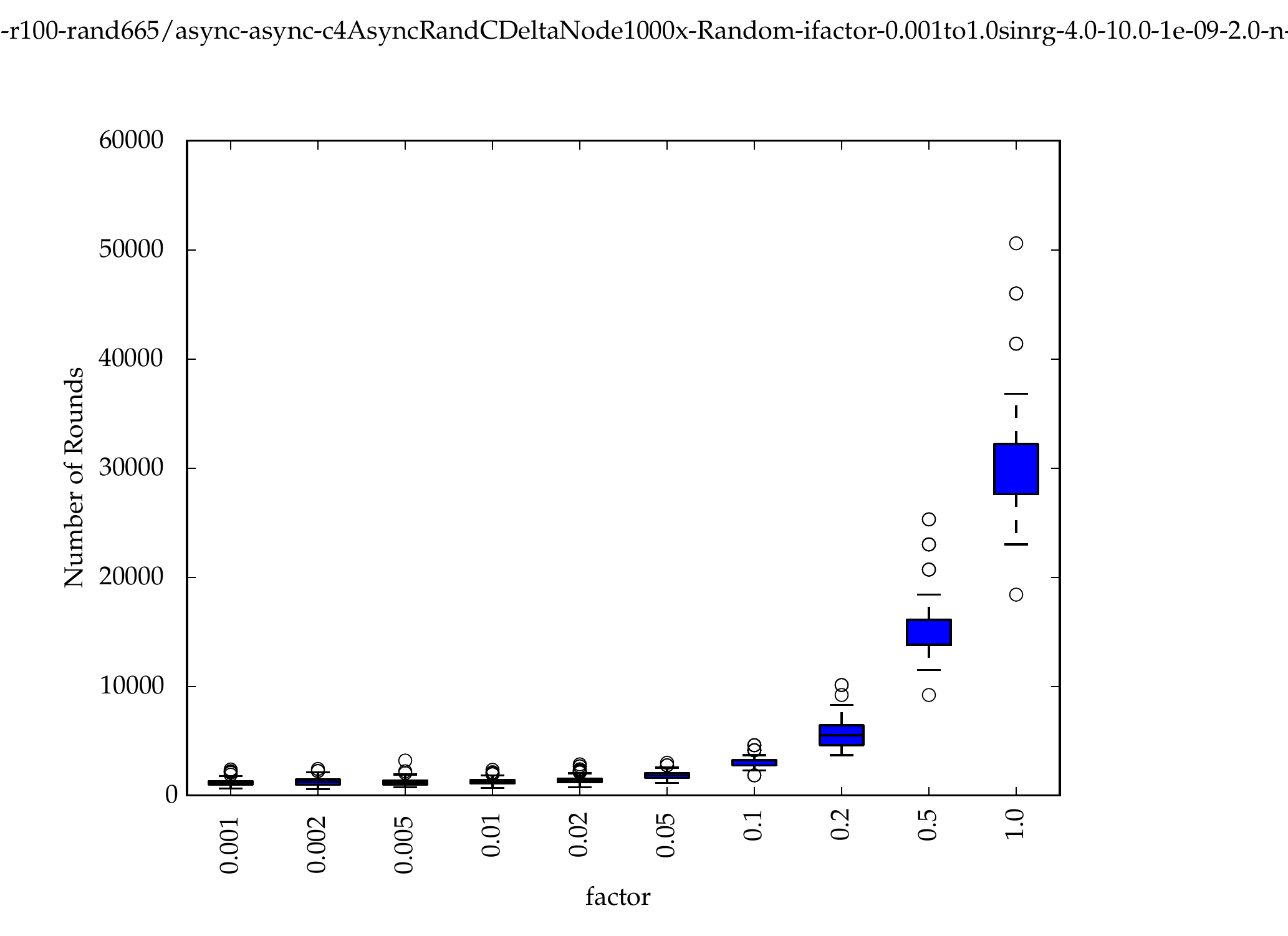}
   \caption{Runtime}
 \end{subfigure}
  \begin{subfigure}[b]{0.4\textwidth}
    \centering
    \includegraphics[width=1\textwidth, trim=5mm 0mm 30mm 20mm,
    clip]{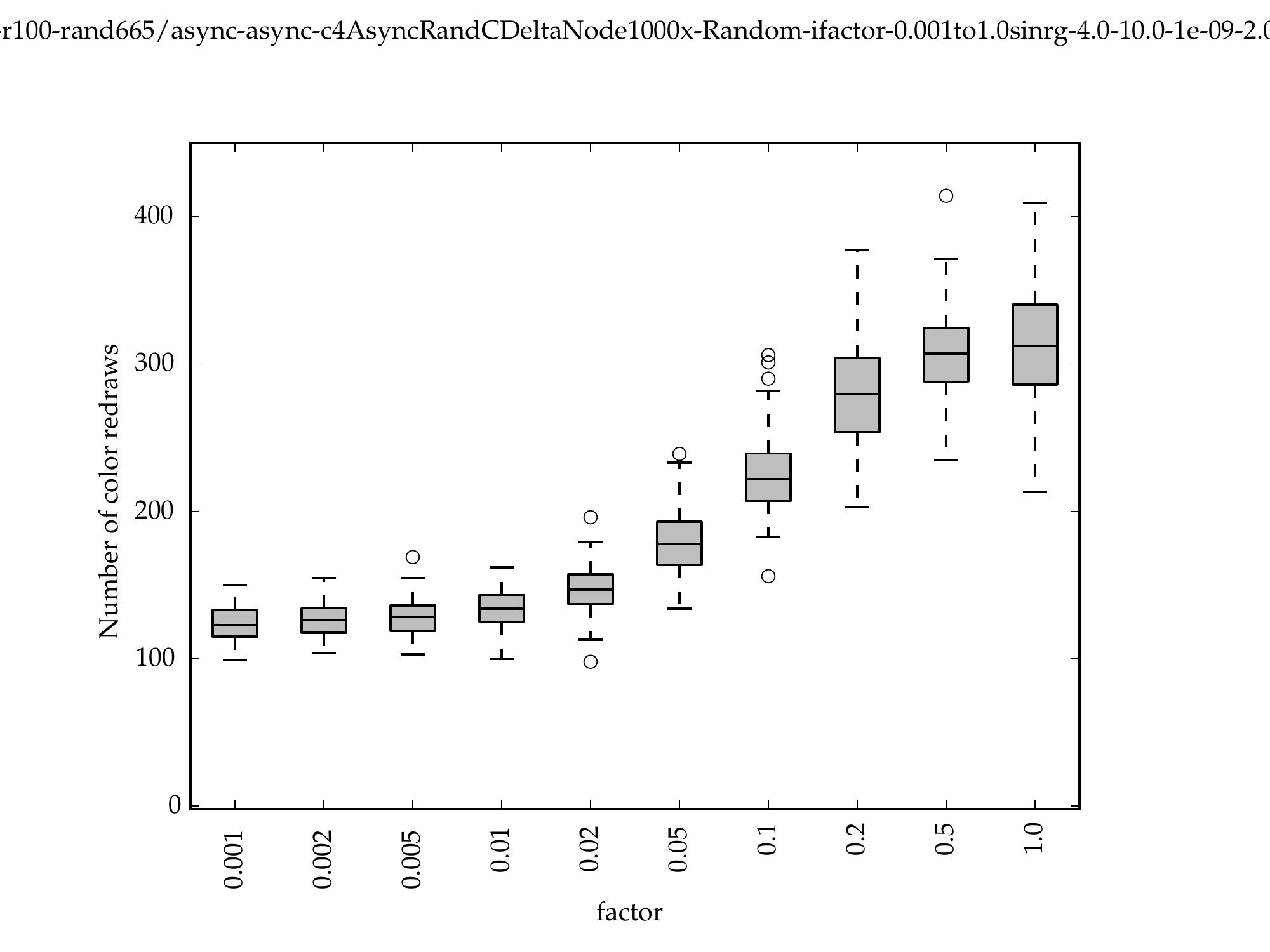}
   \caption{Total number of color redraws}
\label{ch:exp:fig:rand-factor-redraws}
 \end{subfigure}
  \hfill
 \caption{Determining the parameter \factor\ for {\sc
     RandCDeltaColoring}, which influences the length of each
   phase. Both the runtime (left) and the number of color redraws
   (right) increase with increasing \factor.}
\label{ch:exp:fig:rand-factor}
\end{figure}

Let us study the parameter \factor\ using the random deployment in our
first experiment. We use values ranging from $\num{0.001}$ to
$\num{1}$, corresponding to phase-lengths between $\num{5}$ and
$\num{4600}$ time slots. Our results are depicted in
\cref{ch:exp:fig:rand-factor}.

We observe that both the runtime and the number of color redraws
increases with the phase-length. Especially for the runtime this was
expected, as our theoretical analysis guarantees the runtime of
$\O(\Delta \log n)$ time slots only for a phase-length
of~$\O(\Delta)$, while \factor\ $=1$ sets the phase-length to one
round of local broadcasting, which is asymptotically in
$\O(\Delta \log n)$.  However, there is also a less formal intuition
justifying the decreasing runtime for the decreasing
phase-length. Once a node has detected a conflict (by receiving a
message from a neighbor), it is not beneficial if the node must wait
for the end of the phase before changing its color.  Assume the node
waits until the phase ends. It may happen that the node transmits its
color to its neighbors, which may become aware of a conflict. If this
happens, the respective neighbors also reset their color at the end of
their phase, although this conflict would be resolved without the
neighbors intervention with significant probability at the end of the
phase.  On the other hand, dealing with the conflict directly does not
introduce any penalty, as it is already determined that the detecting
node resets its color. Hence, the shorter the phases are, the lower
the runtime of the coloring algorithm.

Regarding the number of color redraws, we can observe something
interesting. The longer the phases are, the more redraws are required,
cf. \cref{ch:exp:fig:rand-factor-redraws}. This corresponds to the
fact that the longer the phases are, the higher the probability that
all conflicts are detected by the nodes. If a conflict is only
detected by one of the conflicting nodes, the probability that it is
resolved is already significant. Therefore, using phases of minimal
length intuitively reduces the number of color redraws by a factor of
two compared to phases of length \duration. As each color redraw leads
to some possibility of selecting the color of a neighbor we observe a
factor of even slightly more than two in the total number of color
redraws for long phases. Note that even longer phases do not lead to a
further increase, as almost all conflicts are detected after
phase-lengths that correspond to \factor\ $=1$.

\section{Illustration of the Algorithms}
\label{sec:illustr-algos}
\renewcommand{\figpath}{images/}

\begin{figure}[htb!]
  \centering
  \begin{subfigure}[b]{0.3\textwidth}
    \centering
    \includegraphics[width=1\textwidth, trim=0mm 0mm 0mm 0mm,
    clip]{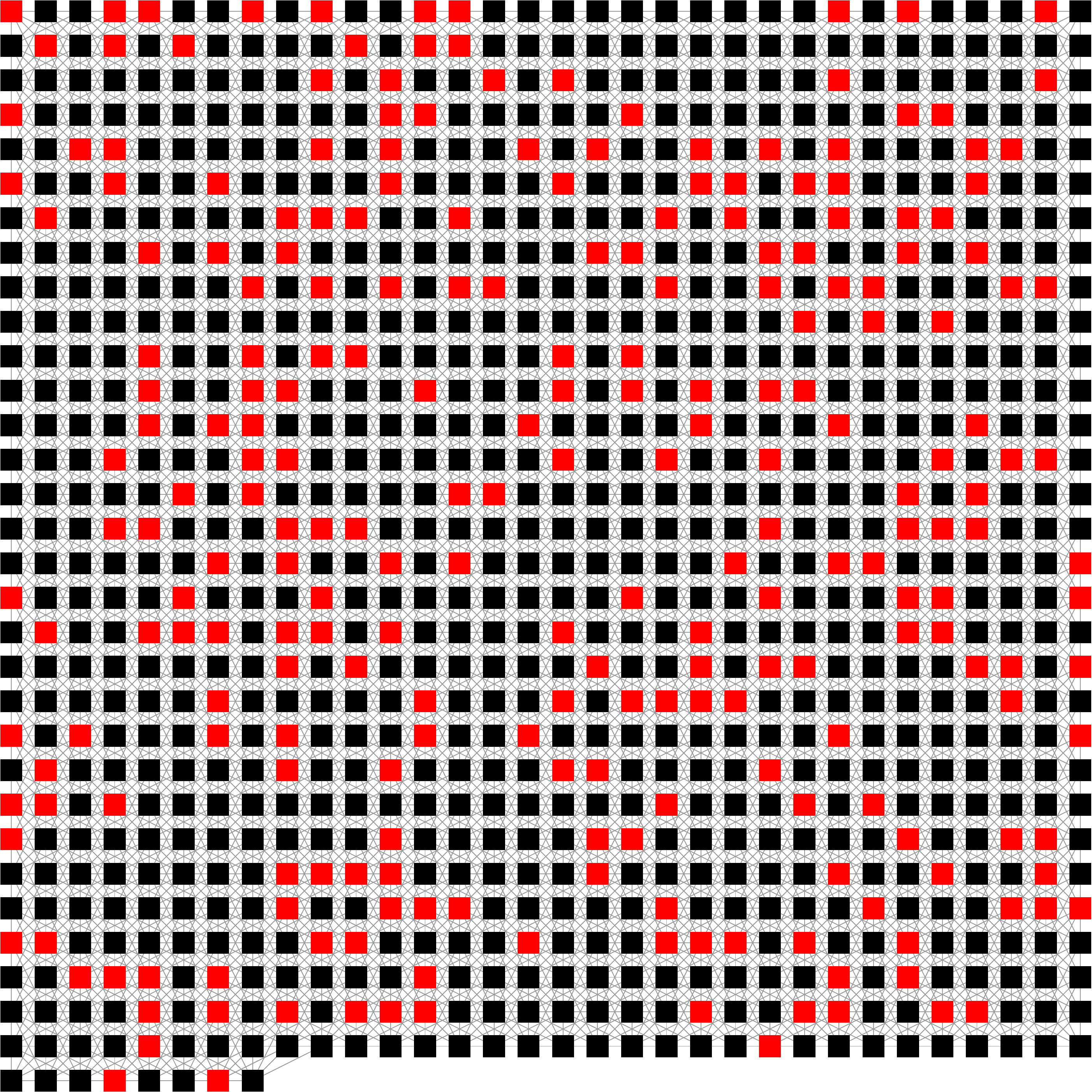}
   \caption{After deployment}
 \end{subfigure}
  \hfill
  \begin{subfigure}[b]{0.3\textwidth}
    \centering
    \includegraphics[width=1\textwidth, trim=0mm 0mm 0mm 0mm,
    clip]{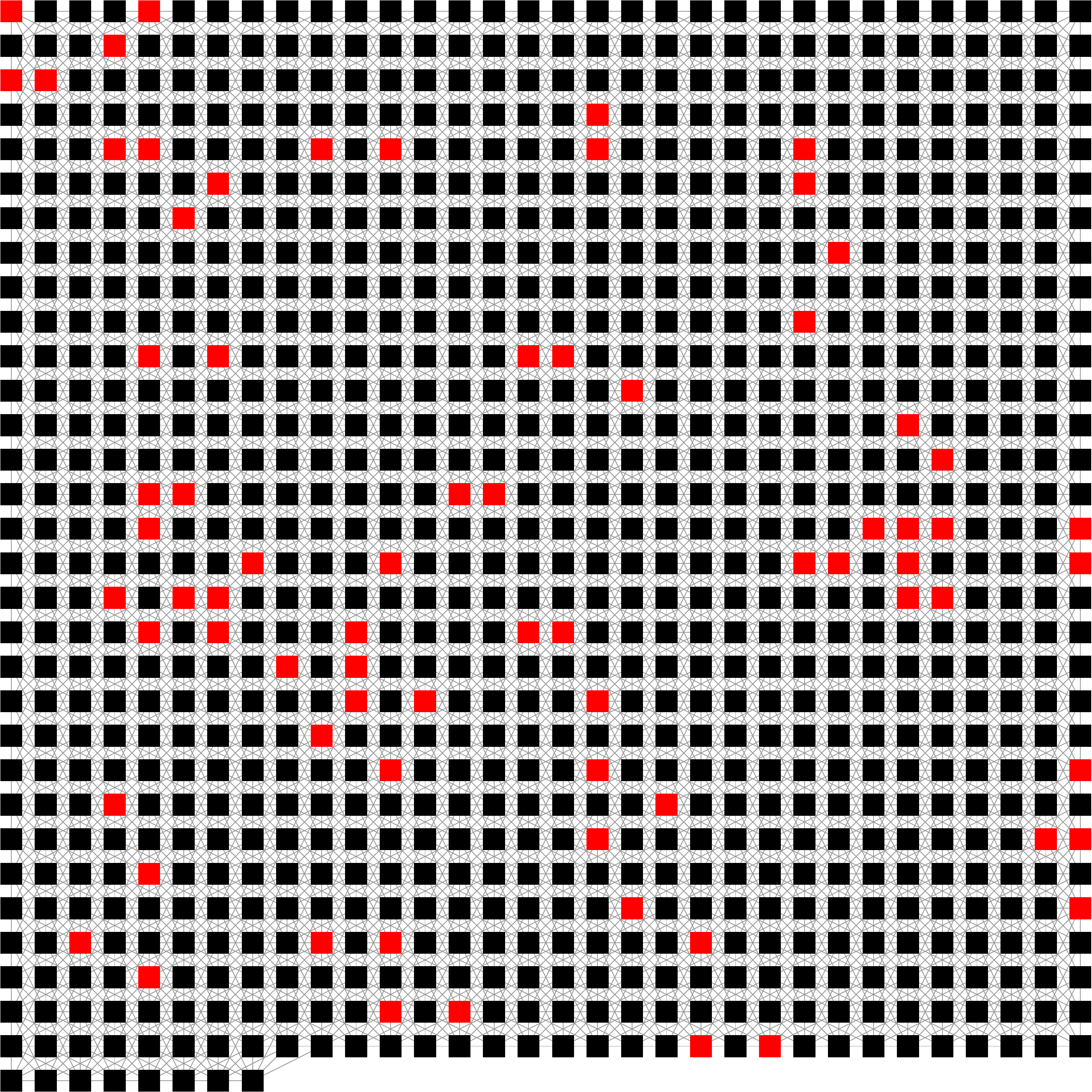}
   \caption{Resolving conflicts}
 \end{subfigure}
 \hfill
 \begin{subfigure}[b]{0.3\textwidth}
    \centering
  \includegraphics[width=1\textwidth, trim=0mm 0mm 0mm 0mm,
  clip=true]{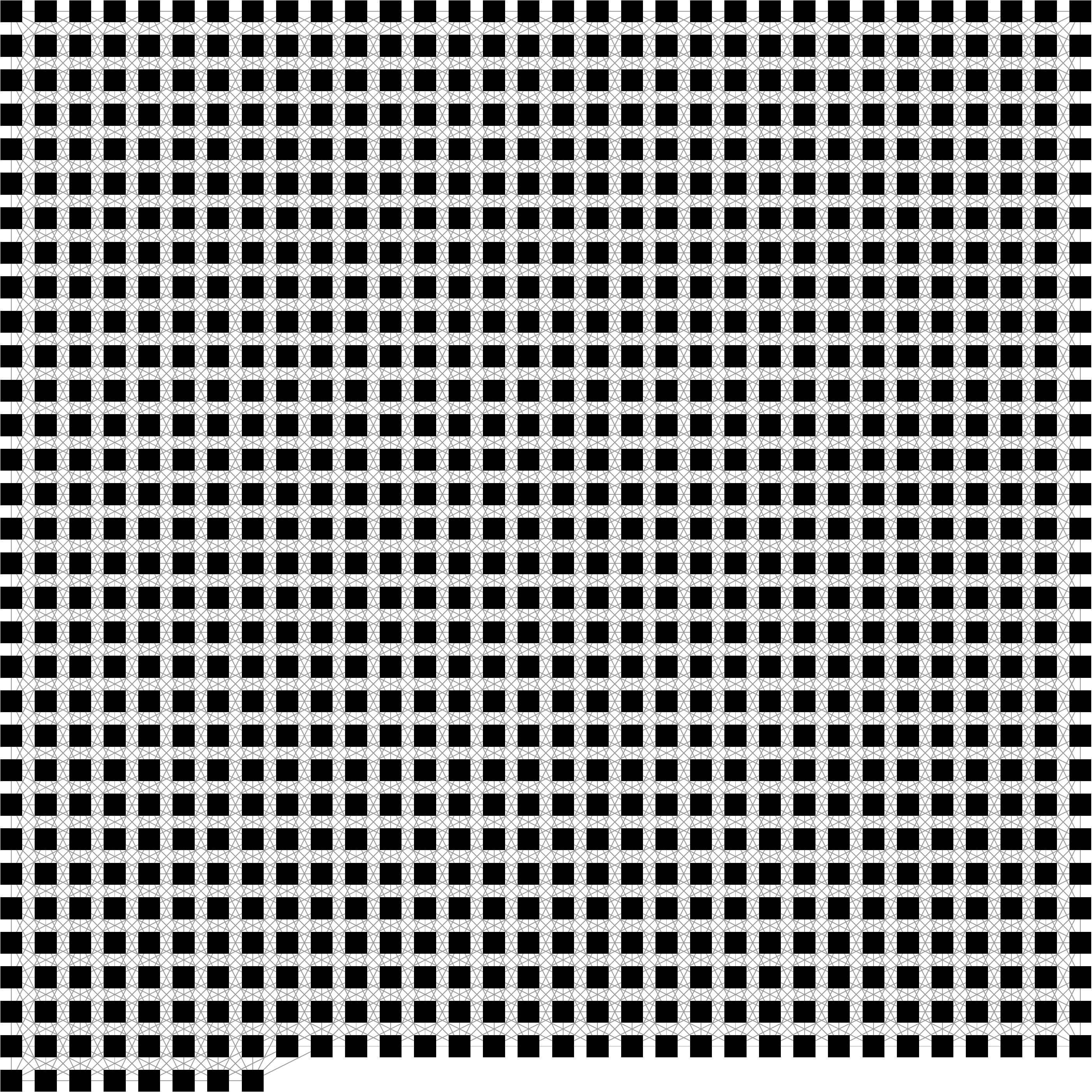}
  \caption{All colors are valid}
\end{subfigure}
 \caption{Illustration of an execution of \randvd. Black nodes have a
   valid color, red nodes are in conflict with one of their
   neighbors. We observe that even directly after deployment not too
   many conflicts exist (left). Furthermore, they get gradually eliminated
   (center) until all nodes have a valid color (right) }
\label{ch:exp:fig:randvd-images}
\end{figure}

\begin{figure}[htb!]
  \centering
  \begin{subfigure}[b]{0.3\textwidth}
    \centering
    \includegraphics[width=1\textwidth, trim=0mm 0mm 0mm 0mm,
    clip]{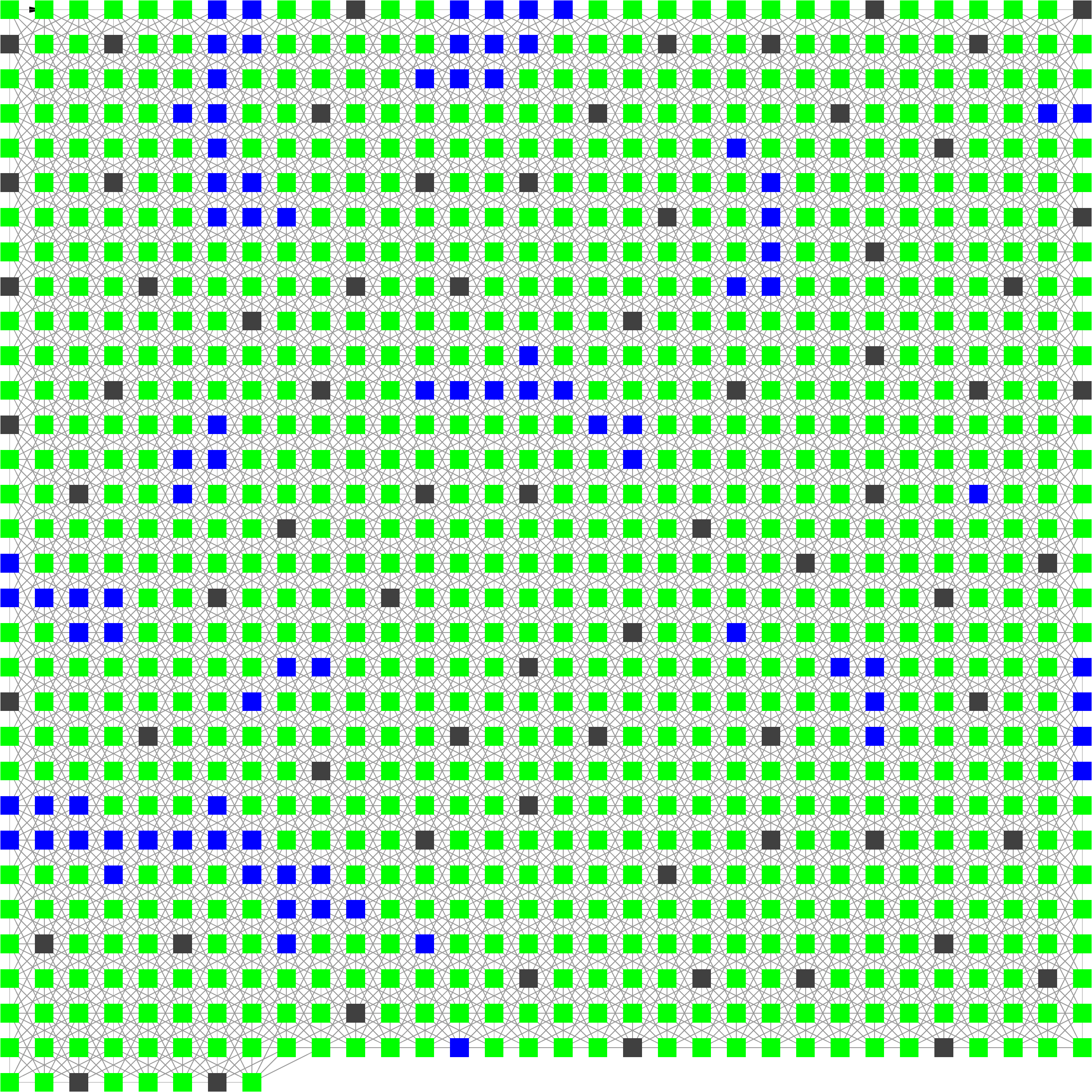}
   \caption{After leader election}
 \end{subfigure}
  \hfill
  \begin{subfigure}[b]{0.3\textwidth}
    \centering
    \includegraphics[width=1\textwidth, trim=0mm 0mm 0mm 0mm,
    clip]{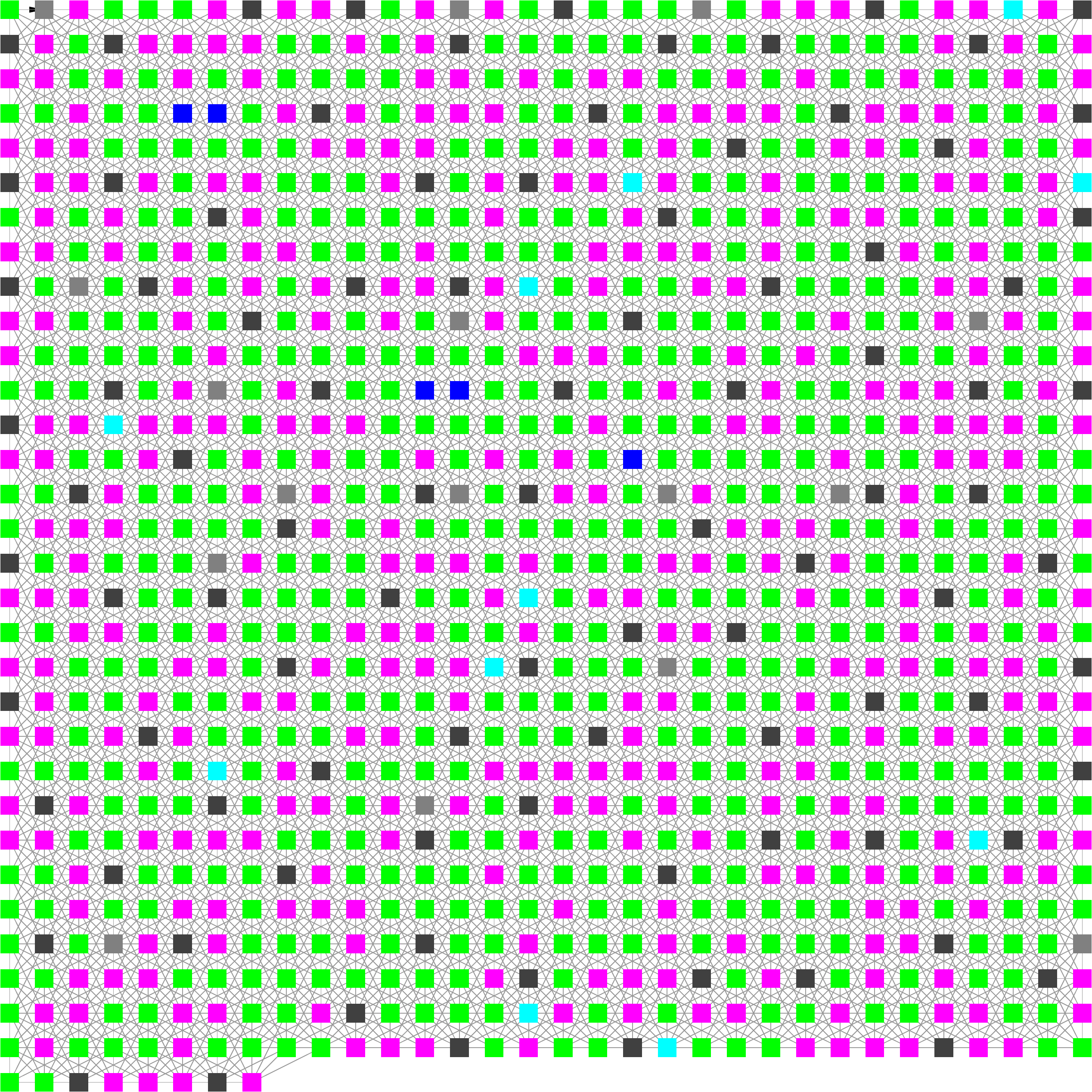}
   \caption{Second level starts}
 \end{subfigure}
 \hfill
 \begin{subfigure}[b]{0.3\textwidth}
    \centering
  \includegraphics[width=1\textwidth, trim=0mm 0mm 0mm 0mm,
  clip=true]{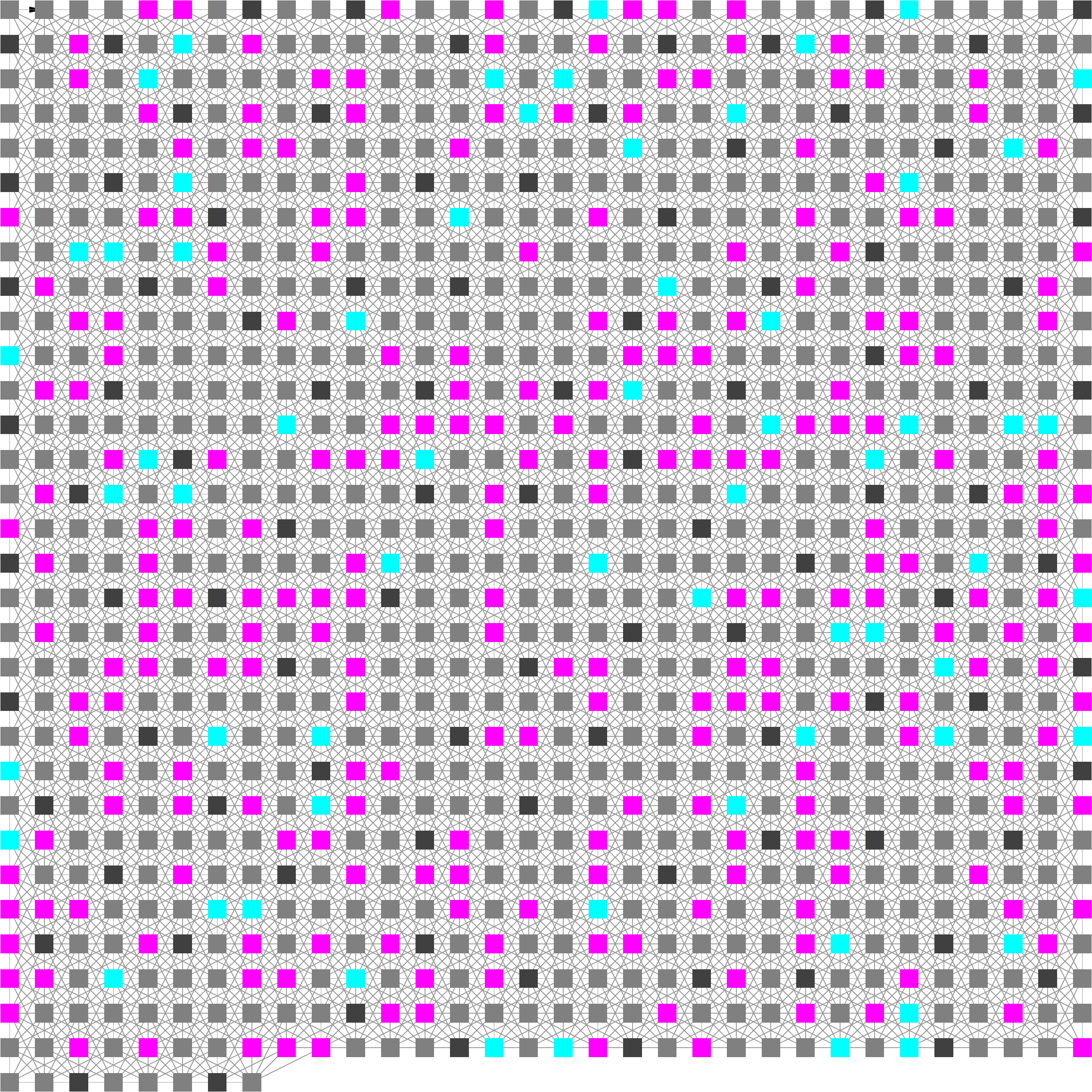}
  \caption{Most nodes finished}
\end{subfigure}
\caption{The flow of \colorred. On the left most leaders (black) are
  computed. Green nodes are dominated and blue nodes still execute the
  first level MIS. In the center many dominated nodes received their
  active interval (magenta nodes), some cyan nodes compete in the
  second level MIS for a color. Very few gray nodes have already
  finalized their color by winning a second level MIS. On the right,
  most nodes selected their final color (gray), while others wait for
  their active interval or actively compete for selecting a final
  color.}
\label{ch:exp:fig:colorred-images}
\end{figure}

\begin{figure}[htb!]
  \centering
  \begin{subfigure}[b]{0.3\textwidth}
    \centering
    \includegraphics[width=1\textwidth, trim=0mm 0mm 0mm 0mm,
    clip]{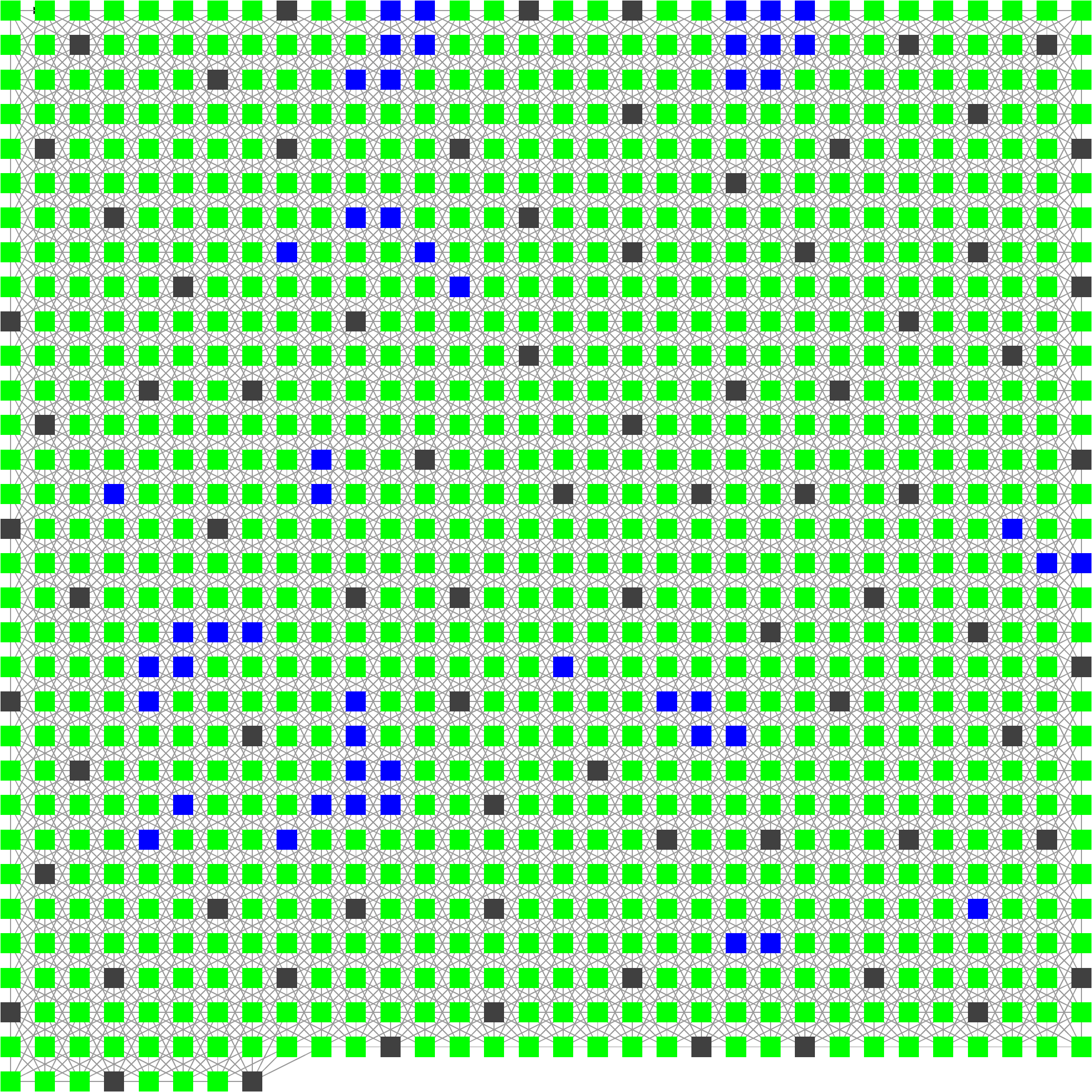}
   \caption{After leader election}
 \end{subfigure}
  \hfill
  \begin{subfigure}[b]{0.3\textwidth}
    \centering
    \includegraphics[width=1\textwidth, trim=0mm 0mm 0mm 0mm,
    clip]{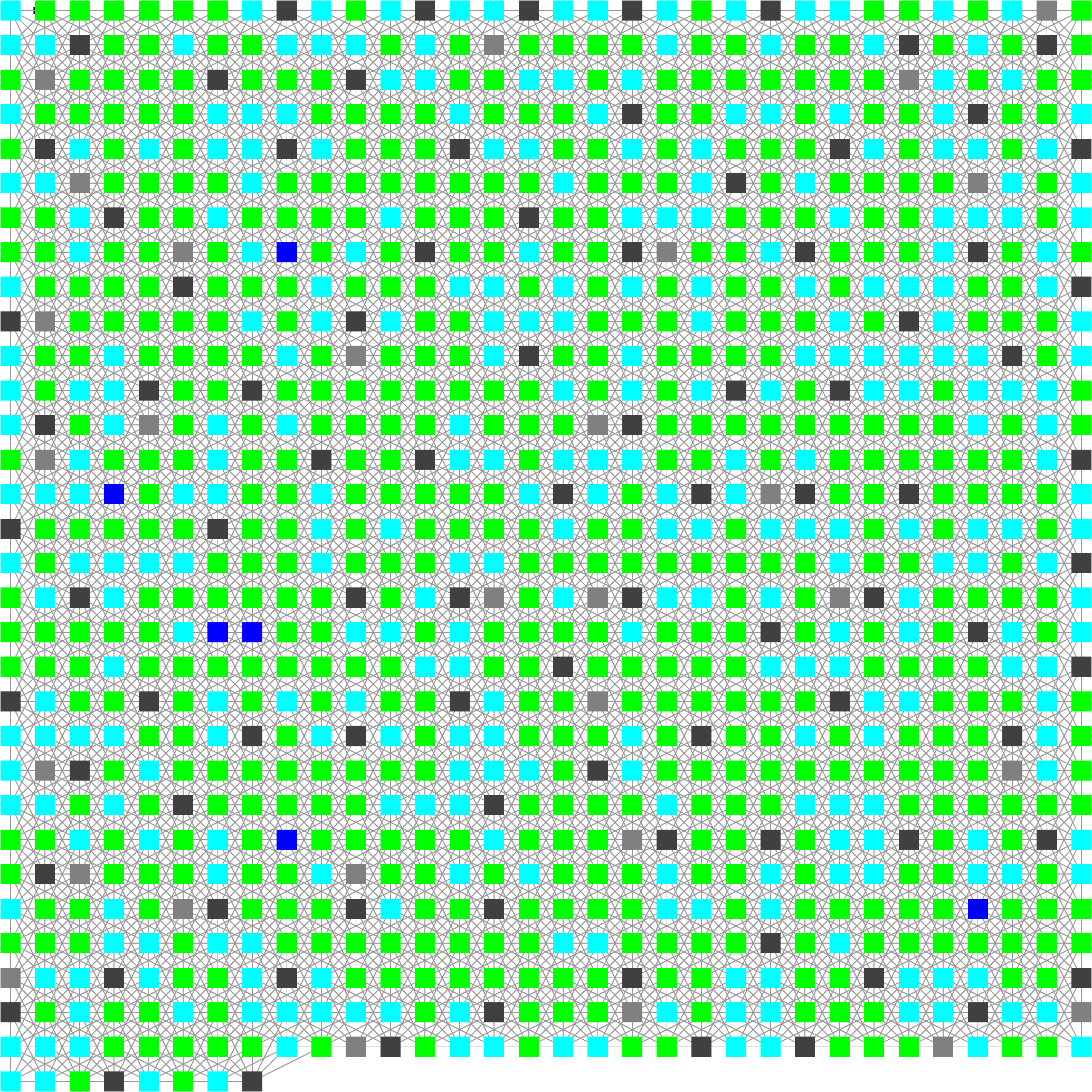}
   \caption{Compete for color}
 \end{subfigure}
 \hfill
 \begin{subfigure}[b]{0.3\textwidth}
    \centering
  \includegraphics[width=1\textwidth, trim=0mm 0mm 0mm 0mm,
  clip=true]{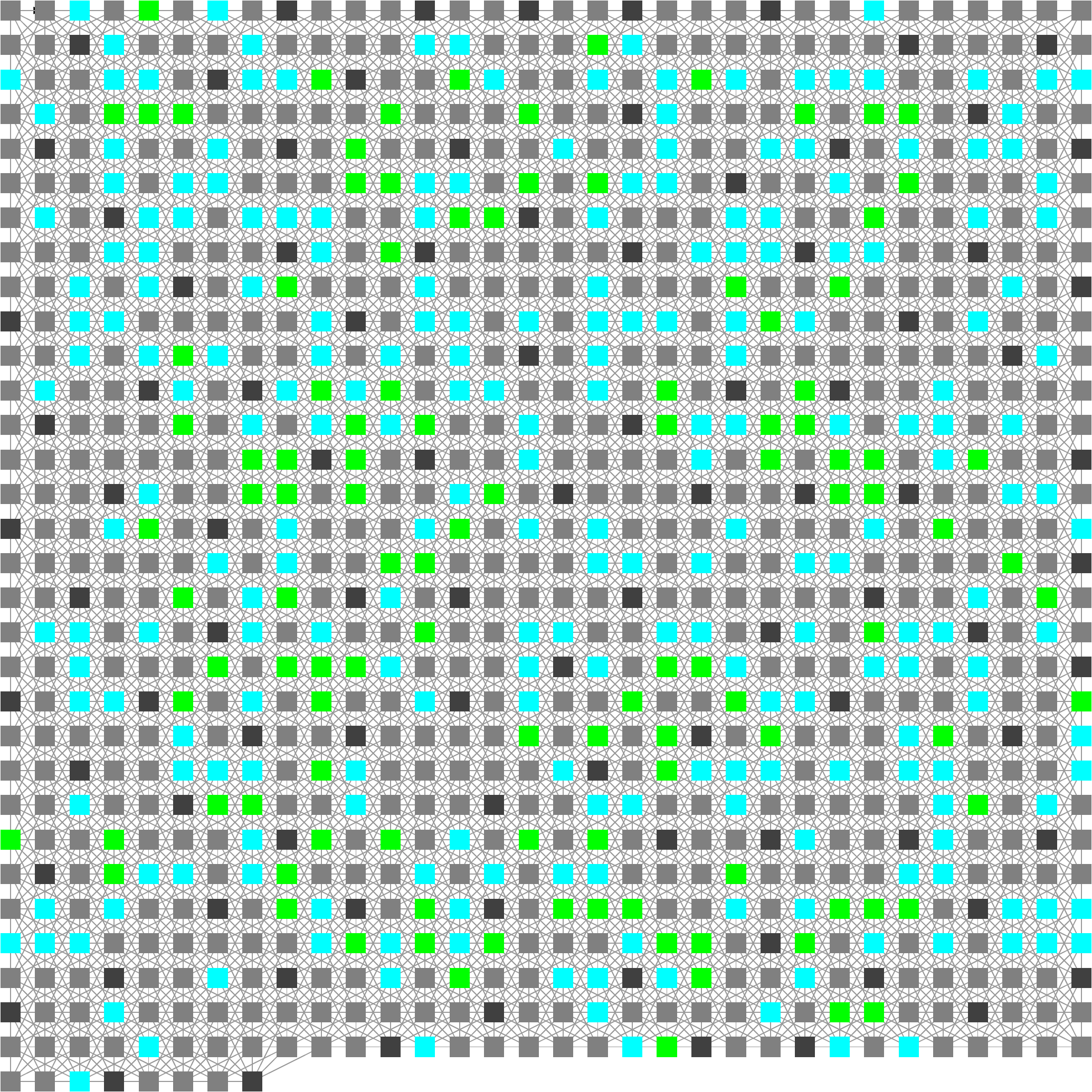}
  \caption{Nodes finalize color}
\end{subfigure}
\caption{One execution of \mwcolor. On the left almost all leaders
  (black) are computed. Green nodes are dominated and request a color
  block, while blue nodes still compete in the MIS to become
  leaders. In the center many nodes received their color blocks to
  compete for a color (cyan) and some nodes already selected their
  final color (gray). On the right all leaders are computed and more
  nodes selected their final color.}
\label{ch:exp:fig:mwcolor-images}
\end{figure}

\begin{figure}[htb!]
  \centering
  \begin{subfigure}[b]{0.3\textwidth}
    \centering
    \includegraphics[width=1\textwidth, trim=0mm 0mm 0mm 0mm,
    clip]{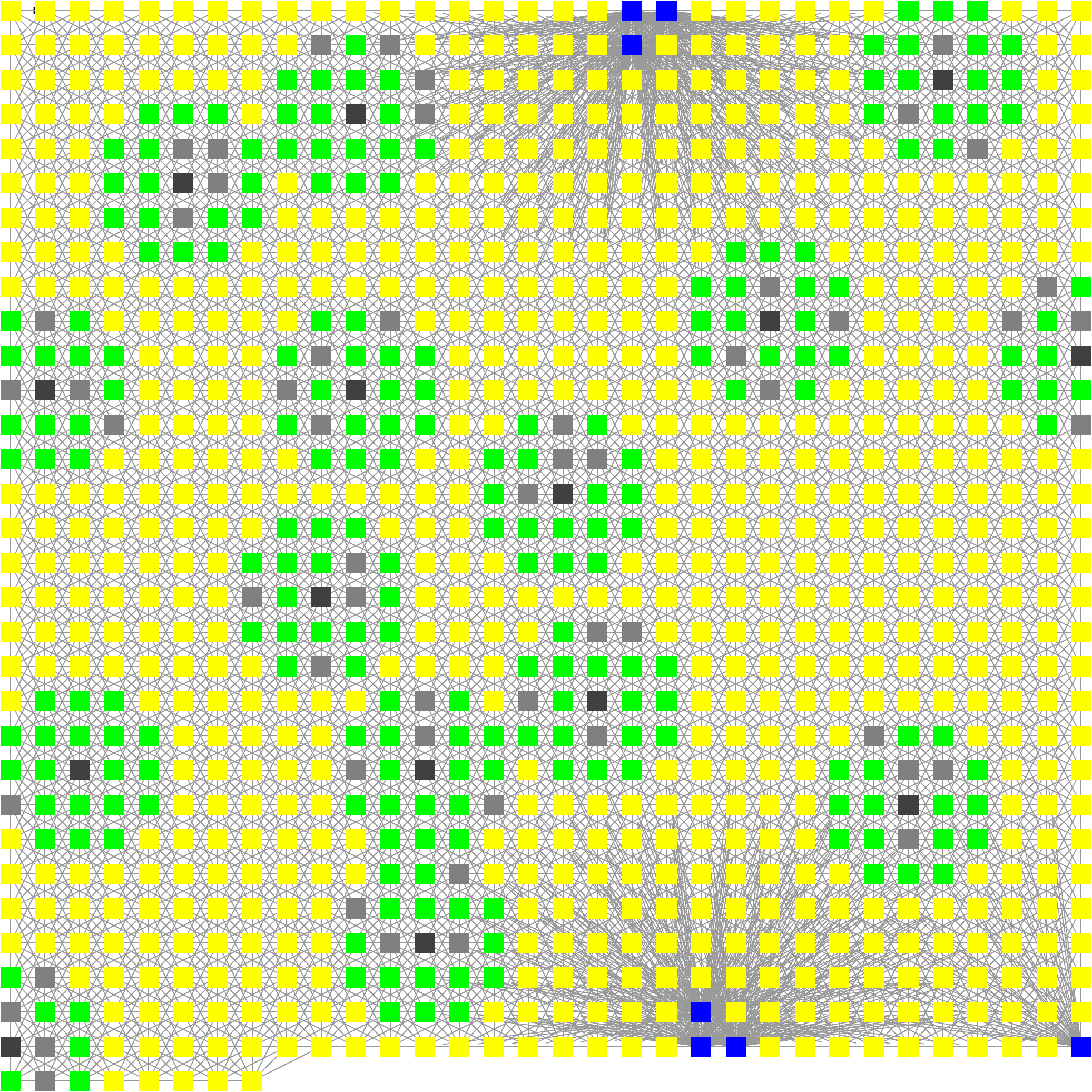}
   \caption{First leaders}
 \end{subfigure}
  \hfill
  \begin{subfigure}[b]{0.3\textwidth}
    \centering
    \includegraphics[width=1\textwidth, trim=0mm 0mm 0mm 0mm,
    clip]{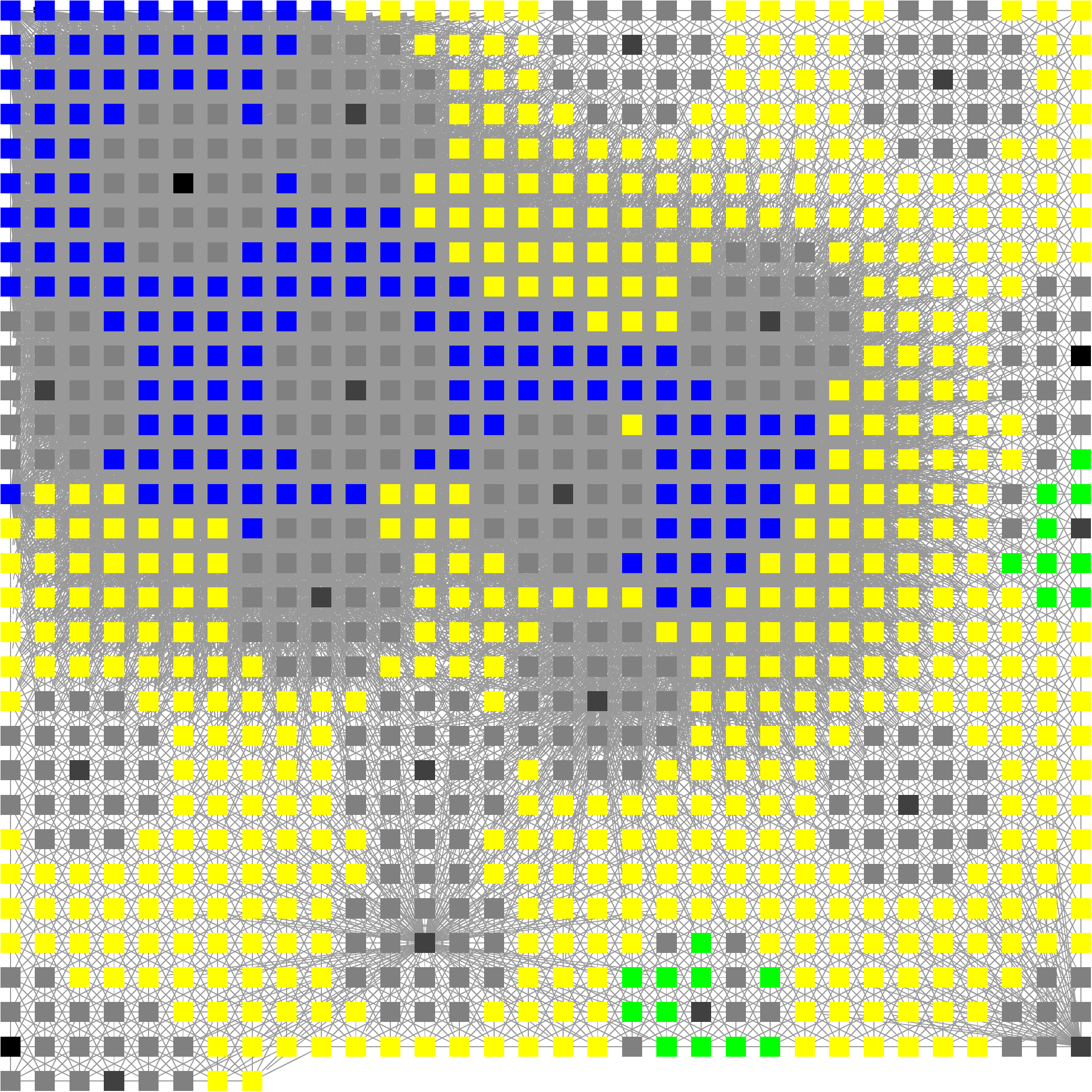}
   \caption{Second MIS starts}
 \end{subfigure}
 \hfill
 \begin{subfigure}[b]{0.3\textwidth}
    \centering
  \includegraphics[width=1\textwidth, trim=0mm 0mm 0mm 0mm,
  clip=true]{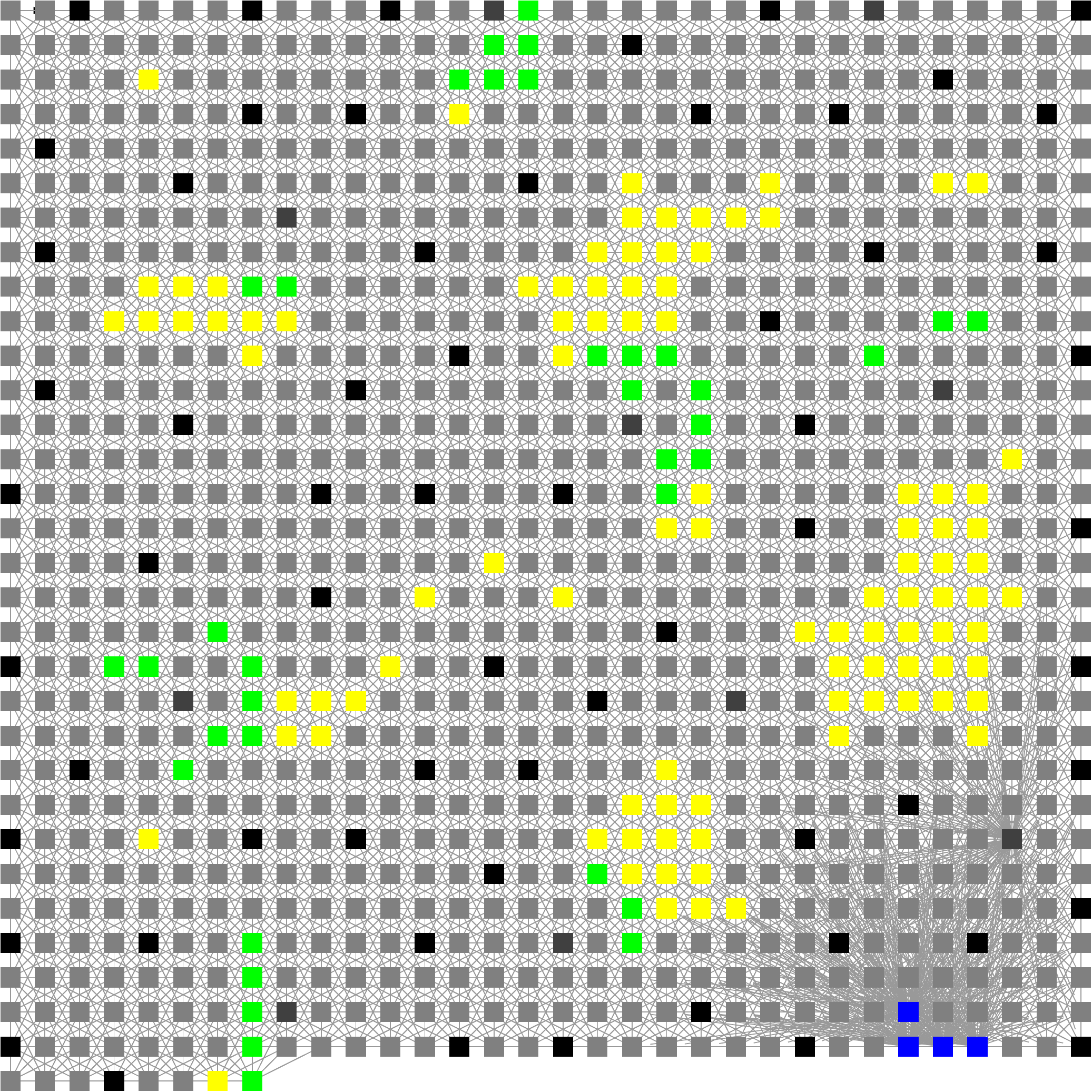}
  \caption{Many nodes colored}
\end{subfigure}
\caption{In the illustration of \yucolor\ the MIS is regarding a
  larger range. Thus, on the left only few leaders (black) dominate
  green nodes that request permission to select a color. Some nodes
  already selected their final color (gray). Most nodes are blocked
  (yellow), while some blue nodes still compete to enter the MIS. In
  the center, the first MIS nodes start to resign as leaders, allowing
  formerly blocked node to compete in the MIS again (blue nodes).  On
  the right, only few nodes remain to be colored, some wait to receive
  the permission by their leader, most others are currently blocked.}
\label{ch:exp:fig:yucolor-images}
\end{figure}

In this section we illustrate the execution of \randvd, \colorred,
\mwcolor, and \yucolor in
\cref{ch:exp:fig:randvd-images,ch:exp:fig:colorred-images,ch:exp:fig:mwcolor-images,ch:exp:fig:yucolor-images}
on the grid deployment to increase the readability.  We use the
parameters as described in \cref{ch:coloringexp:sec:setup} and refer
to \cref{app:illustr-depl} for a more detailed
visualization of a network using the grid deployment.

\newcommand{\algofontsize}{\footnotesize}

In the first section we describe the required modifications to Sinalgo
to correct a flaw in the simulators SINR model implementation. In
\cref{app:exp:sec:all-distributions} we report results of experiments
for the deployment strategies not presented in the paper

\section{Sinalgo - Patch for SINR Model}
\label{app:sec:sinr-patch}

In this section we report on a modification of the SINR interference
model that is delivered with Sinalgo version 0.75.3. The modifications are required to ensure that the SINR
interference constraints are correctly evaluated.  Without
modification, the simulation framework consideres for a transmitted
packet p the signal emitted during the transmission of the same
packet p both as desired signal and interference. To correct this
issue we equip each packet with a broadcast ID (or transmission ID),
and ensure that we do only consider the interference of other packets
(i.e., other transmissions). We state the modification in detail in
\cref{patch:sinr-java,patch:packet-java,patch:node-java}. The name of
the algorithms gives the path to the respective file relative to the
src folder of Sinalgo. The line number before and after the code marks
the lines between which the code should be added (all line numbers are
relative to the unchanged file).

\LinesNumberedHidden 
\begin{algorithm} \algofontsize
\DontPrintSemicolon
\setcounter{AlgoLine}{115}
\ShowLn \;
 \verb+/* detect if a packet is from the same broadcast as this packet.+\;
 \verb+* If so, ignore the active packet. +\;
 \verb+* If the broadcast id is -1, the packet is not from a broadcast, +\;
 \verb+* and duplicate packets are found via pack == p+\;
 \verb+*/+\;
\verb+if (p.broadcastId != -1 && (pack.origin.ID == p.origin.ID && +\;
\verb+          && pack.broadcastId == p.broadcastId))+\;
\verb+    continue;+\;
\verb+}+\;
\setcounter{AlgoLine}{116}
\ShowLn
\caption{projects/defaultProject/models/interferenceModels/SINR.java}
\label{patch:sinr-java}
\end{algorithm} 

\begin{algorithm} \algofontsize
\DontPrintSemicolon
\setcounter{AlgoLine}{103}
\ShowLn \;
\verb+/** broadcast id, allows to determine whether+\;
\verb+*  2 packets origined from the same broadcast +\;
\verb+*/+\;
\verb+public int broadcastId;+\;
\setcounter{AlgoLine}{104}
\ShowLn \;
\vspace*{3mm}
\setcounter{AlgoLine}{193}
\ShowLn \;
\verb+pack.broadcastId = -1;+\;
\setcounter{AlgoLine}{194}
\ShowLn \;
\vspace*{3mm}
\setcounter{AlgoLine}{258}
\ShowLn \;
 \verb+broadcastId = -1;+\;
\setcounter{AlgoLine}{259}
\ShowLn 
\caption{sinalgo/nodes/messages/Packet.java}
\label{patch:packet-java}
\end{algorithm}

\begin{algorithm} \algofontsize
\DontPrintSemicolon
\setcounter{AlgoLine}{76}
\ShowLn \; \verb+import sinalgo.tools.Tools;+\;
\setcounter{AlgoLine}{77}
\ShowLn \;
\vspace*{3mm}
\setcounter{AlgoLine}{1486}
\ShowLn \; \verb+int broadcastId = Tools.getRandomNumberGenerator().nextInt(100000);+\;
\setcounter{AlgoLine}{1487}
\ShowLn \;
\vspace*{3mm}
\setcounter{AlgoLine}{1494}
\ShowLn \; \verb+sentP.broadcastId = broadcastId;+\;
\setcounter{AlgoLine}{1495}
\ShowLn 
\caption{sinalgo/nodes/Node.java}
\label{patch:node-java}
\end{algorithm} 

\FloatBarrier

\section{Experiments: Other Distributions}
\label{app:exp:sec:all-distributions}

In the paper we showed detailed data only for the
random deployment strategy, for other deployments we restricted
ourselves to show only the overall best results. In this section we
present additional data obtained from the experiments described in the
referenced chapter. This data justifies our selection of the
parameters as used to obtain the overall best results.

Let us briefly describe the contents of the tables. In each table
we report the average number of conflicts and the average runtime and
mark the best or the selected combination as bold. In
\cref{ch:expapp:tab:randcolor-all} we consider variants of our
phase-based $(4\Delta)$-coloring algorithm \randvd, in which nodes
simply select a new random color at the end of a phase if a conflict
was detected during the phase. For \colorred\ we need to determine the
parameter \factor, the results for different values are given in
\cref{ch:expapp:tab:cr-factor-all}. For all deployment strategies we
selected the same \factor\ of $0.6$ to be an optimal balance between
number of conflicts and runtime. In
\cref{ch:expapp:tab:colorrand-all-distributions} we consider
\colorred\ and its variant \colorrand, which replaces the valid color
of each node with a random color. We observe that, despite differences
in the average runtime and the number of conflicts, the results are
similar and as for $2\Delta$ available colors good results are
achieved for all deployments. For the algorithms \mwcolor\ and
\yucolor\ we show the results for various values of the parameter
\factor\ in
\cref{ch:expapp:tab:mw-factor-all,ch:expapp:tab:yu-factor-all},
respectively. We select \factor\ $= 0.2$ as optimal value for both
algorithms and all deployment strategies.

The results of our heuristic improvements of \colorrand, \mwcolor, and
\yucolor\ for different values of \duration' are given in
\cref{ch:expapp:tab:correcting-duration-cluster-etc,ch:expapp:tab:correcting-duration-g-pg-r}. In
these heuristics we reduce the number of conflicts by allowing nodes
to reset to certain points in the algorithms once a conflict is
detected. This increases the runtime for the standard \duration,
however, using a decreased \duration' we can decrease both the average
runtime and the average number of conflicts. The values selected as
optimal are marked bold and use a $1/16$ or $1/8$ fraction of
\duration\ as \duration'.


\begin{table}[hbt]
\centering
\caption{Runtime and number of conflicts for \randvd\ and its variants.}
\label{ch:expapp:tab:randcolor-all}
\begin{tabular}{llrr}
\toprule
Deployment & Algorithm & Runtime & Conflicts\\ 
\midrule
\multirow{4}{*}{Cluster}& \randvd & \num{3321}& \num{0.00}\\  
& \randrespect & \num{15639}& \num{0.00}\\  
& \randfinal & \num{16110}& \num{0.00}\\  
& \randed & \num{11822}& \num{0.00}\\  
\midrule
\multirow{4}{*}{Cluster\&Grid}& \randvd & \num{2186}& \num{0.00}\\  
& \randrespect & \num{10172}& \num{0.00}\\  
& \randfinal & \num{10456}& \num{0.00}\\  
& \randed & \num{8211}& \num{0.00}\\  
\midrule
\multirow{4}{*}{Cluster\&PGrid}& \randvd & \num{2016}& \num{0.00}\\  
& \randrespect & \num{9847}& \num{0.00}\\  
& \randfinal & \num{10199}& \num{0.00}\\  
& \randed & \num{6627}& \num{0.00}\\  
\midrule
\multirow{4}{*}{Cluster\&Random}& \randvd & \num{2316}& \num{0.00}\\  
& \randrespect & \num{10175}& \num{0.00}\\  
& \randfinal & \num{10349}& \num{0.00}\\  
& \randed & \num{8153}& \num{0.00}\\  
\midrule
\multirow{4}{*}{Grid}& \randvd & \num{974}& \num{0.00}\\  
& \randrespect & \num{4244}& \num{0.00}\\  
& \randfinal & \num{4354}& \num{0.00}\\  
& \randed & \num{3358}& \num{0.00}\\  
\midrule
\multirow{4}{*}{PGrid}& \randvd & \num{1372}& \num{0.00}\\  
& \randrespect & \num{6114}& \num{0.00}\\  
& \randfinal & \num{6283}& \num{0.00}\\  
& \randed & \num{4548}& \num{0.00}\\  
\midrule
\multirow{4}{*}{Random}& \randvd & \num{1256}& \num{0.00}\\  
& \randrespect & \num{5668}& \num{0.00}\\  
& \randfinal & \num{5865}& \num{0.00}\\  
& \randed & \num{4174}& \num{0.00}\\  
 \bottomrule
\end{tabular}
\end{table}


\begin{table}[hbt]
\centering
\caption{Average number of conflicts and average runtime for \colorred\ using different
  parameters \factor. We report the values for each deployment strategy.}
\label{ch:expapp:tab:cr-factor-all}
\begin{tabular}{lrrrrrrr}
\toprule
\factor      & {\bf \num{0.05}} & {\bf \num{0.1}} & {\bf \num{0.2}} & {\bf \num{0.3}} & {\bf \num{0.4}} & {\bf \num{0.6}}& {\bf \num{0.8}} \\ \midrule
Cluster \\
\;\;\;\; conflicts & \num{0.00}& \num{0.04}& \num{0.02}& \num{0.17}& \num{0.18}& $\mathbf{1.04}$& \num{2.74}\\ 
\;\;\;\; runtime & \num{902421}& \num{468046}& \num{240713}& \num{163781}& \num{126039}& $\mathbf{88094}$& \num{69837}\\ 
Cluster\&Grid\\
\;\;\;\; conflicts & \num{0.00}& \num{0.00}& \num{0.06}& \num{0.17}& \num{0.20}& $\mathbf{1.05}$& \num{3.15}\\ 
\;\;\;\; runtime & \num{590431}& \num{297464}& \num{151579}& \num{103595}& \num{80001}& $\mathbf{55760}$& \num{44176}\\ 
Cluster\&PGrid\\
\;\;\;\; conflicts & \num{0.00}& \num{0.04}& \num{0.02}& \num{0.18}& \num{0.63}& $\mathbf{3.32}$& \num{9.55}\\ 
\;\;\;\; runtime & \num{583200}& \num{294533}& \num{150570}& \num{103126}& \num{78824}& $\mathbf{55157}$& \num{43619}\\ 
Cluster\&Random\\
\;\;\;\; conflicts & \num{0.00}& \num{0.04}& \num{0.04}& \num{0.04}& \num{0.30}& $\mathbf{1.31}$& \num{4.31}\\ 
\;\;\;\; runtime & \num{578623}& \num{292450}& \num{150245}& \num{102685}& \num{78946}& $\mathbf{55005}$& \num{43716}\\ 
Grid\\
\;\;\;\; conflicts & \num{0.00}& \num{0.08}& \num{0.12}& \num{0.11}& \num{0.08}& $\mathbf{0.30}$& \num{1.17}\\ 
\;\;\;\; runtime & \num{272122}& \num{137497}& \num{70855}& \num{48354}& \num{37137}& $\mathbf{25770}$& \num{20258}\\ 
PGrid\\
\;\;\;\; conflicts & \num{0.02}& \num{0.02}& \num{0.00}& \num{0.04}& \num{0.06}& $\mathbf{0.14}$& \num{0.52}\\ 
\;\;\;\; runtime & \num{375157}& \num{190144}& \num{97660}& \num{66541}&\num{51139}& $\mathbf{35624}$& \num{27964}\\ 
Random \\
\;\;\;\; conflicts & \num{0.00}& \num{0.04}& \num{0.10}& \num{0.00}& \num{0.12}& $\mathbf{0.51}$& \num{2.47}\\ 
\;\;\;\; runtime & \num{339013}& \num{171099}& \num{87924}& \num{59995}& \num{46266}& $\mathbf{32224}$& \num{25384}\\ 
 \bottomrule
\end{tabular}
\end{table}

\begin{table}[hbt]
\centering
\caption{Average runtime for \colorred\ and \colorrand\ for colorings
  of different sizes. The runtimes are almost identical although
  \colorrand\ uses only a random color to replace the valid coloring
  used in \colorred.}
\label{ch:expapp:tab:colorrand-all-distributions}
\renewcommand{\arraystretch}{0.85}
\begin{tabular}{llrrrrrr}
\toprule
\multicolumn{2}{l}{{Number of colors}}       & $\Delta+1$ & $2\Delta$ & $3\Delta$ & $4\Delta$ \\ \midrule
Cluster\\
\multirow{2}{*}{\;\;\;\; \colorred} 
& conflicts & \num{4.99}& \num{0.88}& \num{0.88}& \num{0.64}\\ 
& runtime& \num{51476}& \num{67624}   & \num{88101}& \num{108721}\\ 
\multirow{2}{*}{\;\;\;\; \colorrand} 
& conflicts & \num{13.19}& $\mathbf{1.05}$& \num{0.76}& \num{0.67}\\ 
& runtime& \num{53566}& $\mathbf{67881}$   & \num{88529}& \num{109178}\\ 
Cluster\&Grid\\
\multirow{2}{*}{\;\;\;\; \colorred} 
& conflicts & \num{2.56}& \num{1.08}& \num{0.98}& \num{1.31}\\ 
& runtime& \num{30861}& \num{42682}   & \num{55536}& \num{69465}\\ 
\multirow{2}{*}{\;\;\;\; \colorrand} 
& conflicts & \num{5.67}& $\mathbf{1.03}$& \num{0.63}& \num{0.95}\\ 
& runtime& \num{31713}& $\mathbf{42692}$   & \num{55599}& \num{68964}\\ 
Cluster\&PGrid\\
\multirow{2}{*}{\;\;\;\; \colorred} 
& conflicts & \num{6.00}& \num{3.45}& \num{3.32}& \num{3.49}\\ 
& runtime& \num{32024}& \num{42094}   & \num{55148}& \num{68682}\\ 
\multirow{2}{*}{\;\;\;\; \colorrand} 
& conflicts & \num{10.62}& $\mathbf{3.23}$& \num{3.77}& \num{3.67}\\ 
& runtime& \num{32913}& $\mathbf{42385}$   & \num{55091}& \num{68015}\\ 
Cluster\&Random\\
\multirow{2}{*}{\;\;\;\; \colorred} 
& conflicts & \num{2.63}& \num{1.53}& \num{1.34}& \num{1.17}\\ 
& runtime& \num{30337}& \num{42466}   & \num{54884}& \num{68399}\\ 
\multirow{2}{*}{\;\;\;\; \colorrand} 
& conflicts & \num{5.94}& $\mathbf{1.83}$& \num{1.12}& \num{1.24}\\ 
& runtime& \num{30939}& $\mathbf{42034}$   & \num{55113}& \num{68619}\\ 
Grid\\
\multirow{2}{*}{\;\;\;\; \colorred} 
& conflicts & \num{9.22}& \num{0.42}& \num{0.42}& \num{0.20}\\ 
& runtime& \num{16128}& \num{20310}   & \num{25929}& \num{31297}\\ 
\multirow{2}{*}{\;\;\;\; \colorrand} 
& conflicts & \num{31.01}& $\mathbf{0.82}$& \num{0.42}& \num{0.32}\\ 
& runtime& \num{17555}& $\mathbf{20367}$   & \num{25798}& \num{31257}\\ 
PGrid \\
\multirow{2}{*}{\;\;\;\; \colorred} 
& conflicts & \num{1.54}& \num{0.18}& \num{0.10}& \num{0.04}\\ 
& runtime& \num{20367}& \num{27669}   & \num{35683}& \num{43631}\\ 
\multirow{2}{*}{\;\;\;\; \colorrand}  
& conflicts& \num{5.31}& $\mathbf{0.10}$& \num{0.10}& \num{0.15}\\ 
& runtime& \num{20996}& $\mathbf{27817}$   & \num{35682}& \num{43633}\\ 
Random\\
\multirow{2}{*}{\;\;\;\; \colorred} 
& conflicts &\num{3.29}& \num{0.86}& \num{0.55}& \num{0.61}\\ 
& runtime& \num{18638}& \num{24824}   & \num{32197}& \num{39737}\\ 
\multirow{2}{*}{\;\;\;\; \colorrand} 
& conflicts & \num{6.52}& $\mathbf{0.93}$& \num{0.65}& \num{0.63}\\ 
& runtime& \num{19766}& $\mathbf{24758}$ & \num{32287}& \num{39696}\\ 
 \bottomrule
\end{tabular}
\end{table}


\begin{table}[hbt]
\centering
\caption{Average number of conflicts and average runtime for \mwcolor\ using different
  parameters \factor. We report the values for each deployment strategy.}
\label{ch:expapp:tab:mw-factor-all}
\begin{tabular}{llrrrrrr}
\toprule
\multicolumn{2}{l}{{\factor}}      & {\bf \num{0.05}} & {\bf \num{0.1}} & {\bf \num{0.2}} & {\bf \num{0.3}} & {\bf \num{0.4}} & {\bf \num{0.6}} \\ \midrule
\multirow{2}{*}{Cluster} & conflicts & \num{0.14}   & \num{0.24}   & $\mathbf{0.56}$   & \num{1.22}  & \num{2.03} & \num{3.66} \\ 
                            & runtime   & \num{197790} & \num{111416} & $\mathbf{73670}$ & \num{67658} & \num{65895} & \num{64854} \\

\multirow{2}{*}{Cluster\&Grid} & conflicts & \num{0.02}   & \num{0.20}   & $\mathbf{0.34}$   & \num{0.66}  & \num{1.66} & \num{3.19} \\ 
                            & runtime   & \num{130064} & \num{73089} & $\mathbf{47041}$ & \num{42956} & \num{41848} & \num{40854} \\
\multirow{2}{*}{Cluster\&PGrid} & conflicts & \num{0.16}   & \num{0.28}   & $\mathbf{1.06}$   & \num{1.47}  & \num{2.9} & \num{7.89} \\ 
                            & runtime   & \num{129632} & \num{70384} & $\mathbf{46142}$ & \num{41730} & \num{40823} & \num{38883} \\
\multirow{2}{*}{Cluster\&Random} & conflicts & \num{0.18}   & \num{0.16}   & $\mathbf{0.44}$   & \num{0.84}  & \num{1.33} & \num{3.78} \\ 
                            & runtime   & \num{129266} & \num{74177} & $\mathbf{46363}$ & \num{41599} & \num{39944} & \num{39601} \\
\multirow{2}{*}{Grid} & conflicts & \num{0.02}   & \num{0.06}   & $\mathbf{0.26}$   & \num{1.16}  & \num{2.04} & \num{4.1} \\ 
                            & runtime   & \num{76474} & \num{41798} & $\mathbf{25456}$ & \num{20680} & \num{18918} & \num{17221} \\
\multirow{2}{*}{PGrid} & conflicts & \num{0.04}   & \num{0.06}   & $\mathbf{0.28}$   & \num{0.48}  & \num{0.84} & \num{1.14} \\ 
                            & runtime   & \num{95826} & \num{53807} & $\mathbf{32812}$ & \num{27421} & \num{25652} & \num{23765} \\
\multirow{2}{*}{Random} & conflicts & \num{0.10}   & \num{0.12}   & $\mathbf{0.42}$   & \num{1.02}  & \num{1.48} & \num{2.71} \\ 
                            & runtime   & \num{81195} & \num{44700} & $\mathbf{27982}$ & \num{23995} & \num{22807} & \num{21870} \\
 \bottomrule
\end{tabular}
\end{table}


\begin{table}[hbt]
\centering
\caption{Average number of conflicts and average runtime for \yucolor\ using different
  parameters \factor. We report the values for each deployment
  strategy. C=Cluster, G=Grid, PG=PGrid, R=Random}
\label{ch:expapp:tab:yu-factor-all}
\begin{tabular}{llrrrrrr}
\toprule
\multicolumn{2}{l}{{\factor}}      & {\bf \num{0.05}} & {\bf \num{0.1}} & {\bf \num{0.2}} & {\bf \num{0.3}} & {\bf \num{0.4}} & {\bf \num{0.6}} \\ \midrule
\multirow{2}{*}{C} & conflicts & \num{0.42}   & \num{0.57}   & $\mathbf{0.9}$   & \num{1.55}  & \num{3.84} & \num{13.61} \\ 
                            & runtime   & \num{382860} & \num{233591} & $\mathbf{164839}$ & \num{145760} & \num{135567} & \num{137233} \\

\multirow{2}{*}{C\&G} & conflicts & \num{0.5}   & \num{0.52}   & $\mathbf{0.94}$   & \num{2.19}  & \num{4.2} & \num{15.68} \\ 
                            & runtime   & \num{298736} & \num{189883} & $\mathbf{141003}$ & \num{127842} & \num{127151} & \num{120294} \\
\multirow{2}{*}{C\&PG} & conflicts & \num{0.56}   & \num{1.20}   & $\mathbf{1.30}$   & \num{3.56}  & \num{9.03} & \num{36.42} \\ 
                            & runtime   & \num{286326} & \num{175088} & $\mathbf{126488}$ & \num{112753} & \num{107833} & \num{109203} \\
\multirow{2}{*}{C\&R} & conflicts & \num{0.48}   & \num{0.58}   & $\mathbf{0.88}$   & \num{2.52}  & \num{6.05} & \num{23.38} \\ 
                            & runtime   & \num{277447} & \num{172869} & $\mathbf{122267}$ & \num{110815} & \num{104874} & \num{106367} \\

\multirow{2}{*}{Grid} & conflicts & \num{1.04}   & \num{1.06}   & $\mathbf{2.01}$   & \num{3.7}  & \num{6.16} & \num{20.48} \\ 
                            & runtime   & \num{379283} & \num{195925} & $\mathbf{113105}$ & \num{86592} & \num{73141} & \num{60833} \\
\multirow{2}{*}{PGrid} & conflicts & \num{1.00}   & \num{0.96}   & $\mathbf{1.12}$   & \num{1.83}  & \num{3.04} & \num{10.35} \\ 
                            & runtime   & \num{402421} & \num{214349} & $\mathbf{129054}$ & \num{100451} & \num{88521} & \num{76147} \\
\multirow{2}{*}{Random} & conflicts & \num{0.62}   & \num{0.71}   & $\mathbf{1.39}$   & \num{2.90}  & \num{6.41} & \num{22.43} \\ 
                            & runtime   & \num{286167} & \num{160088} & $\mathbf{99946}$ & \num{82707} & \num{72660} & \num{67131} \\
 \bottomrule
\end{tabular}
\end{table}


\begin{table}[hbt]
\centering
\caption{Average number of conflicts and average runtime for the
  correcting variants \colorcor, \mwcor, and \yucor\ for different
  values of \duration'. In this table: Deployments involving the cluster deployment}
\label{ch:expapp:tab:correcting-duration-cluster-etc}
\begin{tabular}{lllrrrrrr}
\toprule
  \multicolumn{3}{l}{Fraction of \duration} & $1/32$  & $1/16$ & $1/8$   & $1/4$   & $1/2$   & $1$     \\ 
  \multicolumn{3}{l}{Resulting \duration'} & $143$  & $287$ & $575$
                                                                         & $1150$   & $2300$   & $4600$     \\ \midrule

\multicolumn{2}{l}{Cluster}\\
\;\;& \multirow{2}{*}{\colorcor} 
& conflicts       & \num{0.00}& \num{0.00}& $\mathbf{0.00}$& \num{0.00}& \num{0.00}& \num{0.00}\\ 	
& & runtime & \num{24425}& \num{17860}& $\mathbf{17802}$& \num{25023}& \num{40185}& \num{68493}\\ 
& \multirow{2}{*}{\mwcor} 
& conflicts & \num{0.63}& \num{0.94}& $\mathbf{0.45}$& \num{0.10}& \num{0.00}& \num{0.00}\\ 
& & runtime & \num{30931}& \num{21637}& $\mathbf{23531}$& \num{33245}& \num{50780}& \num{105200}\\ 
& \multirow{2}{*}{\yucor} 
& conflicts& \num{3.13}& $\mathbf{0.74}$& \num{0.14}& \num{0.02}& \num{0.00}& \num{0.00}\\ 	
& & runtime & \num{12843}& $\mathbf{20849}$& \num{31481}& \num{49630}& \num{92327}& \num{176551}\\ 

\multicolumn{2}{l}{Cluster\&Grid}\\
\;\;& \multirow{2}{*}{\colorcor} 
& conflicts       & \num{0.00}& \num{0.00}& $\mathbf{0.00}$& \num{0.00}& \num{0.02}& \num{0.00}\\ 	
& & runtime & \num{17008}& \num{12473}& $\mathbf{11333}$& \num{15642}& \num{24069}& \num{43139}\\ 
& \multirow{2}{*}{\mwcor} 
& conflicts& \num{0.32}& \num{0.59}& $\mathbf{0.36}$& \num{0.04}& \num{0.00}& \num{0.00}\\ 	
& & runtime & \num{20963}& \num{14189}& $\mathbf{13535}$& \num{18601}& \num{30008}& \num{54896}\\ 
& \multirow{2}{*}{\yucor} 
& conflicts & \num{3.13}& $\mathbf{1.23}$& \num{0.22}& \num{0.02}& \num{0.02}& \num{0.00}\\ 	
& & runtime & \num{7615}& $\mathbf{15835}$& \num{25747}& \num{42543}& \num{78202}& \num{154572}\\ 

\multicolumn{2}{l}{Cluster\&PGrid}\\
\;\;& \multirow{2}{*}{\colorcor} 
& conflicts       & \num{0.00}& $\mathbf{0.02}$& \num{0.02}& \num{0.00}& \num{0.00}& \num{0.00}\\ 	
& & runtime & \num{13609}& $\mathbf{10896}$& \num{11133}& \num{15440}& \num{23863}& \num{42658}\\ 
& \multirow{2}{*}{\mwcor} 
& conflicts & \num{0.38}& $\mathbf{0.42}$& \num{0.41}& \num{0.02}& \num{0.02}& \num{0.00}\\ 	
& & runtime & \num{19031}& $\mathbf{13353}$& \num{13402}& \num{17490}& \num{30973}& \num{55544}\\ 
& \multirow{2}{*}{\yucor} 
& conflicts & \num{3.11}& $\mathbf{0.64}$& \num{0.17}& \num{0.02}& \num{0.00}& \num{0.00}\\ 	
& & runtime & \num{7681}& $\mathbf{14620}$& \num{22982}& \num{38887}& \num{71253}& \num{139299}\\ 

\multicolumn{2}{l}{Cluster\&Random}\\
\;\;& \multirow{2}{*}{\colorcor} 
& conflicts       & \num{0.00}& \num{0.00}& $\mathbf{0.00}$& \num{0.00}& \num{0.00}& \num{0.00}\\ 	
& & runtime & \num{16903}& \num{12789}& $\mathbf{11884}$& \num{16009}& \num{24272}& \num{42773}\\ 
& \multirow{2}{*}{\mwcor} 
& conflicts & \num{0.23}& \num{0.33}& $\mathbf{0.44}$& \num{0.04}& \num{0.00}& \num{0.00}\\ 
& & runtime & \num{22256}& \num{14631}& $\mathbf{12780}$& \num{17341}& \num{28919}& \num{53017}\\ 
& \multirow{2}{*}{\yucor} 
& conflicts & \num{3.74}& $\mathbf{0.81}$& \num{0.20}& \num{0.08}& \num{0.00}& \num{0.00}\\ 
& & runtime & \num{7391}& $\mathbf{15060}$& \num{24173}& \num{38689}& \num{69636}& \num{135511}\\ 
 \bottomrule
\end{tabular}
\end{table}

\begin{table}[hbt]
\centering
\caption{Average number of conflicts and average runtime for the
  correcting variants \colorcor, \mwcor, and \yucor\ for different
  values of \duration'. In this table: Grid, PGrid and Random deployment}
\label{ch:expapp:tab:correcting-duration-g-pg-r}
\begin{tabular}{lllrrrrrr}
\toprule
  \multicolumn{3}{l}{Fraction of \duration} & $1/32$  & $1/16$ & $1/8$   & $1/4$   & $1/2$   & $1$     \\ 
  \multicolumn{3}{l}{Resulting \duration'} & $143$  & $287$ & $575$
                                                                         & $1150$   & $2300$   & $4600$     \\ \midrule

\multicolumn{2}{l}{Grid}\\
\;\;& \multirow{2}{*}{\colorcor} 
& conflicts         & \num{0.00}& \num{0.00}& $\mathbf{0.00}$& \num{0.00}& \num{0.00}& \num{0.00}\\ 
& &  runtime & \num{7538}& \num{5343}& $\mathbf{5113}$& \num{7013}& \num{11839}& \num{20756}\\ 
& \multirow{2}{*}{\mwcor}
& conflicts & \num{0.04}& $\mathbf{0.12}$& \num{0.21}& \num{0.02}& \num{0.02}& \num{0.00}\\ 	
& & runtime & \num{8300}& $\mathbf{5740}$& \num{5741}& \num{8287}& \num{15636}& \num{27958}\\ 
& \multirow{2}{*}{\yucolor }
& conflicts & \num{5.86}& $\mathbf{1.77}$& \num{0.26}& \num{0.09}& \num{0.02}&\num{0.04}\\ 	
& & runtime & \num{3625}& $\mathbf{8654}$& \num{16819}& \num{33089}& \num{68253}& \num{135904}\\ 

\multicolumn{2}{l}{PGrid}\\
& \multirow{2}{*}{\colorcor} 
& conflicts & \num{0.00}& \num{0.00}& $\mathbf{0.00}$& \num{0.00}& \num{0.00}& \num{0.00}\\ 	
&& runtime & \num{10804}& \num{7904}& $\mathbf{7017}$& \num{9638}& \num{15267}&\num{27889}\\ 
& \multirow{2}{*}{\mwcor}
&  conflicts & \num{0.08}& \num{0.12}& $\mathbf{0.12}$& \num{0.06}& \num{0.00}& \num{0.00}\\ 	
& & runtime & \num{11831}& \num{7925}& $\mathbf{7567}$& \num{10558}& \num{18263}& \num{35860}\\ 
& \multirow{2}{*}{\yucor} 
 & conflicts & \num{7.09}& $\mathbf{2.15}$& \num{0.63}& \num{0.14}& \num{0.07}& \num{0.02}\\ 	
& & runtime & \num{4770}& $\mathbf{11479}$& \num{20947}& \num{38641}& \num{77174}& \num{155428}\\ 

\multicolumn{2}{l}{Random}\\
& \multirow{2}{*}{\colorcor} 
& conflicts & \num{0.00}& \num{0.00}& $\mathbf{0.00}$& \num{0.00}& \num{0.00}& \num{0.00}\\ 
& & runtime & \num{8965}& \num{6984}& $\mathbf{6489}$& \num{8883}& \num{14348}& \num{25218}\\ 
& \multirow{2}{*}{\mwcor}
& conflicts & \num{0.13}& \num{0.23}& $\mathbf{0.10}$& \num{0.02}& \num{0.02}& \num{0.00}\\ 	
& & runtime & \num{11065}& \num{7688}& $\mathbf{6834}$& \num{9105}& \num{15762}& \num{31027}\\ 
& \multirow{2}{*}{\yucor}
& conflicts & \num{4.62}& $\mathbf{1.23}$& \num{0.28}& \num{0.09}& \num{0.00}& \num{0.00}\\ 	
& & runtime & \num{4370}& $\mathbf{9807}$& \num{16652}& \num{29635}& \num{58189}& \num{116646}\\ 
 \bottomrule
\end{tabular}
\end{table}


\end{document}